\documentclass[12pt, a4paper]{article}

\usepackage{amsmath, amssymb, amsthm}
\usepackage{geometry}
\usepackage{setspace}
\usepackage{indentfirst}
\usepackage{graphicx}
\usepackage[round,abbr]{harvard}
\citationstyle{apsr}
\usepackage{tikz}
\usepackage{xcolor}
\usepackage{booktabs}
\usepackage{enumitem}
\usepackage{threeparttable}
\usepackage{longtable}
\usepackage{xltabular}
\usepackage{float} 
\usepackage{placeins}
\usepackage{needspace}
\usepackage{longtable}
\usetikzlibrary{arrows.meta,calc,positioning}
\usepackage[hidelinks]{hyperref}
\title{\textbf{Delayed Retirement, Intergenerational Transfers, and Fertility}}
\author{
Yi Fan\thanks{Business School, National University of Singapore, Singapore.}
\and
Shuangshuang Song\thanks{College of Public Finance and Investment, Shanghai University of Finance and Economics, China.}
\and
Yewen Yu\thanks{School of Economics, Beijing Technology and Business University, China.}
}
\date{\today}

\begin{document}

\maketitle

\begin{abstract}
Population aging and fertility decline are occurring simultaneously worldwide, straining labor supply and pension systems. China began a gradual delayed-retirement (GDR) reform in 2025, raising statutory retirement ages on different schedules for men, female cadres, and female workers across birth cohorts. To evaluate its effects on older individuals and within the family, we develop a model of older adults' retirement, intergenerational transfers, and adult children's fertility decisions, with parameters estimated via fuzzy regression discontinuity at the pre-2025 statutory retirement ages. Counterfactual simulations show that, for mature reform cohorts, the GDR reform increases older employment and monetary transfers to adult children but reduces grandparental childcare. Because the reduction in childcare outweighs the increase in monetary transfers, fertility and older adults’ welfare decline, with particularly large losses among lower-income women. Fully flexible retirement, mandatory partial retirement, and mandatory full-time work do not raise employment without similarly reducing fertility and welfare. Childcare subsidies can offset the declines in fertility and welfare, while restricting eligibility to women or lower-income women requires higher subsidy rates but reduces average fiscal expenditure. These findings suggest that evaluations of retirement policy should consider not only its direct effects on older adults’ labor supply, but also its fertility effects through intergenerational transfers; childcare support can help mitigate these negative intergenerational effects of delayed retirement.

\end{abstract}




\section{Introduction}
\label{sec:introduction}

Population aging and declining fertility are reducing the relative size of the working-age population and increasing pressure on public pension finances in many countries \cite{vangroezenSocialSecurityEndogenous2003}\footnote{According to the United Nations, the number of people aged 65 years or older worldwide is projected to increase from 761 million in 2021 to approximately 1.6 billion in 2050, while the global total fertility rate fell from 3.31 births per woman in 1990 to 2.25 in 2024, and more than half of all countries and areas now have fertility below the replacement level of 2.1. See United Nations, \textit{World Social Report 2023: Leaving No One Behind in an Ageing World}, \url{https://www.un.org/en/node/193107}; and United Nations Department of Economic and Social Affairs, Population Division, \textit{World Population Prospects 2024: Summary of Results}, \url{https://desapublications.un.org/publications/world-population-prospects-2024-summary-results}.}. Raising statutory retirement ages and adjusting pension eligibility conditions are important policy responses to these demographic changes\footnote{The United States has gradually raised the Social Security normal retirement age; several European countries have raised statutory retirement ages, increased penalty-free retirement ages, or linked early-retirement conditions to life expectancy; Japan requires employers to secure employment opportunities for Senior workers up to age 65 and further encourages employment opportunities up to age 70; Korea requires employers to set the retirement age at 60 or older while gradually raising the pensionable age under the National Pension Scheme; and Singapore operates a retirement and re-employment system under which eligible employees can continue working beyond the statutory retirement age. See U.S. Social Security Administration, Normal Retirement Age, \url{https://www.ssa.gov/oact/progdata/nra.html}; OECD, \textit{Pensions at a Glance 2025}, \url{https://www.oecd.org/en/publications/pensions-at-a-glance-2025_e40274c1-en.html}; Japan Ministry of Health, Labour and Welfare, \textit{Outline of Measures for Employment of the Elderly}, \url{https://www.mhlw.go.jp/content/10500000/001086776.pdf}; Korea National Pension Service, Old-age Pension, \url{https://www.nps.or.kr/eng/ntnlpnsplan/bnft/getOHAI0014M0.do}; and Singapore Ministry of Manpower, Responsible Re-employment, \url{https://www.mom.gov.sg/employment-practices/re-employment}.}. Research on retirement age and pension reforms has extensively examined retirement decisions \cite{frenchEffectsHealthWealth2005,mastrobuoniLaborSupply2009,seibold2021reference}, household saving \cite{attanasioRohwedder2003,attanasioBrugiavini2003,garciaMirallesLeganza2024,heNingZhu2026}, consumption in retirement \cite{kolsrud2024retirement}, and public pension expenditure and related fiscal consequences \cite{staubli_zweimuller_2013,laliveMagesanStaubli2023}. When older parents also contribute to resource allocation within their adult children's families, however, delayed retirement can affect not only their own labor supply but also their adult children's fertility through intergenerational transfers. These fertility consequences have received comparatively limited attention.

China, the world's largest developing country, provides an important social and institutional setting for studying this question. More than 40\% of Chinese Seniors provide grandparental childcare, and most serve as primary daily caregivers \cite{xieAssociationGrandchildCare2024,yuFertilityPolicyFamilyPension2021,daiImpactDelayedRetirement2025}. This participation rate is substantially higher than in the United States and most European countries, including Sweden and Germany \cite{aparicioFenollWorkingWomen2015,garciaMoranStringsAttached2017}. Together with the limited availability of formal public childcare,\footnote{ See \textit{Ensuring access to affordable, reliable childcare for public benefit}, August 2, 2024, published on the website of the State Council of the People's Republic of China: \url{https://english.www.gov.cn/news/202408/02/content_WS66ac9240c6d0868f4e8e9b2c.html}.} this extensive involvement in childcare makes Seniors' time availability particularly important for household fertility decisions. China also has broad basic pension insurance coverage, with a participation rate exceeding 95\% reported in 2025, and employee pension eligibility is linked to the statutory retirement age.\footnote{ The Ministry of Human Resources and Social Security reported a basic pension insurance participation rate above 95\% in 2025; see \url{https://chinajob.mohrss.gov.cn/h5/c/2025-10-09/453709.shtml}. This figure covers basic pension insurance as a whole, including employee and resident schemes, and should not be interpreted as the share of the population subject to employee statutory retirement rules.} From 1978 through 2024, the statutory retirement age was generally 60 for men, 55 for female cadres (white collar), and 50 for female workers (blue-collar). On January 1, 2025, China began implementing its gradual delayed retirement (GDR) reform. Over a 15-year transition period, statutory retirement ages rise across birth cohorts from 60 to 63 for men, from 55 to 58 for female cadres, and from 50 to 55 for female workers. The reform also introduces flexible retirement: workers who meet the minimum contribution requirement may retire up to three years before the new statutory retirement age, but not before the pre-reform statutory retirement age; with their employer's agreement, they may also delay retirement by up to three years beyond the new statutory retirement age.\footnote{ See \textit{Decision of the Standing Committee of the National People's Congress on Implementing a Gradual Increase in the Statutory Retirement Age} and \textit{Interim Measures for Implementing Flexible Retirement}. \url{https://chinajob.mohrss.gov.cn/h5/c/2025-01-03/422128.shtml}.} Whether actual retirement occurs later therefore depends on worker category, birth cohort, and the flexible retirement provisions, allowing the effects of delayed retirement to vary across groups.

We evaluate the effects of China's 2025 GDR reform on older adults' labor force participation, annual deferred pension payments (DPP), and welfare, as well as its consequences for adult children's fertility through intergenerational transfers. We first develop a family decision model comprising an older parent (the Senior), an adult child (the Junior), and grandchildren. The model jointly incorporates retirement, intergenerational monetary transfers, grandparental childcare, and Junior fertility. Retirement releases time for childcare but also changes monetary support through the gap between labor earnings and pension income; its net effect on fertility depends on the relative importance of the two transfers. We then use the 2011, 2013, 2015, and 2018 CHARLS waves to implement a fuzzy regression discontinuity(RD) design . Discontinuities in retirement probability at the pre-reform statutory retirement ages identify the local average treatment effect (LATE) of retirement on the number of grandchildren, intergenerational monetary transfers, and grandparental childcare hours. Combining this evidence with supplementary information from the China Family Panel Studies (CFPS), we estimate the model parameters and conduct counterfactual analyses of retirement policies and childcare subsidies.

The RD estimates show that retirement increases the number of grandchildren by 0.613, reduces annual net monetary transfers to adult children by approximately RMB 4177, and increases annual grandparental childcare by approximately 786 hours. These results are consistent with the model's prediction when time transfers dominate: additional childcare may offset the effect of reduced monetary support on fertility, yielding a positive net fertility response. Subgroup estimates are positive and statistically significant for women and lower-income older adults, whereas the estimates for men and the higher-income group are not statistically significant.

The counterfactual analysis shows that the employment gains and family consequences of GDR depend on the constraints on retirement choice. For cohorts reaching the pre-reform statutory retirement age in 2025--2030, flexible retirement still permits retirement at that age. The simulated Senior labor force participation rate is 2.8\%, with only small increases in the average number of grandchildren and mean Senior welfare. For cohorts reaching the pre-reform statutory retirement age in 2031--2039, women previously subject to age 50 face a higher minimum retirement age. Senior labor force participation rises to 41.8\%, and annual DPP reach RMB 10221 per Senior.\footnote{ As defined in Section~\ref{sec:institution_gradual}, cohorts are indexed by the year in which a Senior reaches the applicable pre-reform statutory retirement age. Counterfactual outcomes are evaluated at that age relative to the pre-reform rules. DPP measure the annual amount of pension payments deferred at the evaluation age. Monetary amounts are expressed in 2018 RMB.} Additional monetary support does not cover the market replacement cost of reduced grandparental childcare: the average number of grandchildren falls by 0.043, and mean Senior welfare declines by 1.578\%. These losses are concentrated among female and lower-income Seniors and their adult children.

We next compare three alternative retirement regimes to assess whether changes in retirement choice and work intensity can increase Senior employment and DPP without reducing Senior welfare or Junior fertility: fully flexible retirement (FFR), mandatory partial retirement (MPR), and mandatory full-time work (MFT). FFR allows Seniors to choose between retirement and continued work at the pre-reform statutory retirement age. It therefore avoids the fertility and welfare losses generated by GDR for the 2031--2039 cohorts, but produces only a small increase in labor force participation. MPR and MFT generate larger increases in participation and DPP, together with larger fertility and welfare losses. The effects of partial retirement depend on both working hours and pension eligibility: when pension receipt is deferred until full retirement, the additional childcare made possible by shorter working hours does not offset the reduction in monetary support. None of the three alternatives eliminates the declines in Junior fertility and Senior welfare while maintaining the employment and DPP gains generated by GDR for the later cohorts. Adjusting retirement choice and work intensity alone thus appears insufficient to achieve these objectives jointly, motivating complementary policies that address families' childcare costs.

We therefore examine whether childcare subsidies can mitigate the fertility and welfare losses caused by GDR by lowering the cost of substituting formal childcare for grandparental childcare. As the subsidy rate rises, Senior welfare and the number of grandchildren increase, while grandparental childcare declines further. This pattern reflects improved family resource allocation as formal childcare becomes a less costly substitute for grandparental childcare. For the 2031--2039 cohorts, a universal childcare subsidy of approximately 5.7\% eliminates the negative effect of GDR on average Junior fertility, while a subsidy of approximately 12.2\% eliminates the loss in mean Senior welfare caused by GDR. The corresponding average annual fiscal expenditures are RMB 836 and RMB 1905 per Senior, respectively, without weakening the reform's effect on Senior labor force participation.\footnote{ The fertility target is the full-sample average number of grandchildren; the welfare target is the full-sample mean Senior welfare index. Both are evaluated relative to the pre-reform benchmark. The baseline fiscal calculation assumes that formal childcare meets one half of childcare demand remaining after grandparental childcare, with the Junior providing the other half. Expenditure is averaged over the full sample. Section~\ref{sec:sensitivity} examines alternative assumptions.}

Given the concentration of GDR-induced fertility and welfare losses among female and lower-income Seniors, we compare universal subsidies with two targeted schemes: one covers households with female Seniors, and the other covers households with lower-income female Seniors. Restricting eligibility to either group raises the subsidy rate required to eliminate the negative effect of GDR on full-sample average Junior fertility or mean Senior welfare, but lowers the corresponding average fiscal expenditure. Subsidy design therefore involves not only the subsidy rate but also coverage and the specific outcome to be compensated.

We also conduct robustness checks of the RD estimates and sensitivity analyses of parameter calibration and other model assumptions. The results support the robustness of our main conclusions.

This paper makes three contributions. First, we add evidence on the fertility consequences of delayed retirement. Research on higher retirement ages and later pension eligibility has extensively examined older workers' labor supply, household saving, and fiscal consequences \cite{gustmanSocialSecurityEarly2005,mastrobuoniLaborSupply2009,staubli_zweimuller_2013,laliveMagesanStaubli2023,garciaMirallesLeganza2024}, and some studies have begun to examine the effects on adult children's fertility \cite{ilciukasFertilityParentalRetirement2023,akyol_atalay_2025}. We combine microeconomic evidence on how retirement affects intergenerational transfers and the number of grandchildren to quantify the fertility consequences of China's 2025 GDR reform. Our findings reveal a potential trade-off between increasing older adults' labor force participation and deferring pension payments, on the one hand, and avoiding a decline in fertility, on the other. Evaluations of delayed retirement should therefore consider not only current older-worker labor supply and pension payments but also fertility as a determinant of future labor supply. If a reform-induced decline in fertility further reduces the future working-age population and the pension contribution base, omitting this intergenerational consequence may overstate the reform's contribution to long-run labor supply and its ability to ease pressure on public pension finances.

Second, we contribute to research on childcare policy and subsidy design by examining whether childcare support can mitigate the family losses caused by delayed retirement. Existing studies examine the effects of childcare provision and subsidies on maternal labor supply \cite{LefebvreMerrigan2008ChildCarePolicy,havnesMoneyNothingUniversal2011,bauernschusterPublicChildCare2015,duletzkiCanEarlyPublic2026} and family welfare \cite{bakerGruberMilligan2008,reviewGuner2020}, their effects on fertility \cite{bauernschuster_hener_rainer_2016}, and the relationship between childcare demand, supply, and policy design \cite{blauDemandQualityChild1998,hoEfficientChildCare2020,berlinskiChildcareMarketsParental2024a}. In China, where formal childcare provision for children under age three remains limited,\footnote{ See \textit{Report of the State Council on Progress in Developing Childcare Services}, September 10, 2024, \url{https://www.npc.gov.cn/npc/c2/c30834/202409/t20240911_439363.html}.} we show how childcare subsidies can complement delayed retirement. Lower formal childcare prices can mitigate the declines in Junior fertility and Senior welfare caused by GDR without reducing its employment gains. We then compare the subsidy rates and fiscal expenditures required under different compensation objectives and eligibility rules. These comparisons can inform childcare support in other economies facing both population aging and limited formal childcare provision for young children.

Third, we contribute to research on retirement and intergenerational transfers by clarifying the relative roles of monetary and time transfers in fertility decisions and the conditions that determine them. Previous research incorporates both types of transfers in a common framework \cite{cardiaNgIntergenerationalTime2003} and examines how grandparental childcare affects younger families' decisions \cite{garciaMoranStringsAttached2017,talamasmarcos2023grandmothers}. We build on this work by showing how the formal childcare price, pre-retirement labor income, and the pension replacement rate jointly determine the fertility effect of retirement. Holding other factors constant, a higher formal childcare price raises the market replacement value of the time released by retirement, while lower pre-retirement labor income or a higher pension replacement rate reduces the associated income loss. Under these conditions, additional time support is more likely to outweigh reduced monetary support, making retirement more likely to increase Junior fertility.

The remainder of the paper is organized as follows. Section~\ref{sec:institution} describes the institutional background. Sections~\ref{sec:Theoretical Framework} and~\ref{sec:econometric_framework} develop the theoretical and econometric frameworks. Section~\ref{sec:data_sample} describes the data and sample, and Section~\ref{sec:RD_results} presents the regression discontinuity results. Section~\ref{sec:identification_estimation} discusses the identification and estimation of the structural parameters. Section~\ref{sec:counterfactual} evaluates retirement policies and childcare subsidies. Section~\ref{sec:sensitivity} reports sensitivity analyses, and Section~\ref{sec:conclusion} concludes.

\section{Institutional Background}
\label{sec:institution}

\subsection{1978--2024: Mandatory Statutory Retirement}

\begin{figure}[H]
\centering
\includegraphics[width=0.72\textwidth]{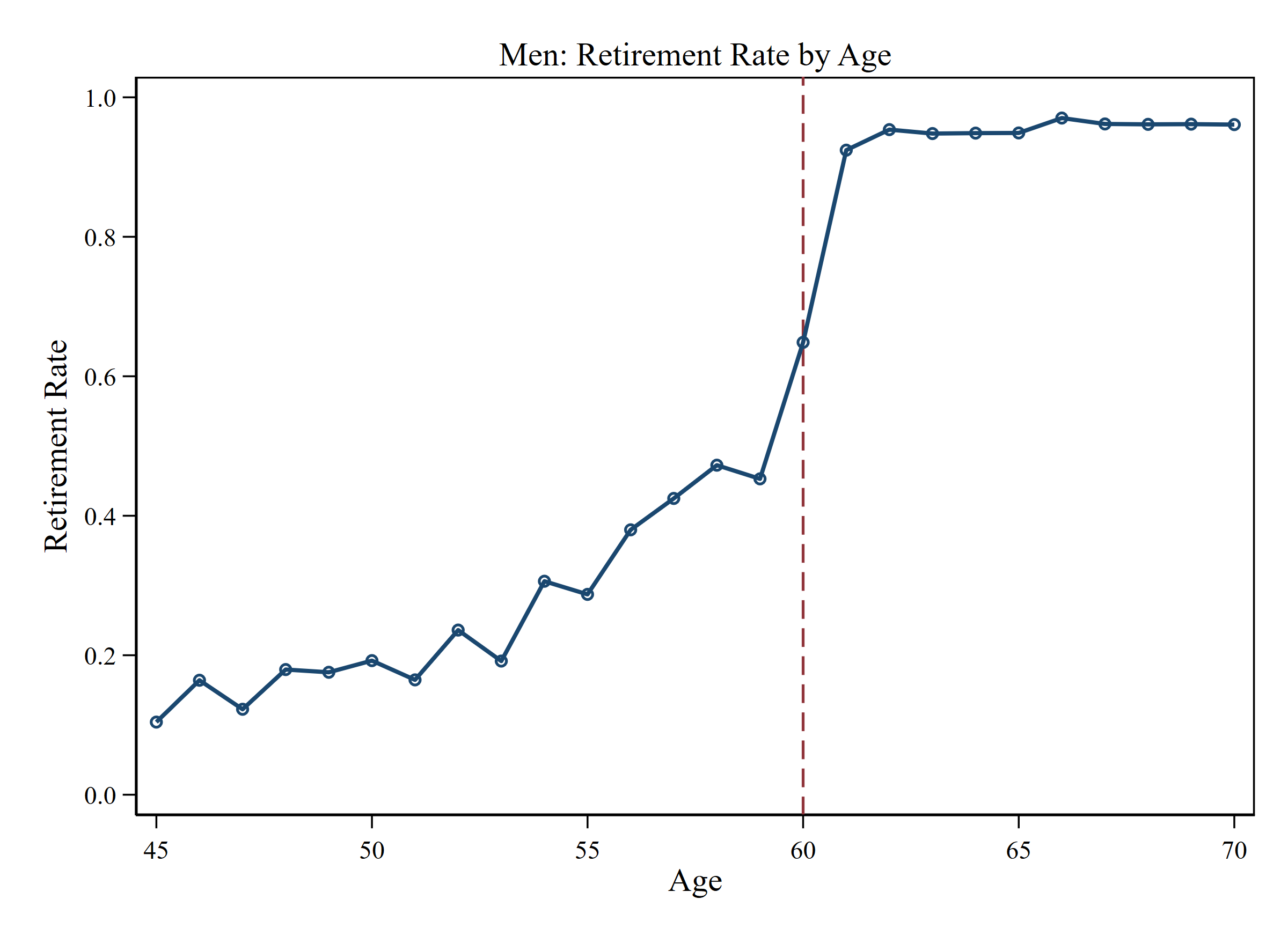}
\caption{Retirement Rate by Age for Men under the Mandatory Retirement Regime}
\label{fig:mandatory_retirement_men}
\begin{minipage}{0.85\textwidth}
\footnotesize
\textit{Notes:} The figure plots the average retirement rate by age for men. The dashed vertical line marks age 60, the statutory retirement age for men under the pre-2025 mandatory retirement system. The figure is constructed using data from the 2011, 2013, 2015, and 2018 waves of the CHARLS.
\end{minipage}
\end{figure}

\begin{figure}[H]
\centering
\includegraphics[width=0.72\textwidth]{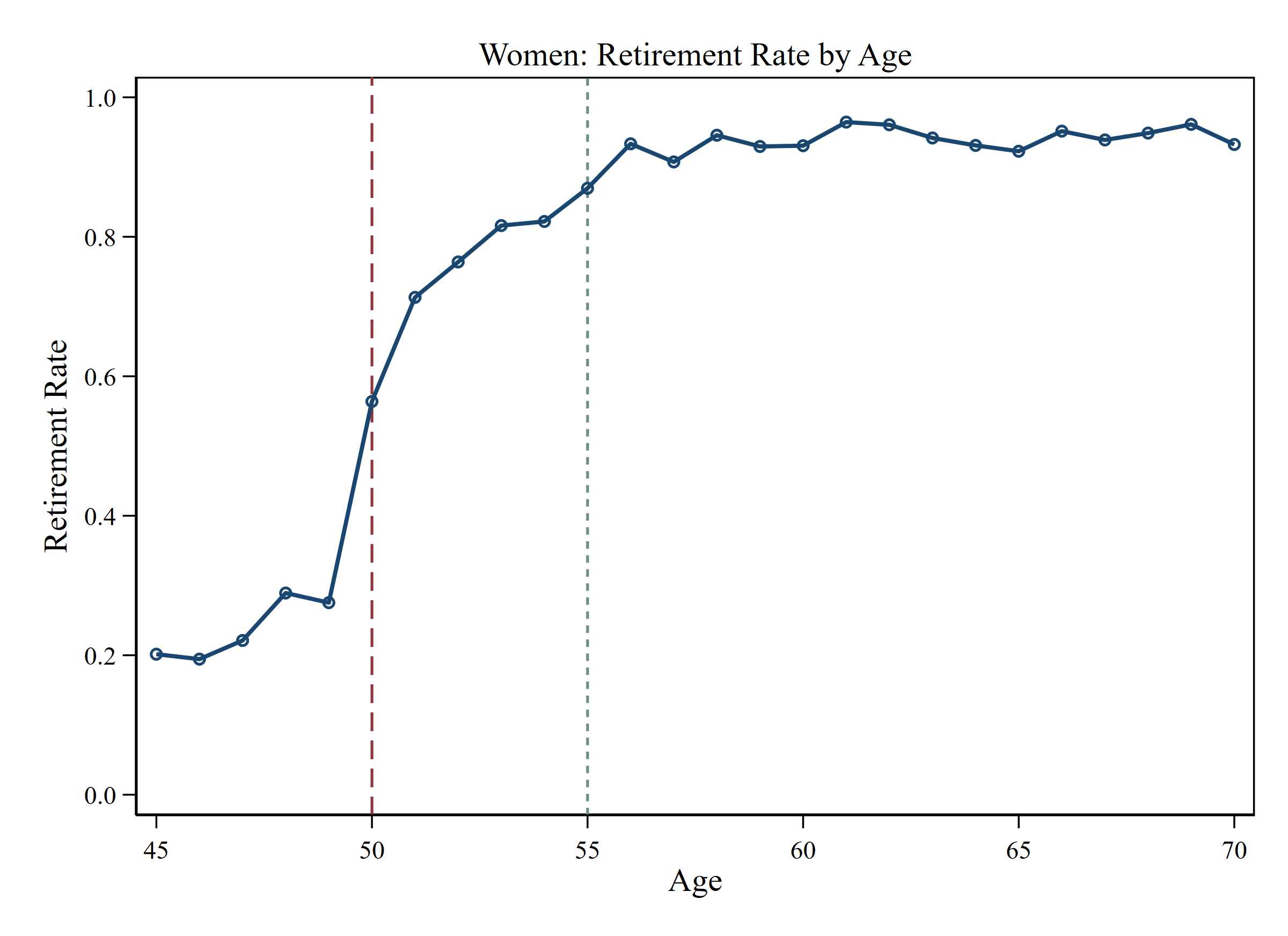}
\caption{Retirement Rate by Age for Women under the Mandatory Retirement Regime}
\label{fig:mandatory_retirement_women}
\begin{minipage}{0.85\textwidth}
\footnotesize
\textit{Notes:} The figure plots the average retirement rate by age for women. The dashed vertical lines mark ages 50 and 55, corresponding to the statutory retirement ages for female workers and female cadres, respectively, under the pre-2025 mandatory retirement system. The figure is constructed using data from the 2011, 2013, 2015, and 2018 waves of the CHARLS.
\end{minipage}
\end{figure}

From 1978 through 2024, statutory retirement ages governed retirement in China. The statutory retirement age was generally 60 for men, 55 for female cadres, and 50 for female workers. Reaching the applicable statutory age was closely associated with retirement eligibility and pension receipt, and formal-sector workers typically retired at the applicable threshold. Figure~\ref{fig:mandatory_retirement_men} shows a sharp increase in the retirement rate around age 60 for men. For women, Figure~\ref{fig:mandatory_retirement_women} shows a sharp increase around age 50 but no comparable change around age 55. This pattern may reflect the relatively small share of women subject to the age-55 rule: approximately 93\% of women are subject to the retirement age of 50 in practice \cite{FangFeng2018}.

The statutory retirement ages strongly constrained retirement timing and generated discontinuities in retirement probabilities. We use this variation in a fuzzy RD design to estimate the LATE of retirement on intergenerational transfers and the number of grandchildren.

\subsection{Since 2025: GDR Reform}
\label{sec:institution_gradual}

China began implementing its GDR reform on January 1, 2025. Unlike Greece's 2012 pension reform, which raised the general eligibility age for a full pension directly from 65 to 67 on January 1, 2013,\footnote{European Commission (2017), \textit{The 2018 Ageing Report: Underlying Assumptions and Projection Methodologies}, European Economy Institutional Paper 065, p.~40.} China's reform phases in higher statutory retirement ages across birth cohorts over the 15-year transition period from 2025 to 2039. For men and for women cadres, retirement ages increase by one month across successive four-month birth cohorts, rising from 60 to 63 and from 55 to 58, respectively. For women workers, the retirement age increases by one month across successive two-month birth cohorts, rising from 50 to 55. Appendix Tables~\ref{tab:official_schedule_men}--\ref{tab:official_schedule_women50} report the complete official monthly retirement-age schedules.

The reform also introduces flexible retirement provisions. Workers covered by these provisions who meet the minimum contribution requirement may voluntarily retire up to three years before the new statutory retirement age, but not before the pre-reform statutory age. They may also postpone retirement by up to three years beyond the new statutory age by mutual agreement with their employer. Let $a_i^{\mathrm{old}}$ denote individual $i$'s pre-reform statutory retirement age and $a_{it}^{\mathrm{new}}$ the new statutory retirement age applicable to the individual's birth cohort, where $t$ denotes the year in which the individual reaches the pre-reform statutory retirement age. These rules imply the retirement-age interval
\[
[\underline a_{it},\overline a_{it}]
=
\left[
\max\{a_i^{\mathrm{old}},a_{it}^{\mathrm{new}}-3\},
a_{it}^{\mathrm{new}}+3
\right],
\]
where $\underline a_{it}$ and $\overline a_{it}$ denote the earliest retirement age and the latest retirement age permitted under the policy, respectively.

\begin{figure}[H]
\centering
\includegraphics[width=0.98\textwidth]{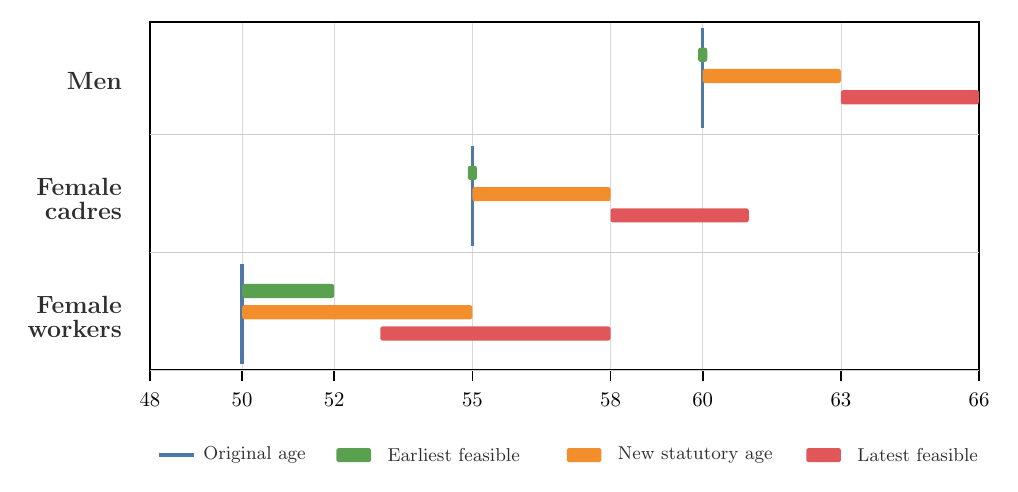}
\caption{China's GDR Reform: Feasible Retirement Windows by Worker Type}
\label{fig:gradual_retirement_policy}
\begin{minipage}{0.85\textwidth}
\footnotesize
\textit{Notes:} The figure illustrates the feasible retirement age window under the 2025 reform for different types of workers. The blue marker indicates the original statutory retirement age; the green segment indicates the earliest feasible retirement age; the orange segment indicates the new statutory retirement age; and the red segment indicates the latest feasible retirement age. Under the reform, individuals may retire up to three years earlier than the new statutory retirement age, but not earlier than the original statutory retirement age. After reaching the new statutory retirement age, retirement may be postponed by up to three years with employer consent. Because the increase in the statutory retirement age for men and female cadres does not exceed three years, their earliest feasible retirement age coincides with the original statutory retirement age throughout the transition. In contrast, for female workers, whose statutory retirement age rises from 50 to 55, the earliest feasible retirement age eventually exceeds 50 for later birth cohorts.
\end{minipage}
\end{figure}

These rules imply different retirement windows for female cadres, female workers, and men. As shown in Figure~\ref{fig:gradual_retirement_policy}, the earliest retirement age, $\underline a_{it}$, remains 60 for men and 55 for women whose pre-reform statutory retirement age was 55. Women whose pre-reform statutory retirement age was 50 face a different retirement constraint. Those reaching age 50 between January 2025 and December 2030 may still choose to retire at age 50. For those reaching age 50 from January 2031 onward, the earliest retirement age exceeds 50 and rises across successive cohorts until the new statutory retirement age reaches 55. Thereafter, the earliest retirement age remains at 52 under the three-year early-retirement provision. Accordingly, the effects of the GDR reform may differ by gender and birth cohort.

\section{Theoretical Framework}
\label{sec:Theoretical Framework}

The theoretical framework characterizes how retirement affects intergenerational monetary transfers, grandparental childcare, and Junior fertility through changes in Senior income and non-work time. It also examines the implications of retirement for Senior welfare.


\subsection{Model Setup}
Consider an economy populated by overlapping families, each comprising a Senior (denoted by $s$), a Junior (denoted by $j$), and potentially grandchildren (Baby). The family structure and transfer channels build on existing models of intergenerational support \cite{reviewRupert2018}. Both generations derive utility from consumption $c$ and leisure $l$. We assume altruistic preferences \cite{beckerReformulationEconomicTheory1988,barroBecker1989,cordobaRipollIntergenerational2019}: the Junior values the number of children, denoted by $n_j$, while the Senior values own consumption and leisure as well as the Junior's welfare. Utility is strictly increasing and concave, with separable logarithmic forms:

\begin{equation}
U_s=\ln c_s+\alpha_s \ln l_s+\rho U_j,
\label{eq:elderly_utility}
\end{equation}
and
\begin{equation}
U_j=\ln c_j+\alpha_j \ln l_j+\beta_j \ln n_j,
\end{equation}
where $c_s>0$ and $c_j>0$ denote the consumption of the Senior and the Junior, respectively; $l_s>0$ and $l_j>0$ denote leisure; $\alpha_s>0$ and $\alpha_j>0$ measure preferences for leisure; $\beta_j>0$ measures the Junior's preference for children; and $\rho>0$ is the Senior's intergenerational altruism parameter.

The Senior is endowed with one unit of time and allocates it according to retirement status, denoted by $R\in\{0,1\}$. If not retired $(R=0)$, the Senior supplies a fixed amount of labor $\bar h_s$ and earns labor income $w_s\bar h_s$. The remaining time is allocated between leisure, $l_s$, and grandparental childcare, $b_s$. If retired $(R=1)$, the Senior withdraws from the labor market and receives pension income $P_s$, where $0<P_s<w_s\bar h_s$. The entire time endowment is then available for leisure and grandparental childcare. The Senior allocates total income between own consumption, $c_s$, and monetary transfers to the Junior, $m_s$. The Senior's budget constraint is therefore
\begin{equation}
c_s+m_s=w_s\bar h_s(1-R)+P_sR \equiv I_s(R).
\end{equation}
The Senior's time constraint is
\begin{equation}
l_s+b_s=1-\bar h_s(1-R)\equiv \Omega_s(R).
\end{equation}
Here, $I_s(R)$ denotes the Senior's income, and $\Omega_s(R)$ denotes the Senior's non-work time available for leisure and grandparental childcare. Both depend on retirement status.

The Junior is endowed with one unit of time, allocated among market work $h_j$, leisure $l_j$, and own childcare $b_j$. The time constraint is therefore
\begin{equation}
h_j+l_j+b_j=1.
\end{equation}
The Junior earns labor income $w_jh_j$ and receives monetary transfers $m_s$ from the Senior. Each of the Junior's $n_j$ children requires a fixed amount of childcare time $\bar h$, giving a total childcare requirement of $\bar h n_j$. Three sources can meet this requirement: the Junior's own childcare time $b_j$, the Senior's grandparental childcare time $b_s$, and formal childcare purchased at a constant price $p$ per unit of time, given by $f_j=\bar h n_j-b_j-b_s$. The Junior allocates total monetary resources, comprising labor income and transfers from the Senior, between own consumption $c_j$ and formal childcare expenditure $pf_j$. The Junior's budget constraint is therefore
\begin{equation}
c_j+pf_j=w_jh_j+m_s.
\end{equation}

The Senior chooses monetary transfers $m_s$ and grandparental childcare time $b_s$ to maximize utility. Equations (1), (3), and (4) yield the following first-order conditions with respect to $m_s$ and $b_s$:
\begin{equation}
-\frac{1}{I_s(R)-m_s}+\rho\frac{1}{c_j}=0,
\end{equation}
and
\begin{equation}
-\frac{\alpha_s}{\Omega_s(R)-b_s}+\rho\frac{p}{c_j}=0.
\end{equation}
The first condition equates the marginal utility cost of reducing the Senior's consumption with the altruistic marginal benefit of increasing the Junior's consumption. The second equates the marginal utility cost of reducing the Senior's leisure with the altruistic marginal benefit of reducing the Junior's formal childcare expenditure.

The Junior chooses market work $h_j$, own childcare time $b_j$, and the number of children $n_j$. Equations (2), (5), and (6) yield the following first-order conditions with respect to $h_j$, $b_j$, and $n_j$:
\begin{equation}
\frac{w_j}{c_j}-\frac{\alpha_j}{l_j}=0,
\end{equation}
\begin{equation}
\frac{p}{c_j}-\frac{\alpha_j}{l_j}=0,
\end{equation}
and
\begin{equation}
\frac{\beta_j}{n_j}-\frac{p\bar h}{c_j}=0.
\end{equation}
The first condition equates the marginal consumption benefit of market work with the marginal utility cost of forgone leisure. The second equates the marginal benefit of saving on formal childcare through own care with the marginal utility cost of forgone leisure. The third equates the marginal utility of an additional child with the marginal consumption cost of the childcare required.

Equations (9) and (10), together with the Karush-Kuhn-Tucker (KKT) conditions, imply that the Junior's optimal allocation of time between market work and childcare typically yields a corner solution. When $w_j \ge p$, the return to market work weakly exceeds the savings from own childcare, so the Junior specializes in market work, with $b_j = 0$ and $h_j > 0$. Conversely, when $w_j < p$, own childcare saves formal childcare expenditure at an implicit return above the market wage, yielding a corner solution with $b_j > 0$ and $h_j = 0$.

Combining the first-order conditions, the time and budget constraints, and the corner-solution conditions yields the Senior's and the Junior's optimal choices:
\begin{equation}
c_j^\ast(R)=
\frac{\max\{p,w_j\}+I_s(R)+p\Omega_s(R)}
{1+\alpha_j+\beta_j+(1+\alpha_s)/\rho}=\frac{Y_j+I_s(R)+p\Omega_s(R)}{\Lambda}.
\label{eq:junior_consumption_optimal}
\end{equation}

\begin{equation}
h_j^\ast(R)
=
\left(1-\frac{\alpha_j}{w_j}c_j^\ast(R)\right)
\mathbb{I}\{w_j\ge p\}.
\end{equation}

\begin{equation}
b_j^\ast(R)
=
\left(1-\frac{\alpha_j}{p}c_j^\ast(R)\right)
\mathbb{I}\{w_j<p\}.
\end{equation}

\begin{equation}
l_j^\ast(R)=\frac{\alpha_j}{Y_j}c_j^\ast(R),
\label{eq:junior_leisure_optimal}
\end{equation}

\begin{equation}
n^\ast(R)=\frac{\beta_j}{p\bar h}c_j^\ast(R),
\label{eq:reaction_N}
\end{equation}

\begin{equation}
m_s^\ast(R)=I_s(R)-\frac{c_j^\ast(R)}{\rho},
\end{equation}

\begin{equation}
b_s^\ast(R)=\Omega_s(R)-\frac{\alpha_s c_j^\ast(R)}{\rho p}.
\end{equation}

\subsection{Retirement, Transfers, and Fertility}

Define the Senior's effective transferable resources as
 $T_s(R)\equiv I_s(R)+p\Omega_s(R)$. When the Senior retires, $T_s(1)=P_s+p$. When the Senior does not retire, $T_s(0)=w_s\bar h_s+p(1-\bar h_s)$. Retirement changes Senior income by $\Delta I_s=P_s-w_s\bar h_s$. The effect of retirement on the Senior's effective transferable resources is therefore $\Delta T_s
=
T_s(1)-T_s(0)
=
p\bar h_s-\left(w_s\bar h_s-P_s\right)$. Equation (12) then gives the effect of the Senior's retirement on the Junior's consumption:
\begin{equation}
\Delta c_j^\ast
=
c_j^\ast(1)-c_j^\ast(0)
=
\frac{p\bar h_s-\left(w_s\bar h_s-P_s\right)}{\Lambda}.
\end{equation}

Retirement also changes the composition of the Senior's intergenerational transfers. Its effect on monetary transfers is
\begin{equation}
\Delta m^\ast
=
-\left(w_s\bar h_s-P_s\right)
-
\frac{p\bar h_s-\left(w_s\bar h_s-P_s\right)}{\rho\Lambda}=
\frac{1}{\rho\Lambda}
\left[
\left(w_s\bar h_s-P_s\right)(1-\rho\Lambda)-p\bar h_s
\right].
\label{eq:theory_delta_m}
\end{equation}
Since $\rho\Lambda=\rho(1+\alpha_j+\beta_j)+1+\alpha_s>1$, $p>0$, and $w_s\bar h_s-P_s>0$, the term in brackets is strictly negative, implying $\Delta m^\ast<0.$ In other words, retirement reduces monetary transfers.

The effect on grandparental childcare is
\begin{equation}
\Delta b_s^\ast
=
\bar h_s
-
\frac{\alpha_s\left[p\bar h_s-\left(w_s\bar h_s-P_s\right)\right]}{\rho p\Lambda}=
\frac{1}{\rho p\Lambda}
\left[ p\bar h_s[\rho(1+\alpha_j+\beta_j)+1]+\alpha_s\left(w_s\bar h_s-P_s\right)
\right].
\label{eq:theory_delta_b}
\end{equation}
Therefore, $\Delta b_s^\ast>0$: retirement increases grandparental childcare.

The effect on fertility is therefore
\begin{equation}
\Delta n^\ast
=
\frac{\beta_j}{p\bar h(1+\alpha_j+\beta_j)}
\left(\Delta m^\ast+p\Delta b_s^\ast\right)
=
\frac{\beta_j}{p\bar h\Lambda}
\left[p\bar h_s-\left(w_s\bar h_s-P_s\right)\right].
\label{eq:reaction_hetero}
\end{equation}
This expression captures the model's central mechanism. Senior retirement affects Junior fertility through two opposing changes in intergenerational transfers. Retirement reduces Senior income from $w_s\bar h_s$ to $P_s$, lowering monetary transfers and the Junior's resources for childrearing. It also releases non-work time, allowing the Senior to provide more grandparental childcare and reducing the Junior's reliance on costly formal childcare.

The net fertility effect depends on the relative importance of these changes. The income cost of retirement is $w_s\bar h_s-P_s$, while the additional non-work time has a market value of $p\bar h_s$. Retirement raises Junior fertility if and only if $p\bar h_s>w_s\bar h_s-P_s$: the fertility benefit of additional time support then exceeds the cost of reduced monetary support. If $p\bar h_s<w_s\bar h_s-P_s$, reduced monetary support dominates, and retirement lowers Junior fertility. Retirement is therefore more likely to raise fertility when formal childcare is more expensive, pre-retirement labor income is lower, or the pension replacement rate is higher, holding other factors constant.

\subsection{The Senior's Welfare}
\label{subsec:senior_welfare}

Retirement also affects Senior welfare. Substituting the optimal choices in equations (12) and (15)--(18) into equation~\eqref{eq:elderly_utility} yields the Senior's indirect utility function conditional on retirement status $R$:
\begin{equation}
V_s^\ast(R)
=
A\ln c_j^\ast(R)+C(p,Y_j)
=
A\ln\left[
\frac{Y_j+T_s(R)}{\Lambda}
\right]+C(p,Y_j),
\label{eq:senior_indirect_utility}
\end{equation}
where $A=1+\alpha_s+\rho(1+\alpha_j+\beta_j)>0$, and $C(p,Y_j)$ collects terms that do not depend on retirement status. At a fixed childcare price, the welfare effect of retirement is the difference in indirect utility between retirement and continued employment:
\begin{equation}
\begin{aligned}
\Delta V_s^\ast
&\equiv
V_s^\ast(1)-V_s^\ast(0)
=
A\left[
\ln c_j^\ast(1)-\ln c_j^\ast(0)
\right]\\
&=
A\ln
\left[
\frac{Y_j+T_s(1)}
{Y_j+T_s(0)}
\right]
=
A\ln
\left[
1+
\frac{\Delta T_s}
{Y_j+T_s(0)}
\right].
\end{aligned}
\label{eq:delta_V_s}
\end{equation}

Equation~\eqref{eq:senior_indirect_utility} shows that Senior welfare is increasing in effective transferable resources. Since retirement changes these resources by $\Delta T_s=p\bar h_s-(w_s\bar h_s-P_s)$, it improves Senior welfare when the market value of the additional non-work time exceeds the income loss:
\begin{equation}
p\bar h_s>w_s\bar h_s-P_s.
\end{equation}
Retirement reduces Senior welfare when the income loss dominates.

\section{Econometric Framework}
\label{sec:econometric_framework}

The retirement effects $\Delta m^\ast$, $\Delta b_s^\ast$, and $\Delta n^\ast$ in equations (20)--(22) are differences in optimal choices between retirement and continued employment. Let $R_s$ denote the Senior's retirement status, and let $n_s(0)$, $m_s(0)$, and $b_s(0)$ denote the corresponding optimal choices under continued employment. The observed outcomes can then be written as
\begin{align}
n_s
&=
n_s(0)+R_s\Delta n^\ast+\varepsilon_{n,s},
\label{eq:econ_outcome_n_R}
\\
m_s
&=
m_s(0)+R_s\Delta m^\ast+\varepsilon_{m,s},
\label{eq:econ_outcome_m_R}
\\
b_s
&=
b_s(0)+R_s\Delta b_s^\ast+\varepsilon_{b,s},
\label{eq:econ_outcome_b_R}
\end{align}
where the error terms capture unobserved determinants of the outcomes. For $Y_s\in\{n_s,m_s,b_s\}$, equations~\eqref{eq:econ_outcome_n_R}--\eqref{eq:econ_outcome_b_R} have the general form
\begin{equation}
Y_s
=
Y_s(0)+R_s\Delta Y_s^\ast+\varepsilon_{Y,s},
\label{eq:structural_outcome_general}
\end{equation}
where $\Delta Y_s^\ast$ denotes the corresponding retirement effect derived in the theoretical model.

Retirement, fertility, and intergenerational transfers are jointly determined within the family. Families with stronger preferences for grandchildren may induce the Senior to retire earlier and provide more childcare. Such preferences may also affect fertility and monetary transfers, creating a correlation between retirement status and $\varepsilon_{Y,s}$. A direct comparison of retired and non-retired individuals therefore need not identify the causal effects of retirement.

To address this endogeneity concern, we exploit the discontinuous increase in retirement probability at the statutory retirement age under China's mandatory retirement regime to identify these effects. Because crossing the cutoff increases retirement probability without fully determining retirement status, we use a fuzzy regression discontinuity (RD) design. During the sample period, the statutory retirement age was 60 for men, 50 for female workers, and 55 for female cadres. We use age 50 as the baseline female cutoff because retirement probability rises sharply at age 50 but not at age 55 (Figure~\ref{fig:mandatory_retirement_women}), consistent with evidence that approximately 93\% of women are subject to the age-50 rule \cite{FangFeng2018}. As a robustness check, we assign women with junior college education or above a cutoff of 55, using education as a proxy for cadre status, and retain a cutoff of 50 for other women.

Let $a_s=age_s-a_s^R$ denote age relative to the applicable statutory retirement cutoff, and let $Z_s=\mathbf{1}\{a_s\geq0\}$ indicate whether individual $s$ is at or above the cutoff. The cutoff indicator serves as an instrument for retirement status in the first-stage equation
\begin{equation}
R_s
=
f(a_s)+\gamma_z Z_s+\gamma_X X_s+\nu_s,
\label{eq:econ_first_stage_bridge}
\end{equation}
where $X_s$ contains predetermined covariates and fixed effects. The covariates include indicators for female, urban residence, rural hukou status\footnote{ Hukou is China's household registration system. It historically distinguished agricultural (rural) from non-agricultural (urban) status, with different entitlements to public services and social benefits. Individuals with rural hukou may reside in urban areas.}, and junior college education or above, as well as the number of living Juniors. The full specification also includes province and survey-wave fixed effects. The function \(f(a_s)\) is a first-degree polynomial in age relative to the cutoff, with coefficients allowed to differ across the threshold. The coefficient $\gamma_z$ measures the discontinuity in retirement probability at the cutoff.

For each outcome $Y_s\in\{n_s,m_s,b_s\}$, we estimate the reduced-form equation
\begin{equation}
Y_s
=
f(a_s)+\delta_Z Z_s+\delta_X X_s+\epsilon_s,
\label{eq:econ_reduced_form_bridge}
\end{equation}
where $\delta_Z$ measures the discontinuity in the outcome at the cutoff.

Under the standard fuzzy RD assumptions of continuity, exclusion, and monotonicity, and with a nonzero first-stage discontinuity, the ratio of the reduced-form coefficient to the first-stage coefficient identifies the local average treatment effect (LATE) of retirement:
\begin{equation}
LATE_Y
=
\frac{\delta_Z}{\gamma_z},
\qquad
\widehat{LATE}_Y
=
\frac{\widehat{\delta}_Z}{\widehat{\gamma}_z}.
\label{eq:econ_wald_bridge}
\end{equation}
This effect is local to compliers at the cutoff, whose retirement status changes in response to crossing the statutory retirement-age threshold.

Our baseline estimation is based on local linear regressions that are allowed to vary on either side of the threshold. We use a triangular kernel, which places greater weight on observations closer to the cutoff. The selection of the bandwidth is informed by the rule-of-thumb procedure established by \citeasnoun{calonicoCattaneoTitiunik2014}. Optimal bandwidth values for our primary outcome is 5 years on both sides of the statutory cutoff. For inference, we adhere to the approach proposed by \citeasnoun{calonicoCattaneoTitiunik2014} and \citeasnoun{calonicoCovariates2019}, adopting cluster-robust confidence intervals centered on bias-corrected estimates as an alternative to conventional confidence intervals. In this framework, we derive standard errors from a specification, which, in our case, includes a second-degree polynomial for the running variable. Thus, in addition to the standard coefficient estimates and their associated cluster-robust standard errors, we provide bias-corrected estimates that are used to construct these cluster-robust confidence intervals. To enhance the reliability of our standard error estimations, we implement clustering, using age from the retirement age as the clustering unit, following the methodology proposed by \citeasnoun{leeCard2008}.


\section{Data and Sample}
\label{sec:data_sample}

\subsection{Data}
\label{sec:data}

We use the 2011, 2013, 2015, and 2018 waves of CHARLS, a nationally representative household survey of Chinese individuals aged 45 and above and their families. The 2011 baseline covers 150 county-level units and 450 village-level units, with approximately 17000 individuals in about 10000 households. Detailed information on retirement, intergenerational transfers, and family outcomes makes the survey well suited to our analysis.

We restrict the analysis to waves before 2020 for two reasons. The 2020 wave does not contain the full set of intergenerational transfer measures needed to examine the mechanisms. In addition, the COVID-19 pandemic may have affected household decisions, labor supply, and health, complicating the interpretation of discontinuities around the statutory retirement ages.

To focus on respondents plausibly covered by the retirement system, we retain individuals who have completed retirement procedures, participate in the urban employee basic pension insurance program, or participate in the urban employee basic medical insurance program.\footnote{ These are employment-based social insurance programs.} We exclude observations with missing values for key variables. Because net intergenerational monetary transfers can be positive or negative, we trim this variable at the 1st and 99th percentiles. For non-negative variables, including time transfers and the number of grandchildren, we trim only the upper 1 percent. All income and expenditure measures are expressed in 2018 prices using the CPI for the respondent's city of residence. The resulting estimation samples are described below.

\subsection{Sample}

The RD estimation sample contains 9523 observations after imposing the retirement-system coverage criteria, excluding observations with missing values for key variables, and trimming outliers. Table~\ref{tab:full_sample_summary} summarizes the main variables. The mean retirement rate is 0.656, and respondents have 1.480 grandchildren on average. Mean annual net monetary transfers to Juniors are RMB 1153.0, while mean annual time transfers are 508.5 hours.

Table~\ref{tab:sample_summary} compares respondents below the statutory retirement cutoff, \(r_s\in(-5,0)\), with those at or above it, \(r_s\in[0,5)\), within the baseline RD window (\(N=3542\)). The retirement rate rises from 0.325 below the cutoff to 0.809 at or above it. Respondents at or above the cutoff have more grandchildren, provide more grandparental childcare, and make smaller net monetary transfers to Juniors. Panel B shows similar shares of urban residents and rural hukou holders on either side. The shares of women and respondents with junior college education or above are lower at or above the cutoff, whereas the mean number of living Juniors is higher. These comparisons describe the local sample; formal tests of covariate continuity at the cutoff are reported in Appendix~\ref{app:rd_robustness}.


\begin{table}[H]
\centering
\caption{Full-Sample Summary Statistics}
\label{tab:full_sample_summary}
\small
\begin{threeparttable}
\setlength{\tabcolsep}{5pt}
\begin{tabular}{lrrrrr}
\toprule
Variable & Obs. & Mean & SD & Min & Max \\
\midrule
Retirement & 9,523 & 0.656 & 0.475 & 0.000 & 1.000 \\
Number of grandchildren & 9,523 & 1.480 & 1.799 & 0.000 & 9.000 \\
Net monetary transfer & 9,523 & 1,153.04 & 7,034.46 & -21,255.31 & 58,348.44 \\
Time transfer & 9,523 & 508.47 & 1,186.93 & 0.00 & 6,720.00 \\
Female & 9,523 & 0.394 & 0.489 & 0.000 & 1.000 \\
Urban & 9,523 & 0.808 & 0.394 & 0.000 & 1.000 \\
Rural hukou & 9,523 & 0.209 & 0.407 & 0.000 & 1.000 \\
Junior college or above & 9,523 & 0.105 & 0.307 & 0.000 & 1.000 \\
Living Juniors & 9,523 & 1.760 & 1.007 & 0.000 & 10.000 \\
\bottomrule
\end{tabular}
\begin{tablenotes}
\footnotesize
\item \textit{Notes:} This table reports summary statistics for the full RD estimation sample. SD denotes standard deviation. Net monetary transfers to Juniors are measured as one half of annual household net transfers and expressed in 2018 RMB. Time transfers are measured as annual hours of grandparental childcare provided by the respondent.
\end{tablenotes}
\end{threeparttable}
\end{table}

\begin{table}[H]
\centering
\renewcommand{\arraystretch}{1.12}
\setlength{\tabcolsep}{8pt}
\caption{Summary Statistics Around the Statutory Retirement Cutoff}
\label{tab:sample_summary}
\begin{tabular}{@{}lccc@{}}
\toprule
& \multicolumn{1}{c}{Below cutoff} & \multicolumn{1}{c}{Above cutoff} & \multicolumn{1}{c}{Difference} \\
& \multicolumn{1}{c}{$(-5,0)$} & \multicolumn{1}{c}{$[0,5)$} & \multicolumn{1}{c}{(Above $-$ Below)} \\
& \multicolumn{1}{c}{$N = 1{,}483$} & \multicolumn{1}{c}{$N = 2{,}059$} & \\
\midrule
\multicolumn{4}{l}{\textit{Panel A: Key Variables}} \\
Retirement & 0.325 & 0.809 & 0.484*** \\
Number of grandchildren & 0.769 & 1.378 & 0.608*** \\
Net monetary transfer & 1,965.11 & 868.65 & $-1{,}096.45$*** \\
Time transfer & 373.70 & 607.11 & 233.41*** \\
\addlinespace[0.5ex]
\multicolumn{4}{l}{\textit{Panel B: Predetermined Covariates}} \\
Female & 0.477 & 0.437 & $-0.040$** \\
Urban & 0.801 & 0.812 & 0.011 \\
Rural hukou & 0.233 & 0.220 & $-0.014$ \\
Junior college or above & 0.118 & 0.085 & $-0.033$*** \\
Living Juniors & 1.496 & 1.651 & 0.155*** \\
\bottomrule
\multicolumn{4}{p{0.92\linewidth}}{\footnotesize \textit{Notes:} The sample comprises 3,542 observations with \(r_s\in(-5,5)\), where \(r_s\) is age in years relative to the statutory retirement cutoff. The first two columns report sample means for \(r_s<0\) and \(r_s\geq0\), respectively. The final column reports differences in means (above minus below). Net monetary transfers to Juniors are measured as one half of annual household net transfers and expressed in 2018 RMB. Time transfers are measured as annual hours of grandparental childcare provided by the respondent. Differences in means are tested using two-sample \(t\)-tests. *** \(p<0.01\), ** \(p<0.05\), * \(p<0.1\).}
\end{tabular}
\end{table}

\section{RD Results}
\label{sec:RD_results}

We first examine the discontinuity in retirement probability at the statutory retirement cutoff and then estimate the effect of retirement on the number of grandchildren. We assess the credibility of these estimates through robustness and diagnostic checks before examining the two intergenerational transfer mechanisms. Finally, we investigate whether the effects vary by Senior income and gender.

\subsection{First Stage}

\begin{figure}[H]
\centering
\includegraphics[width=0.72\textwidth]{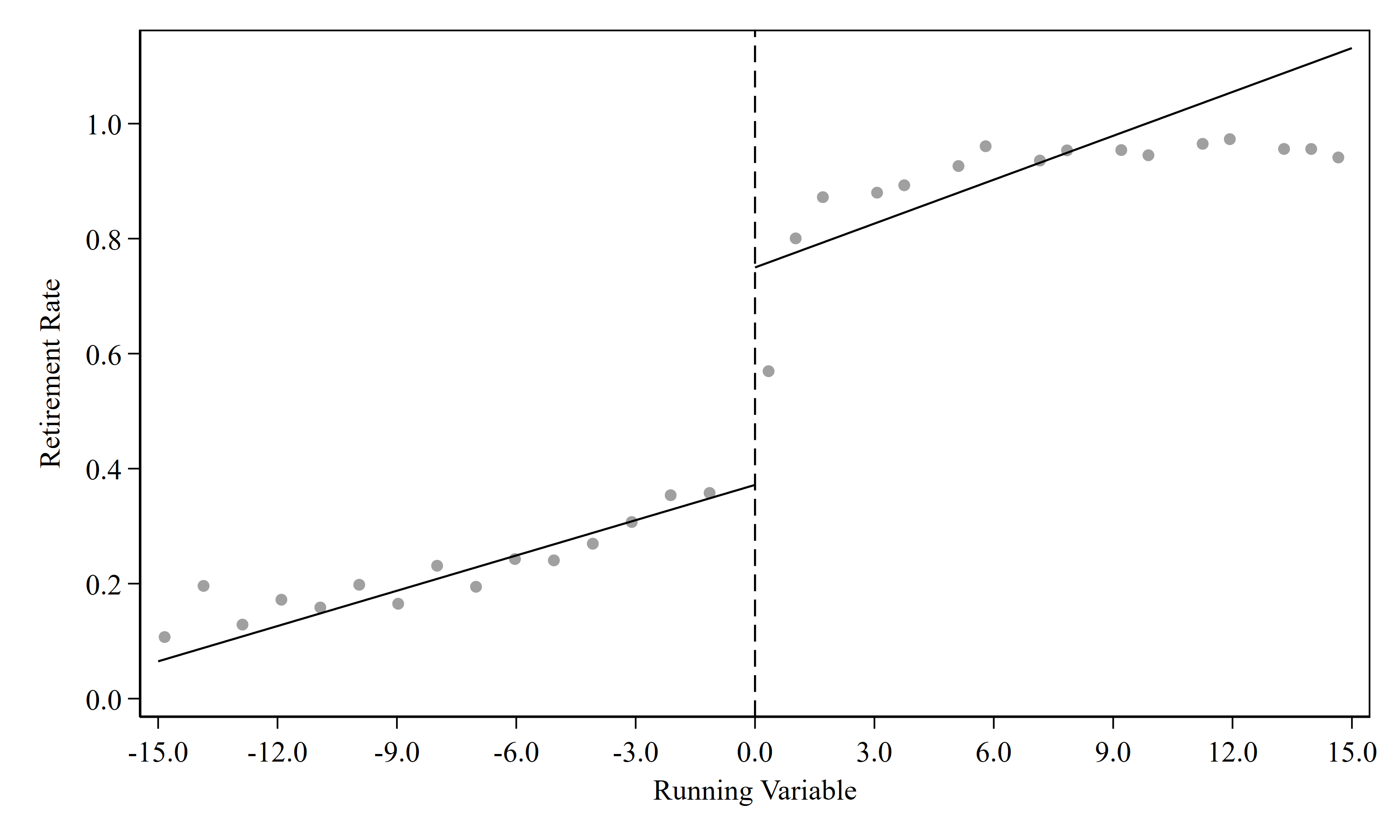}
\caption{Retirement Probability Around the Statutory Retirement Cutoff}
\label{fig:first_stage}
\begin{minipage}{0.85\textwidth}
\footnotesize
\textit{Notes:} The figure plots the probability of retirement against age relative to the relevant statutory retirement cutoff. Dots denote binned sample means, and the solid lines are fitted local linear functions estimated separately on each side of the cutoff. The vertical dashed line marks the cutoff at zero.
\end{minipage}
\end{figure}

We begin by examining how retirement responds to the statutory retirement cutoff. Reaching the statutory retirement age increases the probability of retirement but does not induce all individuals to retire immediately. The design is therefore fuzzy: crossing the cutoff changes the probability of retirement rather than determining retirement status, as in a sharp RD design. Figure~\ref{fig:first_stage} illustrates this relationship by plotting retirement probability against age relative to the statutory retirement cutoff. A nontrivial share of individuals have already retired below the cutoff, while retirement probability rises sharply at the cutoff and remains substantially higher above it.

In the full specification with predetermined covariates and province and survey-wave fixed effects, reaching the statutory retirement cutoff increases retirement probability by 22.4 percentage points (Table~\ref{tab:first_stage}). The discontinuity remains large and statistically significant across all specifications. These estimates establish a substantial first-stage response to the statutory retirement rule, providing the variation in retirement status used to identify the LATE of retirement on family outcomes.

\begin{table}[H]
\centering
\caption{First-Stage Estimates}
\label{tab:first_stage}
\small
\begin{threeparttable}
\setlength{\tabcolsep}{5.5pt}
\begin{tabular}{lccc}
\toprule
& \multicolumn{3}{c}{Dependent variable: Retirement} \\
\cmidrule(lr){2-4}
& (1) & (2) & (3) \\
\midrule
Crossing cutoff & 0.222*** & 0.235*** & 0.224*** \\
& (0.048) & (0.045) & (0.044) \\
\addlinespace
Predetermined covariates & No & Yes & Yes \\
Province fixed effects & No & No & Yes \\
Wave fixed effects & No & No & Yes \\
Observations & 9,523 & 9,523 & 9,523 \\
Bandwidth & 5 & 5 & 5 \\
Eff. obs. left of cutoff & 1,483 & 1,483 & 1,483 \\
Eff. obs. right of cutoff & 2,059 & 2,059 & 2,059 \\
\bottomrule
\end{tabular}
\begin{tablenotes}
\footnotesize
\item \textit{Notes:} The table reports bias-corrected estimates of the discontinuity in retirement probability at the statutory retirement cutoff. Standard errors are reported in parentheses and clustered by age relative to the cutoff. Statistical significance is based on robust bias-corrected inference \cite{calonicoCattaneoTitiunik2014,calonicoCovariates2019}. All specifications use a triangular kernel, a first-order local polynomial, and a symmetric bandwidth of 5 years. Predetermined covariates include gender, urban residence, rural hukou, junior college education or above, and the number of living Juniors. Province and survey-wave fixed effects are included as indicated. *** \(p<0.01\), ** \(p<0.05\), * \(p<0.1\).
\end{tablenotes}
\end{threeparttable}
\end{table}

\subsection{Main Results}

\begin{figure}[H]
\centering
\includegraphics[width=0.72\textwidth]{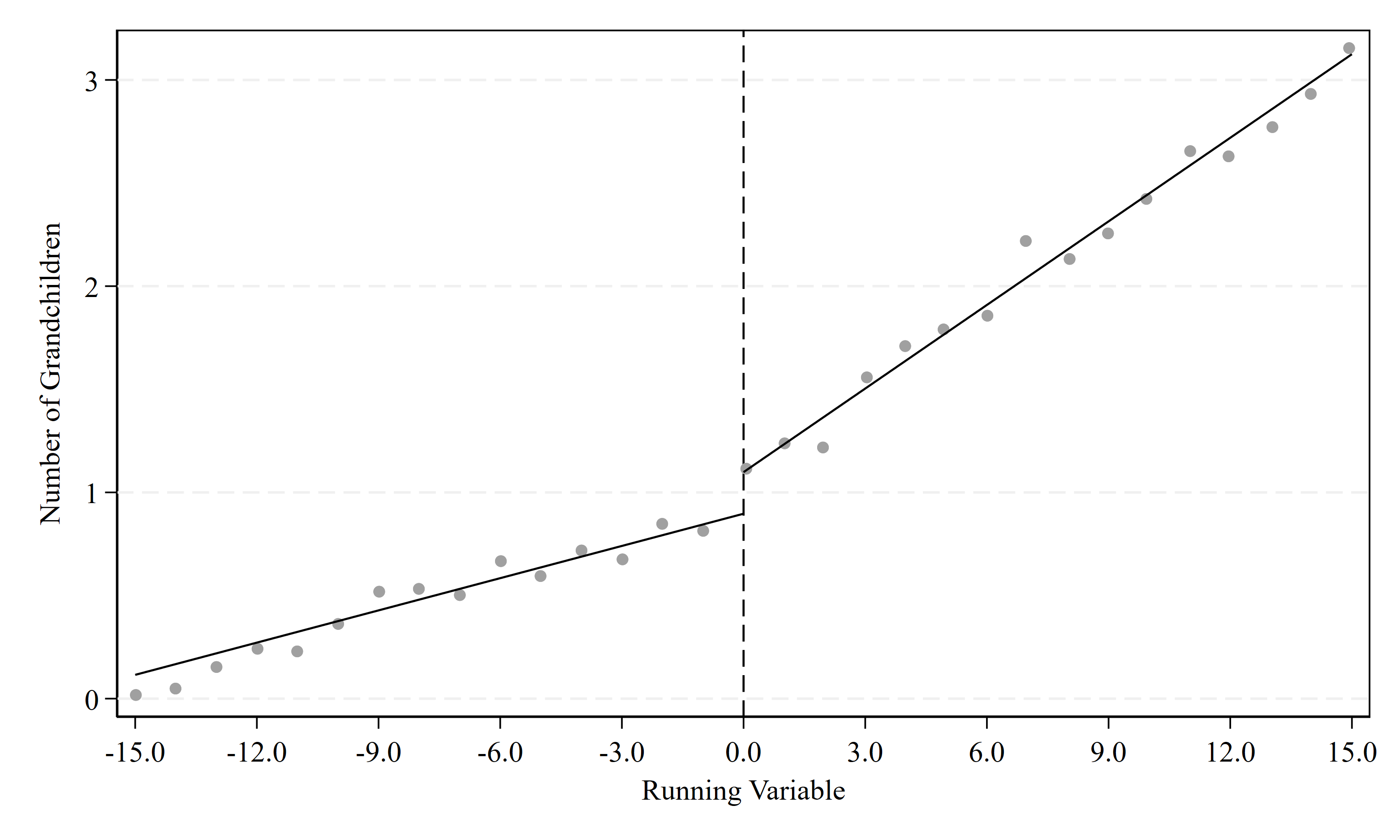}
\caption{Effect of Retirement on the Number of Grandchildren}
\label{fig:main_rd}
\begin{minipage}{0.85\textwidth}
\footnotesize
\textit{Notes:} The figure plots the number of grandchildren against age relative to the relevant statutory retirement cutoff. Dots denote binned sample means, and the solid lines are fitted local linear functions estimated separately on each side of the cutoff. The vertical dashed line marks the cutoff at zero.
\end{minipage}
\end{figure}

Figure~\ref{fig:main_rd} plots the number of grandchildren against age relative to the statutory retirement cutoff. The figure shows conditional means within bins of the running variable and fitted local linear functions on either side of the cutoff. The visible upward discontinuity suggests a positive association between retirement and the realized number of grandchildren among individuals near the cutoff.

Table~\ref{tab:main_results} reports the corresponding bias-corrected fuzzy RD estimates of the LATE of retirement for compliers at the statutory retirement cutoff. In the full specification with predetermined covariates and fixed effects, the estimate is 0.613 grandchildren, with a standard error of 0.311, and is statistically significant at the 5 percent level. Across the three specifications, the estimates decline from 0.958 to 0.613 while remaining statistically significant. Relative to the below-cutoff mean in Table~\ref{tab:sample_summary}, these estimates imply an economically meaningful increase in the number of grandchildren among individuals whose retirement behavior responds to the statutory cutoff.

In the theoretical framework, retirement reduces monetary transfers but increases time transfers through grandparental childcare. These opposing responses make the net effect on fertility theoretically ambiguous. The positive estimate suggests that, near the statutory retirement cutoff, the increase in time support following retirement outweighs the reduction in monetary support. This pattern is consistent with binding childcare constraints, under which the additional time provided by Seniors can facilitate Junior fertility.

\begin{table}[H]
\centering
\caption{LATE of Retirement on the Number of Grandchildren}
\label{tab:main_results}
\small
\begin{threeparttable}
\setlength{\tabcolsep}{8pt}
\begin{tabular}{lccc}
\toprule
& \multicolumn{3}{c}{Dependent variable: Number of grandchildren} \\
\cmidrule(lr){2-4}
& (1) & (2) & (3) \\
\midrule
Retirement & 0.958*** & 0.701** & 0.613** \\
& (0.285) & (0.292) & (0.311) \\
\addlinespace
Predetermined covariates & No & Yes & Yes \\
Province fixed effects & No & No & Yes \\
Wave fixed effects & No & No & Yes \\
Observations & 9,523 & 9,523 & 9,523 \\
Bandwidth & 5 & 5 & 5 \\
Eff. obs. left of cutoff & 1,483 & 1,483 & 1,483 \\
Eff. obs. right of cutoff & 2,059 & 2,059 & 2,059 \\
\bottomrule
\end{tabular}
\begin{tablenotes}
\footnotesize
\item \textit{Notes:} The table reports bias-corrected fuzzy RD estimates of the LATE of retirement on the number of grandchildren. Standard errors are reported in parentheses and clustered by age relative to the statutory retirement cutoff. Statistical significance is based on robust bias-corrected inference \cite{calonicoCattaneoTitiunik2014,calonicoCovariates2019}. All specifications use a triangular kernel, a first-order local polynomial, and a symmetric bandwidth of 5 years. Predetermined covariates include gender, urban residence, rural hukou, junior college education or above, and the number of living Juniors. Province and survey-wave fixed effects are included as indicated. *** \(p<0.01\), ** \(p<0.05\), * \(p<0.1\).
\end{tablenotes}
\end{threeparttable}
\end{table}

\subsection{Robustness Checks}
\label{subsec:rd_robustness_summary}

We assess the credibility of the RD design through robustness and diagnostic checks. The estimated effect on the number of grandchildren remains positive and statistically significant under alternative bandwidths, polynomial specifications, kernels, clustering levels, and an alternative coding of the female retirement cutoff, although the magnitudes vary. Placebo-cutoff estimates are statistically insignificant at the selected thresholds away from the statutory retirement ages. Density and covariate-continuity tests provide no evidence of discontinuities at the statutory cutoff. Together, these results support interpreting the estimates as local causal effects around the statutory retirement ages. Appendix~\ref{app:rd_robustness} reports the checks in detail.

\subsection{Mechanisms}

\begin{figure}[H]
\centering
\includegraphics[width=0.72\textwidth]{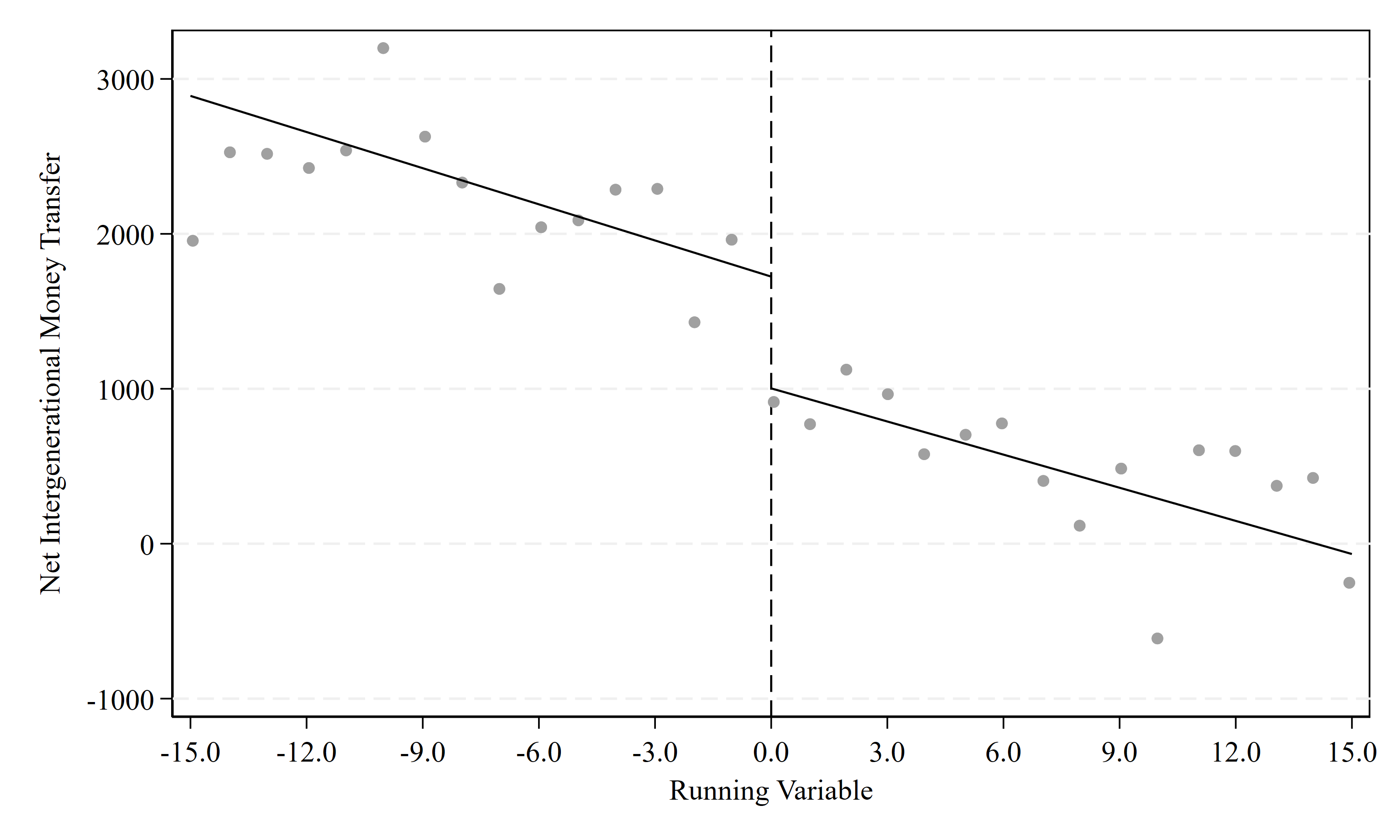}
\caption{Retirement and Net Monetary Transfers to Juniors}
\label{fig:mechanism_financial}
\begin{minipage}{0.85\textwidth}
\footnotesize
\textit{Notes:} The figure plots individual net monetary transfers to Juniors, measured as one half of household net transfers in 2018 RMB per year, against age relative to the relevant statutory retirement cutoff. Dots denote binned sample means, and the solid lines are fitted local linear functions estimated separately on each side of the cutoff. The vertical dashed line marks the cutoff at zero.
\end{minipage}
\end{figure}

\begin{figure}[H]
\centering
\includegraphics[width=0.72\textwidth]{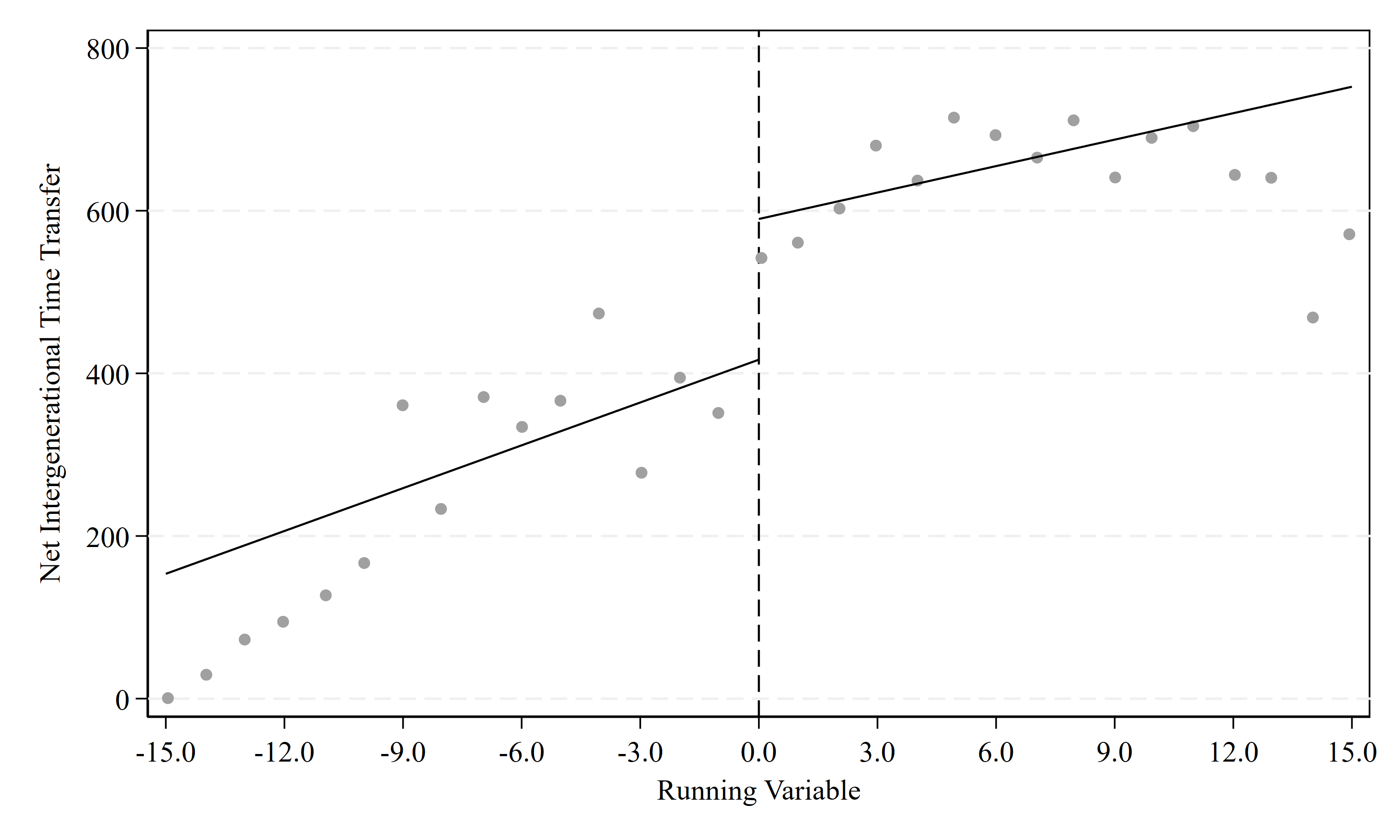}
\caption{Retirement and Time Transfers Through Grandparental Childcare}
\label{fig:mechanism_time}
\begin{minipage}{0.85\textwidth}
\footnotesize
\textit{Notes:} The figure plots the respondent's own annual grandparental childcare hours against age relative to the relevant statutory retirement cutoff. Dots denote binned sample means, and the solid lines are fitted local linear functions estimated separately on each side of the cutoff. The vertical dashed line marks the cutoff at zero.
\end{minipage}
\end{figure}

The theoretical framework highlights two intergenerational channels through which retirement may affect fertility: monetary transfers and time transfers. The empirical evidence is consistent with a shift in the composition of support from money to time. Figure~\ref{fig:mechanism_financial} plots net monetary transfers to Juniors around the statutory retirement cutoff, while Figure~\ref{fig:mechanism_time} plots grandparental childcare hours. The figures show a decline in net monetary transfers and an increase in grandparental childcare at the cutoff.

Table~\ref{tab:mechanisms} reports the fuzzy RD estimates for these two outcomes. Retirement reduces net monetary transfers to Juniors by approximately RMB 4177 per respondent per year and increases the respondent's own grandparental childcare by approximately 786 hours per year. Both effects are statistically significant at the 5 percent level. These estimates suggest that retirement reallocates intergenerational support toward time transfers through grandparental childcare while reducing monetary transfers. This shift helps rationalize the positive effect on the number of grandchildren despite the decline in monetary support from Seniors.

\begin{table}[H]
\centering
\caption{Mechanism Evidence}
\label{tab:mechanisms}
\small
\begin{threeparttable}
\setlength{\tabcolsep}{8pt}
\begin{tabular}{lcc}
\toprule
& \multicolumn{2}{c}{Dependent variables} \\
\cmidrule(lr){2-3}
& (1) & (2) \\
& Net monetary transfer & Time transfer \\
\midrule
Retirement & -4,177.412** & 786.055** \\
& (1,999.106) & (309.424) \\
\addlinespace
Predetermined covariates & Yes & Yes \\
Province fixed effects & Yes & Yes \\
Wave fixed effects & Yes & Yes \\
Observations & 9,523 & 9,523 \\
Bandwidth & 5 & 5 \\
Eff. obs. left of cutoff & 1,483 & 1,483 \\
Eff. obs. right of cutoff & 2,059 & 2,059 \\
\bottomrule
\end{tabular}
\begin{tablenotes}
\footnotesize
\item \textit{Notes:} The table reports bias-corrected fuzzy RD estimates of the effects of retirement on net monetary transfers and grandparental childcare hours. Standard errors are reported in parentheses and clustered by age relative to the statutory retirement cutoff. Both specifications use a triangular kernel, a first-order local polynomial, and an estimation bandwidth of 5 years. Net monetary transfers assign one half of the household net amount to the respondent and are measured in 2018 RMB per year; time transfers are the respondent's own annual grandparental childcare hours. *** \(p<0.01\), ** \(p<0.05\), * \(p<0.1\).
\end{tablenotes}
\end{threeparttable}
\end{table}

\subsection{Heterogeneity Analysis}
Equation~\ref{eq:reaction_hetero} shows that, holding other factors constant, lower pre-retirement labor income reduces the value of continued work relative to the value of the additional time released by retirement, making retirement more likely to increase Junior fertility. This prediction motivates examining whether the effect of retirement on the number of grandchildren varies by Senior income. Grandparental childcare also differs substantially by gender. In Chinese families, grandmothers typically provide more grandparental childcare than grandfathers. Descriptive evidence from \citeasnoun{chenIntergenerationalTiesContext2011} shows that coresident grandmothers provide approximately 21.1 hours of childcare per week on average, compared with approximately 7 hours for grandfathers. We therefore examine heterogeneity by Senior income and gender.\footnote{ Seniors are classified as lower or higher income relative to the median income within each survey wave $\times$ retirement status cell.}

Figure~\ref{fig:heterogeneity} presents the subgroup estimates, which are reported in Appendix Table~\ref{tab:heterogeneity_rd}. Retirement has a positive and statistically significant effect on the number of grandchildren among lower-income Seniors, whereas the estimate for higher-income Seniors is negative and statistically insignificant. This pattern is consistent with the model's opposing effects of time and monetary transfers. For lower-income Seniors, the positive effect of additional time transfers is more likely to outweigh the negative effect of reduced monetary transfers. For higher-income Seniors, the larger income loss associated with retirement strengthens the negative effect of reduced monetary transfers, weakening or offsetting the positive fertility effect of additional grandparental childcare.

The gender results show a positive and statistically significant effect of retirement on the number of grandchildren among female Seniors; the estimate for male Seniors is also positive but statistically insignificant. This pattern is consistent with the time-transfer mechanism. Because grandmothers typically provide more grandparental childcare, the additional non-work time released by retirement is more likely to translate into actual childcare provision among female Seniors. Using German data, \citeasnoun{eibichRetirementIntergenerationalTime2020} find that early retirement increases grandmothers' childcare time, providing evidence consistent with this mechanism.

Together, the income and gender patterns are consistent with the theoretical model and provide further support for the role of intergenerational time transfers in the fertility response to retirement. When formal childcare is relatively costly, the additional grandparental childcare made possible by retirement has greater value, increasing the importance of time transfers for Junior fertility.

\begin{figure}[H]
\centering
\includegraphics[width=0.75\textwidth]{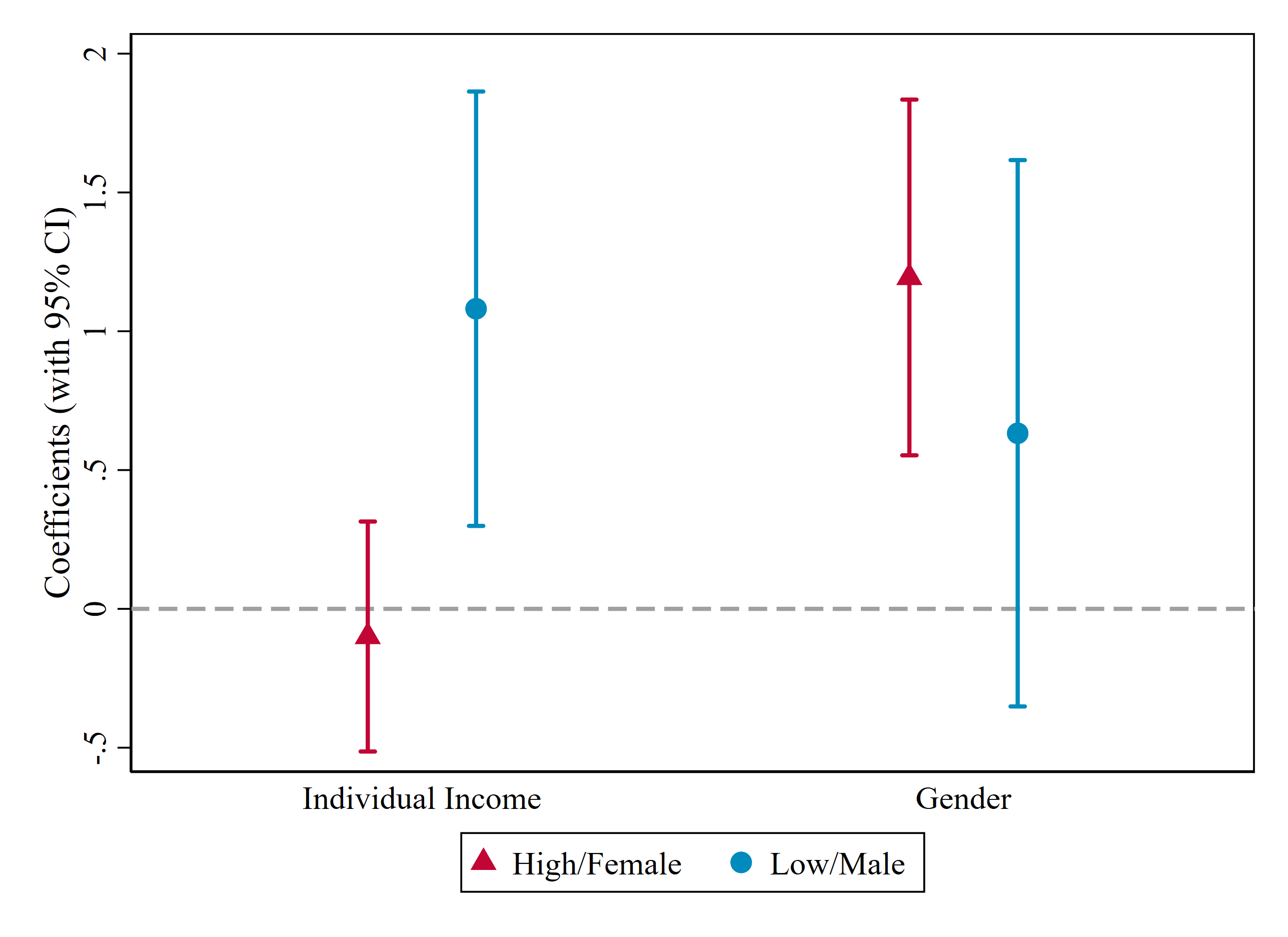}
\caption{Heterogeneous Effects of Retirement on the Number of Grandchildren}
\label{fig:heterogeneity}
\begin{minipage}{0.85\textwidth}
\footnotesize
\textit{Notes:} The figure reports fuzzy RD estimates for subgroups defined by Senior income and gender. All specifications use local linear regressions with a triangular kernel, predetermined covariates, and province and survey-wave fixed effects. Subgroup bandwidths and estimates are reported in Table~\ref{tab:heterogeneity_rd}. Vertical lines represent robust 95\% confidence intervals.
\end{minipage}
\end{figure}

\section{Identification and Estimation}
\label{sec:identification_estimation}

The fuzzy RD estimates presented above show that retirement significantly increases Junior fertility by reshaping intergenerational transfers: retirees provide more grandparental childcare but reduce monetary transfers. These findings are consistent with the theoretical mechanism. We estimate the structural parameters to evaluate the effects of China's GDR reform, alternative retirement regimes, and complementary childcare subsidies on Senior employment, DPP, Junior fertility, and Senior welfare.

The retirement effects in equations~\eqref{eq:reaction_hetero}, \eqref{eq:theory_delta_m}, and~\eqref{eq:theory_delta_b} depend on combinations of the preference parameters. For estimation, we define the following composite parameters:
\begin{equation}
\eta_n=\frac{\beta_j}{\Lambda},
\qquad
\eta_m=\frac{1}{\rho\Lambda},
\qquad
\eta_b=\frac{\alpha_s}{\rho\Lambda},
\qquad
\eta_j=\frac{1+\alpha_j}{\Lambda},
\label{eq:composite_parameters}
\end{equation}
where $\Lambda$ is defined in equation~\eqref{eq:junior_consumption_optimal}. The parameter vector is $\boldsymbol\eta=(\eta_n,\eta_m,\eta_b,\eta_j)'$. These parameters are positive and satisfy the adding-up restriction $\eta_n+\eta_m+\eta_b+\eta_j=1$. The first three parameters govern the responses of fertility, monetary transfers, and grandparental childcare to changes in effective resources.

Equations~\eqref{eq:reaction_hetero}, \eqref{eq:theory_delta_m}, and~\eqref{eq:theory_delta_b} give the retirement effects on fertility, monetary transfers, and grandparental childcare, respectively. Using the composite parameter definitions in equation~\eqref{eq:composite_parameters}, we rewrite these effects as
\begin{align}
\Delta n^\ast
&=
\frac{\eta_n}{p\bar h}\Delta T_s,
\label{eq:econ_delta_n}
\\
\Delta m^\ast
&=
\Delta I_s-\eta_m\Delta T_s,
\label{eq:econ_delta_m}
\\
\Delta b_s^\ast
&=
\bar h_s-\frac{\eta_b}{p}\Delta T_s.
\label{eq:econ_delta_b}
\end{align}

Estimation proceeds in two steps, following \citeasnoun{GourinchasParker2002} and \citeasnoun{Rodriguez2023}. In the first step, we calibrate the childcare price and time parameters, estimate Senior potential earnings $\widehat W_i$ and pension benefits $\widehat P_i$, and construct an additional empirical moment from the China Family Panel Studies (CFPS). In the second step, we estimate the composite parameters using a moment-matching (minimum-distance) estimator, following \citeasnoun{cocciStandardErrorsCalibrated2025}.

Following \citeasnoun{berlinskiChildcareMarketsParental2024a}, we calibrate the time parameters using assumptions about daily time use and annual working days. With the time endowment normalized to $1$, we assume $8$ hours of sleep per day, leaving $16\times365=5{,}840$ hours per year available for work, leisure, and childcare. We assume that the Senior works $8$ hours per day for 250 working days per year, implying 2,000 annual working hours and hence $\bar h_s=2{,}000/5{,}840$. Similarly, assuming that each child requires $8$ hours of care on each of the 250 working days implies an annual childcare requirement of 2,000 hours, yielding $\bar h=2{,}000/5{,}840$.

CHARLS does not report the price of formal childcare, $p$. We therefore calibrate $p$ using external childcare fee data. In 2023, the national average fee for childcare services was RMB 1,978 per child per month, implying $p=10.9114$ RMB per hour in 2018 prices.\footnote{
The 2023 childcare fee is obtained from \textit{Report of the State Council on Progress in Developing Childcare Services}, available at \url{http://www.npc.gov.cn/c2/c30834/202409/t20240911_439363.html}. The report states that the national average fee for childcare services, excluding meals, was RMB 1,978 per child per month in 2023. We first annualize the monthly fee and divide it by the assumed annual childcare requirement of 2,000 hours, yielding an hourly price of RMB 11.868 in 2023 prices. We then deflate this amount to 2018 prices using the cumulative CPI adjustment from 2019 to 2023. Specifically,
\[
p=\frac{1{,}978\times12}{2{,}000}
\times
\frac{1}{1.029\times1.025\times1.009\times1.020\times1.002}
=10.9114
\]
RMB per hour.
}

We next estimate Senior potential income in the work and retirement states. Income in the two states cannot be observed simultaneously. For a retired Senior, we do not observe the labor income that would have been earned had the Senior continued working; for a non-retired Senior, we do not observe the pension income that would have been received upon retirement. We therefore estimate potential annual labor income in the work state, $\widehat W_i$, and potential annual pension income in the retirement state, $\widehat P_i$, which correspond to $w_s\bar h_s$ and $P_s$ in the theoretical model. Because retirement status may be correlated with unobserved determinants of income, estimating these equations directly on the observed samples of workers and retirees may be subject to sample-selection bias. We address this concern using a standard two-step sample-selection correction, following \citeasnoun{heckmanSampleSelectionBias1979}.

We first estimate a probit retirement-selection equation using the same statutory-eligibility variables and predetermined covariates as in equation~\eqref{eq:econ_first_stage_bridge}, augmented by a quadratic term in $a_s$. We use the resulting estimates to construct the inverse Mills ratios for the work and retirement states. We then estimate separate log-income equations for the two states. For non-retired Seniors with positive labor income, we regress log annual labor income on $a_s$, $a_s^2$, $X_s$, province and survey-wave fixed effects, and the inverse Mills ratio for the work state. For retired Seniors with positive pension income, we estimate the analogous specification for log annual pension income using the inverse Mills ratio for the retirement state. We then use these estimates to predict each Senior's potential income in both states.\footnote{
The inverse Mills ratios are included in the second-step income equations to correct for sample selection but are not included when constructing potential-income predictions. We therefore predict each Senior's income in the work and retirement states using the estimated coefficients on the observed covariates. Because the income equations are estimated in logs, we retransform the fitted values into levels using separate Duan smearing corrections for the labor-income and pension-income equations.
}
This yields $\widehat W_i$ and $\widehat P_i$, and hence the retirement-induced income change $\widehat{\Delta I}_i=\widehat P_i-\widehat W_i$.

The first three empirical moments are the robust bias-corrected (RBC) fuzzy RD estimates from CHARLS, $\widehat{\tau}_n^{RBC}$, $\widehat{\tau}_m^{RBC}$, and $\widehat{\tau}_b^{RBC}$, for the retirement effects on the number of grandchildren, net monetary transfers, and grandparental childcare hours, respectively. These estimates identify LATE for compliers at the statutory retirement-age cutoff. Their corresponding model moments, $\tau_n^{model}(\boldsymbol\eta)$, $\tau_m^{model}(\boldsymbol\eta)$, and $\tau_b^{model}(\boldsymbol\eta)$, are the average retirement responses derived from equations~\eqref{eq:econ_delta_n}--\eqref{eq:econ_delta_b} for the same local complier population.

We supplement the three RD estimates with a CFPS moment for $\eta_j$, which captures the share of effective resources allocated to Junior consumption and the opportunity cost of leisure. CHARLS provides detailed information on the Senior but does not contain the Junior consumption and leisure information required to construct this moment. The CFPS allows us to link Juniors to their parents and provides the consumption, wage, leisure, and income information needed to construct this moment, although it does not provide a comparable measure of grandparental childcare hours. Because the 2012 CFPS wave lacks information on time allocation, while the 2014 wave does not report net monetary transfers from the Senior, we use the 2016 and 2018 waves to construct the additional empirical moment for $\eta_j$.\footnote{
The CFPS sample uses the 2016 and 2018 waves and retains Junior observations for which at least one parent satisfies the retirement-system coverage criterion and falls in the interval \((-5,5)\) around the relevant statutory retirement cutoff. The baseline cutoffs are age 60 for fathers and age 50 for mothers. A parent satisfies the coverage criterion if the parent is classified as retired in CFPS, participates in the urban employee basic pension insurance program, or participates in the urban employee basic medical insurance program. If both parents satisfy these conditions, we select the parent closest to the corresponding cutoff as the Senior. Here, \(c_j\) is annual household consumption per capita, \(I_s\) is the Senior's observed labor income when non-retired and observed pension income when retired, and \(\Omega_s=1-\bar h_s(1-R_s)\). Monetary variables are expressed in 2018 prices. Because the model normalizes the annual time endowment to one, the Junior's hourly wage and the hourly childcare price are converted into returns per unit of normalized annual time before constructing \(Y_j\) and \(p\Omega_s\). We exclude observations above the pooled 99th percentile of the Junior's weekly leisure time, the Junior's hourly wage, the Senior's state-specific income, or household consumption per capita.
}
Using the calibrated values of $p$ and $\bar h_s$, we obtain $\widehat{\eta}_j^{CFPS}=0.2553$\footnote{
The optimal choices in equations~\eqref{eq:junior_consumption_optimal} and~\eqref{eq:junior_leisure_optimal} imply
\begin{equation}
c_j^\ast(R)+Y_jl_j^\ast(R)
=
\eta_j
\left[
Y_j+I_s(R)+p\Omega_s(R)
\right].
\label{eq:eta_j_model_moment}
\end{equation}
Taking the sample analogue of equation~\eqref{eq:eta_j_model_moment} gives
\begin{equation}
\widehat{\eta}_j^{CFPS}
=
\frac{
\displaystyle \frac{1}{N}\sum_{i=1}^{N}
\left(c_{j,i}+Y_{j,i}l_{j,i}\right)
}{
\displaystyle \frac{1}{N}\sum_{i=1}^{N}
\left(Y_{j,i}+I_{s,i}+p\Omega_{s,i}\right)
}.
\label{eq:eta_j_cfps_moment}
\end{equation}
},
with a standard error of $0.0111$.\footnote{
Let \(\mathcal G_g\) denote the observations associated with Junior \(g\), and let \(G\) denote the number of Junior clusters. The cluster-robust standard error, clustered by Junior, is
\[
\widehat{\operatorname{SE}}\!\left(\widehat{\eta}_j^{CFPS}\right)
=
\left\{
\frac{G}{G-1}
\sum_{g=1}^{G}
\left[
\sum_{i\in\mathcal G_g}
\frac{
c_{j,i}+Y_{j,i}l_{j,i}
-\widehat{\eta}_j^{CFPS}
\left(Y_{j,i}+I_{s,i}+p\Omega_{s,i}\right)
}{
\displaystyle
\sum_{r=1}^{N}
\left(Y_{j,r}+I_{s,r}+p\Omega_{s,r}\right)
}
\right]^2
\right\}^{1/2}.
\]
}

In the second step, we estimate the composite parameters by matching the model moments to their empirical counterparts. The empirical-moment vector is
$\widehat{\boldsymbol\mu}=(\widehat{\tau}_n^{RBC},\widehat{\tau}_m^{RBC},\widehat{\tau}_b^{RBC},\widehat{\eta}_j^{CFPS})'$,
and the corresponding model-moment vector is
$h(\boldsymbol\eta)=(\tau_n^{model}(\boldsymbol\eta),\tau_m^{model}(\boldsymbol\eta),\tau_b^{model}(\boldsymbol\eta),\eta_j)'$.\footnote{
For each candidate value of $\boldsymbol\eta$, equations~\eqref{eq:econ_delta_n}--\eqref{eq:econ_delta_b}, together with $\widehat{\Delta I}_i$ and the calibrated values of $p$, $\bar h_s$, and $\bar h$, give the retirement responses. We construct the corresponding model moments using the same running variable, retirement treatment, kernel, polynomial order, outcome-specific estimation and bias-correction bandwidths, predetermined covariates, province and survey-wave fixed effects, and clustering procedure as for their empirical CHARLS counterparts.
}
Following \citeasnoun{cocciStandardErrorsCalibrated2025}, we use the minimum-distance estimator:

\begin{equation}
\widehat{\boldsymbol\eta}
=
\arg\min_{\boldsymbol\eta\in\mathcal H}
\left[
\widehat{\boldsymbol\mu}-h(\boldsymbol\eta)
\right]'
\widehat W
\left[
\widehat{\boldsymbol\mu}-h(\boldsymbol\eta)
\right],
\qquad
\mathcal H
=
\left\{
\boldsymbol\eta>0:
\mathbf 1'\boldsymbol\eta=1
\right\}.
\label{eq:minimum_distance_estimator}
\end{equation}
The parameter space $\mathcal H$ imposes positivity and the adding-up restriction. The restriction leaves three independent parameter dimensions, so the four empirical moments imply one overidentifying restriction.

For point estimation, we use the diagonal weighting matrix
$\widehat W=\operatorname{diag}(\widehat\sigma_n^{-2},\widehat\sigma_m^{-2},\widehat\sigma_b^{-2},\widehat\sigma_j^{-2})$,
where $\widehat\sigma_n$, $\widehat\sigma_m$, $\widehat\sigma_b$, and $\widehat\sigma_j$ are the marginal standard errors of the corresponding empirical moments. Because the full covariance matrix of the moments is unavailable, we report the worst-case standard errors proposed by \citeasnoun{cocciStandardErrorsCalibrated2025}. These standard errors maximize the asymptotic variance of each parameter estimate over the admissible correlation structures.

Table~\ref{tab:structural_estimates} reports the calibrated and estimated parameters. Table~\ref{tab:model_fit} compares the empirical and model moments.

\begin{table}[H]
  \centering
  \caption{Calibrated and Estimated Parameters}
  \label{tab:structural_estimates}
  \begin{threeparttable}
    \scriptsize
    \setlength{\tabcolsep}{5pt}
    \begin{tabular}{lp{7.2cm}r}
      \toprule
      Parameter & Description & Value \\
      \midrule
      \multicolumn{3}{l}{\textit{Panel A. Calibrated parameters}} \\
      $p$        & Hourly price of formal childcare, 2018 RMB & 10.91 \\
      $\bar h_s$ & Senior work time                            & 0.342  \\
      $\bar h$   & Childcare requirement per child             & 0.342  \\
      \midrule
      \multicolumn{3}{l}{\textit{Panel B. Estimated composite parameters}} \\
      $\eta_n$ & Junior fertility share
               & 0.123   \\
               & & (0.279) \\
      $\eta_m$ & Senior consumption share
               & 0.122   \\
               & & (0.120) \\
      $\eta_b$ & Senior leisure share
               & 0.500   \\
               & & (0.209) \\
      $\eta_j$ & Junior consumption--leisure share
               & 0.255   \\
               & & (0.012) \\
      \bottomrule
    \end{tabular}
    \begin{tablenotes}[flushleft]
      \footnotesize
      \item \textit{Notes:} Worst-case standard errors for the estimated
      parameters are reported in parentheses
      \cite{cocciStandardErrorsCalibrated2025}.
    \end{tablenotes}
  \end{threeparttable}
\end{table}

\begin{table}[H]
  \centering
  \caption{Model Fit}
  \label{tab:model_fit}
  \begin{threeparttable}
    \scriptsize
    \setlength{\tabcolsep}{4pt}
    \begin{tabular}{lp{5.2cm}rrr}
      \toprule
      Moment
        & Description
        & Data
        & Model\\
      \midrule
      $\widehat\tau_n^{RBC}$
 & Retirement effect on the number of grandchildren
        & 0.613
        & 0.119\\
      $\widehat\tau_m^{RBC}$
        & Retirement effect on net monetary transfers
        & $-4{,}177.400$
        & $-3{,}244.600$\\
      $\widehat\tau_b^{RBC}$
        & Retirement effect on grandparental childcare hours
        & 786.100
        & 1,029.900\\
      $\widehat\eta_j^{CFPS}$
        & Junior consumption--leisure moment
        & 0.255
        & 0.255\\
      \bottomrule
    \end{tabular}
    \begin{tablenotes}[flushleft]
      \footnotesize
      \item \textit{Notes:} The first three empirical moments are
      RBC fuzzy RD estimates from CHARLS. The fourth
      moment is estimated using the 2016 and 2018 waves of the CFPS.
    \end{tablenotes}
  \end{threeparttable}
\end{table}

\section{Counterfactual Analysis}
\label{sec:counterfactual}

Using equations~\eqref{eq:econ_outcome_n_R}--\eqref{eq:econ_outcome_b_R} and the estimated structural parameters, we conduct a series of counterfactual analyses. We first evaluate China's GDR reform implemented in 2025. In the model, the Senior's retirement decision directly affects labor supply and pension receipt and, through changes in intergenerational monetary and time transfers, also affects the Junior's fertility. We therefore examine the reform's effects on Senior employment and DPP, its consequences for Senior welfare, and its spillover effects on Junior fertility.

We next examine how these effects depend on the design of retirement policy. We consider three alternative retirement regimes---FFR, MPR, and MFT---that differ in the retirement choices available to Seniors and work intensity after the pre-reform statutory retirement age. These comparisons allow us to assess whether alternative retirement rules can increase Senior employment and DPP while mitigating the declines in Senior welfare and Junior fertility generated by GDR for the 2031--2039 cohorts.

We then consider childcare subsidies as a complementary policy. Delayed retirement reduces the time that working Seniors can devote to grandparental childcare, and changes in retirement rules alone may not fully offset the resulting declines in Junior fertility and Senior welfare. In the model, a lower formal childcare price reduces the cost of substituting formal childcare for grandparental childcare. We therefore examine the subsidy rates required to eliminate the GDR-induced decline in Junior fertility and, separately, the loss in Senior welfare, while maintaining the employment gains from delayed retirement.

Finally, we examine whether the fiscal cost of achieving these two objectives depends on how childcare subsidies are allocated. The heterogeneity analysis yields larger estimated fertility effects of retirement for female Seniors and lower-income Seniors, while GDR also imposes different retirement constraints across groups. We therefore compare universal subsidies with subsidies targeted to households with female Seniors and to households with lower-income female Seniors. For each subsidy scheme, we calculate the subsidy rate and fiscal expenditure required to eliminate the GDR-induced decline in Junior fertility and, separately, the decline in Senior welfare.

\subsection{The Effects of GDR}
\label{sec:cf_gradual}

 As discussed in Section~\ref{sec:institution_gradual}, the reform phases in higher statutory retirement ages over 2025--2039. We simulate the effects of the GDR reform for cohorts $t=2025-2039$.\footnote{ Following Section~\ref{sec:institution_gradual}, cohort $t$ is defined by the year in which a Senior reaches the pre-reform statutory retirement age. Appendix Tables~\ref{tab:official_schedule_men}--\ref{tab:official_schedule_women50} report the official monthly schedules. For men and female cadres, statutory retirement ages rise from 60 to 63 and from 55 to 58, respectively, across cohorts $t=2025,\ldots,2036$, and remain at these final levels thereafter. For female workers, the statutory retirement age rises from 50 to 55 across cohorts $t=2025,\ldots,2034$ and remains at 55 thereafter. Our analysis therefore includes both transition cohorts and subsequent cohorts subject to the final statutory retirement ages.} At the pre-reform statutory age---60 for men and 50 for women---we compare outcomes under the reform with those that would prevail if all Seniors retired at that age under the pre-reform rules.\footnote{ Using the notation in equation~\eqref{eq:structural_outcome_general}, the benchmark for $Y_s\in\{n_s,m_s,b_s\}$ is $Y_s(1)$. Letting $R_s$ denote retirement status under the reform and holding the childcare price fixed at $p_0$, the individual-level change is
\[
Y_s(R_s)-Y_s(1)
=
(R_s-1)\tau_{Y,s}(\eta,p_0).
\]
The reported changes are averages of these individual-level differences. Appendix~\ref{app:counterfactual} describes the construction of the counterfactual household outcomes and the welfare measure.} Table~\ref{tab:cf_official_tradeoff} and Appendix Figure~\ref{fig:cf_gradual_broad} report the results. 

For the 2025--2030 cohorts, the reform raises Senior labor force participation by 2.8 percentage points and defers RMB 543 in pension payments per Senior-year.\footnote{ DPP are defined as $DPP_i=\widehat P_i(1-R_i)$, where $\widehat P_i$ is the predicted annual pension benefit and $R_i=1$ indicates retirement. DPP measure the annual amount of pension payments deferred at the evaluation age, not a reduction in lifetime pension expenditure.} Senior welfare increases by 0.022\%, and the number of grandchildren increases by 0.002.

These effects reflect the continued availability of retirement at the pre-reform statutory age. Seniors who prefer retirement retain the pre-reform allocation, while those for whom the net income gain from continued employment exceeds the value of the non-work time forgone remain employed. The latter obtain higher utility from continued employment, producing a small increase in mean Senior welfare, as characterized in Section~\ref{subsec:senior_welfare} and equation~\eqref{eq:delta_V_s}. Across these cohorts, annual monetary transfers increase by RMB 840 per Senior-year, while grandparental childcare falls by 68 hours. At the baseline childcare price $p_0$, the increase in monetary transfers exceeds the value of the childcare forgone, resulting in a small increase in Junior fertility. Consistent with this retirement-choice mechanism, higher-income Seniors are more likely to continue working.\footnote{ Holding pension income and working hours fixed, higher wages increase the net income gain from continued employment. Seniors are classified as lower or higher income relative to the median income within each survey wave $\times$ retirement status cell. This empirical median split differs from the model's retirement-choice condition. Differences across income groups may therefore also reflect group composition.}

For the 2031--2039 cohorts, the effects on employment and annual DPP are substantially larger. Senior labor force participation rises to 41.8\%, and DPP reach RMB 10221 per Senior-year. Senior welfare, however, falls by 1.578\%, and the number of grandchildren declines by 0.043. These losses are concentrated primarily among female and lower-income Seniors.

This difference reflects the retirement constraint faced by women whose pre-reform statutory retirement age was 50. For these cohorts, the earliest retirement age exceeds 50, so women who would prefer retirement, $V_i(1)>V_i(0)$, cannot retire at pre-reform statutory retirement age. The increase in employment therefore includes women for whom the income gain from continued work does not compensate for the value of the non-work time forgone, resulting in welfare losses. Men, by contrast, retain the option of retiring at age 60.

The fertility decline reflects the corresponding change in intergenerational support. Across these cohorts, annual monetary transfers increase by RMB 2467 per Senior-year, while grandparental childcare falls by 488 hours. At the baseline childcare price $p_0$, the increase in monetary transfers is insufficient to compensate for the value of the childcare forgone, reducing the effective resources available to the Junior and lowering fertility.

These results show that the effects of delayed retirement on Senior welfare and Junior fertility depend on the constraints imposed on Seniors' retirement choices and on whether the income gain from continued work compensates for the loss of non-work time. Formal childcare prices determine the expenditure required to substitute formal childcare for grandparental childcare and can therefore alter the balance between income gains and time costs in retirement decisions. Based on these results, Section~\ref{sec:cf_regimes} examines whether changes in retirement choice and work intensity can reduce the losses in Senior welfare and Junior fertility while maintaining the employment and DPP gains generated by GDR for the later cohorts. Section~\ref{sec:cf_subsidies} then examines the role of childcare subsidies that reduce the price of formal childcare.

\begingroup
\scriptsize
\setlength{\tabcolsep}{3.0pt}
\renewcommand{\arraystretch}{1.02}
\setlength{\LTleft}{\fill}
\setlength{\LTright}{\fill}
\setlength{\LTcapwidth}{\textwidth}
\setlength{\LTpost}{4pt}
\begin{xltabular}{\textwidth}{@{}>{\raggedright\arraybackslash}Xrrrrrrr@{}}
\caption{GDR by Demographic Group}\label{tab:cf_official_tradeoff}\\
\toprule
Group & Work (\%) & DPP & \(\Delta\mathcal W\) (\%) & \(n^{old}(R=1)\) & \(\Delta n\) & \(\Delta m\) & \(\Delta b\) \\
\midrule
\endfirsthead
\multicolumn{8}{c}{\tablename\ \thetable\ -- continued}\\
\toprule
Group & Work (\%) & DPP & \(\Delta\mathcal W\) (\%) & \(n^{old}(R=1)\) & \(\Delta n\) & \(\Delta m\) & \(\Delta b\) \\
\midrule
\endhead
\midrule
\multicolumn{8}{r}{\footnotesize Continued on next page}\\
\endfoot
\bottomrule
\endlastfoot
\multicolumn{8}{l}{\textit{Panel A. 2025--2030 cohorts}}\\
All & 2.80 & 543 & 0.022 & 1.505 & 0.002 & 840 & \(-68\) \\
Men & 3.90 & 745 & 0.029 & 1.826 & 0.002 & 1,184 & \(-96\) \\
Women & 1.00 & 234 & 0.004 & 1.011 & 0.001 & 313 & \(-25\)\\
Lower income & 1.80 & 248 & 0.020 & 1.538 & 0.001 & 540 & \(-44\) \\
Higher income & 3.70 & 840 & 0.024 & 1.472 & 0.002 & 1,142 & \(-91\) \\
Lower-income men & 3.00 & 362 & 0.030 & 1.893 & 0.001 & 876 & \(-71\) \\
Higher-income men & 4.70 & 1,031 & 0.028 & 1.776 & 0.003 & 1,414 & \(-114\) \\
Lower-income women & 0.70 & 126 & 0.004 & 1.156 & 0.000 & 180 & \(-15\) \\
Higher-income women & 1.50 & 405 & 0.004 & 0.782 & 0.001 & 523 & \(-40\) \\
\addlinespace
\multicolumn{8}{l}{\textit{Panel B. 2031--2039 cohorts}}\\
All & 41.80 & 10,221 & \(-1.578\) & 1.505 & \(-0.043\) & 2,467 & \(-488\) \\
Men & 3.90 & 745 & 0.029 & 1.826 & 0.002 & 1,184 & \(-96\) \\
Women & 100.00 & 24,802 & \(-6.101\) & 1.011 & \(-0.112\) & 4,443 & \(-1,092\)\\
Lower income & 49.80 & 10,479 & \(-2.000\) & 1.538 & \(-0.054\) & 2,443 & \(-556\) \\
Higher income & 33.80 & 9,961 & \(-1.133\) & 1.472 & \(-0.031\) & 2,492 & \(-421\) \\
Lower-income men & 3.00 & 362 & 0.030 & 1.893 & 0.001 & 876 & \(-71\) \\
Higher-income men & 4.70 & 1,031 & 0.028 & 1.776 & 0.003 & 1,414 & \(-114\) \\
Lower-income women & 100.00 & 21,339 & \(-5.596\) & 1.156 & \(-0.114\) & 4,124 & \(-1,076\) \\
Higher-income women & 100.00 & 30,298 & \(-7.314\) & 0.782 & \(-0.108\) & 4,949 & \(-1,119\) \\
\end{xltabular}
\noindent\begin{minipage}{\textwidth}
\scriptsize
\textit{Notes:} Outcomes are evaluated at the statutory retirement ages in effect before the reform: 60 for men and 50 for women in our sample. Panel A reports results for cohorts reaching the pre-reform statutory retirement age in 2025--2030. Panel B reports results for cohorts reaching the pre-reform statutory retirement age in 2031--2039. \(n^{old}(R=1)\) denotes the simulated number of grandchildren in the pre-reform benchmark. Monetary transfers and DPP are measured in RMB per Senior-year, and grandparental childcare in hours per Senior-year. Seniors are classified as lower or higher income relative to the median income within each survey wave $\times$ retirement status cell.
\end{minipage}
\endgroup

\subsection{Alternative Retirement Regimes}
\label{sec:cf_regimes}

To examine how retirement choice and work intensity affect policy outcomes, we compare three alternative retirement regimes: FFR, MPR, and MFT. Under FFR, Seniors in all cohorts may choose freely between full retirement and continued full-time work once they reach the pre-reform statutory retirement age: age 60 for men and age 50 for women. Under MPR and MFT, Seniors must continue working for four and eight hours per day, respectively, and begin receiving pension benefits only after full retirement.

\begin{figure}[H]
\centering
\includegraphics[width=\textwidth]{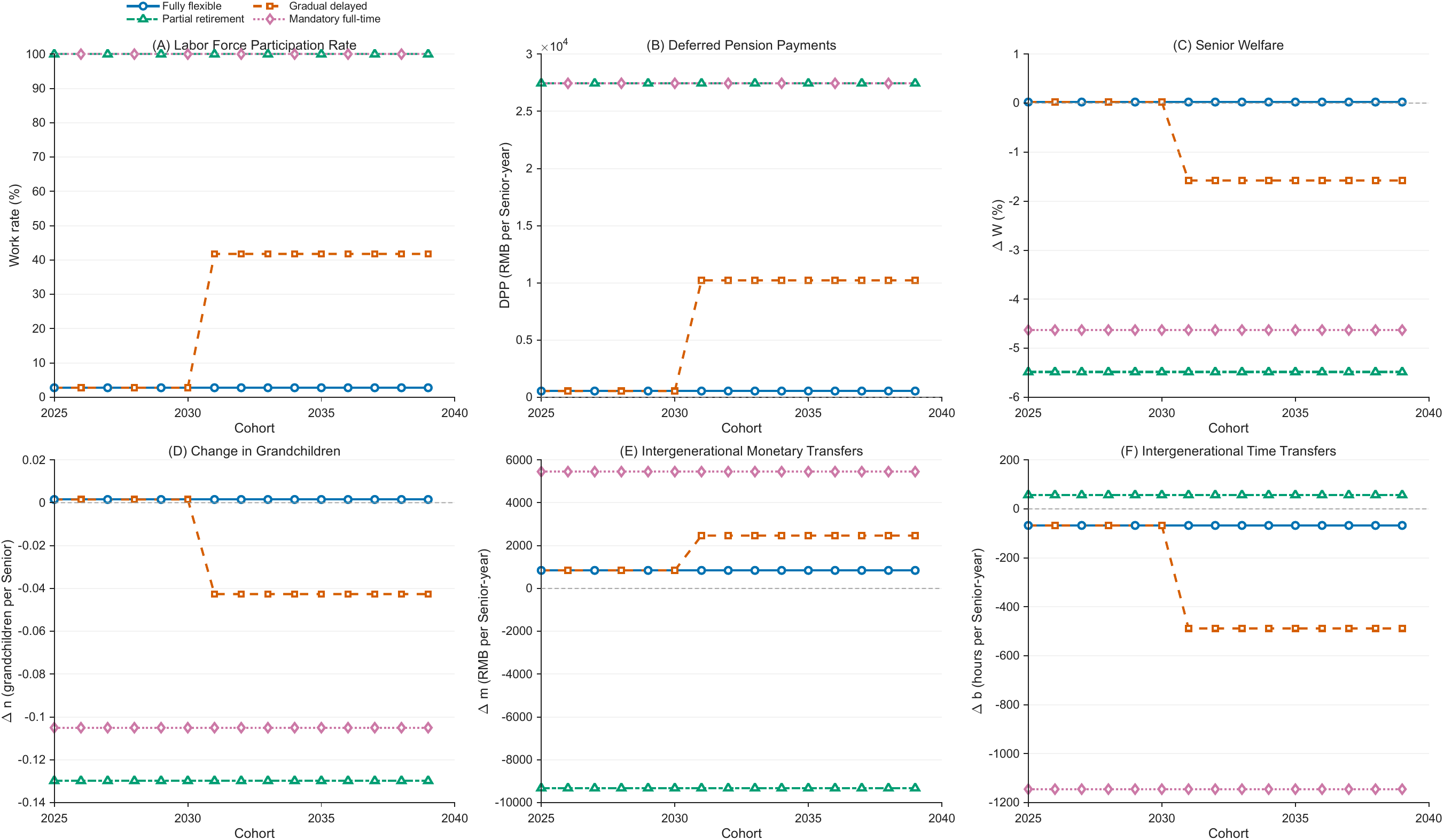}
\caption{Alternative Retirement Regimes: Cohort Effects}
\label{fig:cf_retirement_regimes}
\begin{minipage}{0.94\textwidth}
\footnotesize
\textit{Notes:}
The figure reports counterfactual outcomes under FFR, GDR, MPR, and MFT for cohorts reaching the pre-reform statutory retirement age between 2025 and 2039. Panel A reports the labor force participation rate; Panel B reports DPP; Panels C--F report changes in Senior welfare, the number of grandchildren, annual intergenerational monetary transfers, and annual grandparental childcare, respectively, relative to the pre-reform benchmark. This benchmark assumes that all Seniors retire at the statutory retirement ages applicable to our sample before the reform: 60 for men and 50 for women. Under partial retirement, all Seniors work four hours per day, receive one-half of full-time work-state labor income, and receive no pension until full retirement, so that the 100\% participation rate corresponds to a 50\% full-time-equivalent employment rate. All other working states correspond to an eight-hour workday. Senior welfare is expressed as a percentage change, monetary transfers and DPP are measured in RMB per Senior-year, and grandparental childcare is measured in hours per Senior-year. Appendix Table~\ref{tab:cf_retirement_regimes} reports the 2025--2039 cohort averages for FFR, MPR, and MFT.
\end{minipage}
\end{figure}

Figure~\ref{fig:cf_retirement_regimes} and Table~\ref{tab:cf_retirement_regimes} show that, under FFR, only 2.8\% of Seniors choose to remain employed. DPP amount to RMB 543 per Senior-year, mean Senior welfare increases by 0.02\%, and the average number of grandchildren increases by 0.002. These outcomes are the same as those for the 2025--2030 cohorts under the GDR reform and reflect the role of retirement choice: Seniors who prefer retirement can still retire, while only those who benefit from continued work remain employed. FFR therefore avoids the welfare and fertility losses observed for the 2031--2039 cohorts under the GDR reform but generates much smaller increases in employment and DPP.

MFT requires all Seniors to continue working, raising the labor force participation rate to 100\% and DPP to RMB 27450 per Senior-year. Mean Senior welfare declines by 4.63\%, and the average number of grandchildren falls by 0.105. Intergenerational transfers change in the same direction as for the 2031--2039 cohorts under the GDR reform, but by larger amounts. Relative to the pre-reform benchmark, annual monetary transfers increase by RMB 5452, while grandparental childcare falls by 1145 hours. At the baseline childcare price, the increase in monetary transfers remains insufficient to compensate for the value of the childcare forgone. The larger increase in employment is therefore accompanied by larger losses in Senior welfare and Junior fertility.

MPR also raises the labor force participation rate to 100\%, but the full-time-equivalent employment rate is only 50\%, and pension receipt is deferred until full retirement. Mean Senior welfare declines by 5.48\%, and the average number of grandchildren falls by 0.130, with both declines exceeding those under MFT. Relative to the pre-reform benchmark, MPR increases annual grandparental childcare by 57 hours but reduces annual monetary transfers by RMB 9322. Shorter working hours release more time for childcare, but they also halve labor income without providing contemporaneous pension income. The increase in intergenerational time transfers is therefore insufficient to compensate for the decline in monetary transfers. This result shows that the effects of partial retirement depend not only on working hours but also on the timing of pension receipt \cite{haan_tolan_2019}.

The alternative retirement regimes generate distinct trade-offs between labor supply, DPP, and family outcomes. Greater retirement flexibility avoids the losses in Senior welfare and Junior fertility but produces smaller increases in employment and DPP. Stronger work requirements generate larger employment and DPP gains but are accompanied by greater family losses. Shorter working hours release time for grandparental childcare, but do not eliminate the welfare and fertility losses when pension receipt is also deferred. For the 2031--2039 cohorts, none of the alternative retirement regimes considered here eliminates the losses in Senior welfare and Junior fertility while maintaining the employment and DPP gains generated by GDR. The next section therefore examines whether lower formal childcare prices can reduce these losses by lowering the cost of substituting formal childcare for grandparental childcare.

\subsection{Childcare Subsidies under GDR}
\label{sec:cf_subsidies}

The price of formal childcare determines the market replacement value of the non-work time released by retirement and therefore shapes the trade-off between the income loss and the time gain associated with retirement. Through intergenerational transfers, this trade-off affects not only Senior retirement decisions and welfare but also Junior fertility. Building on the substitution between formal childcare and grandparental childcare emphasized in the childcare and family-policy literature \cite{karademirMultigenerationalImpact2026}, we examine whether childcare subsidies can offset the welfare and fertility losses caused by GDR by reducing the cost of formal childcare.

Let \(s\) denote the subsidy rate and \(z\) the group eligible for the subsidy. The household pays the following price for formal childcare:
\begin{equation}
p_i(s,z)
=
p_0\left[1-\rho_p s\mathbf 1\{i\in z\}\right],
\label{eq:cf_subsidy_price}
\end{equation}
where \(p_0\) denotes the price of formal childcare in the absence of a subsidy. In this section, we consider a universal subsidy, so \(z\) includes all households. Under the baseline specification, the price paid by the household therefore falls to \((1-s)p_0\)\footnote{ The parameter \(\rho_p\in[0,1]\) captures the degree of development of the formal childcare market. A lower value represents a less developed market, characterized by higher search and matching costs and greater information asymmetry. These frictions increase the costs households incur in finding, screening, and matching with reliable childcare providers, limiting the reduction in effective childcare costs achieved by a given subsidy. The baseline specification sets \(\rho_p=1\), representing a fully developed formal childcare market.}.

\begin{figure}[H]
\centering
\includegraphics[width=\textwidth]{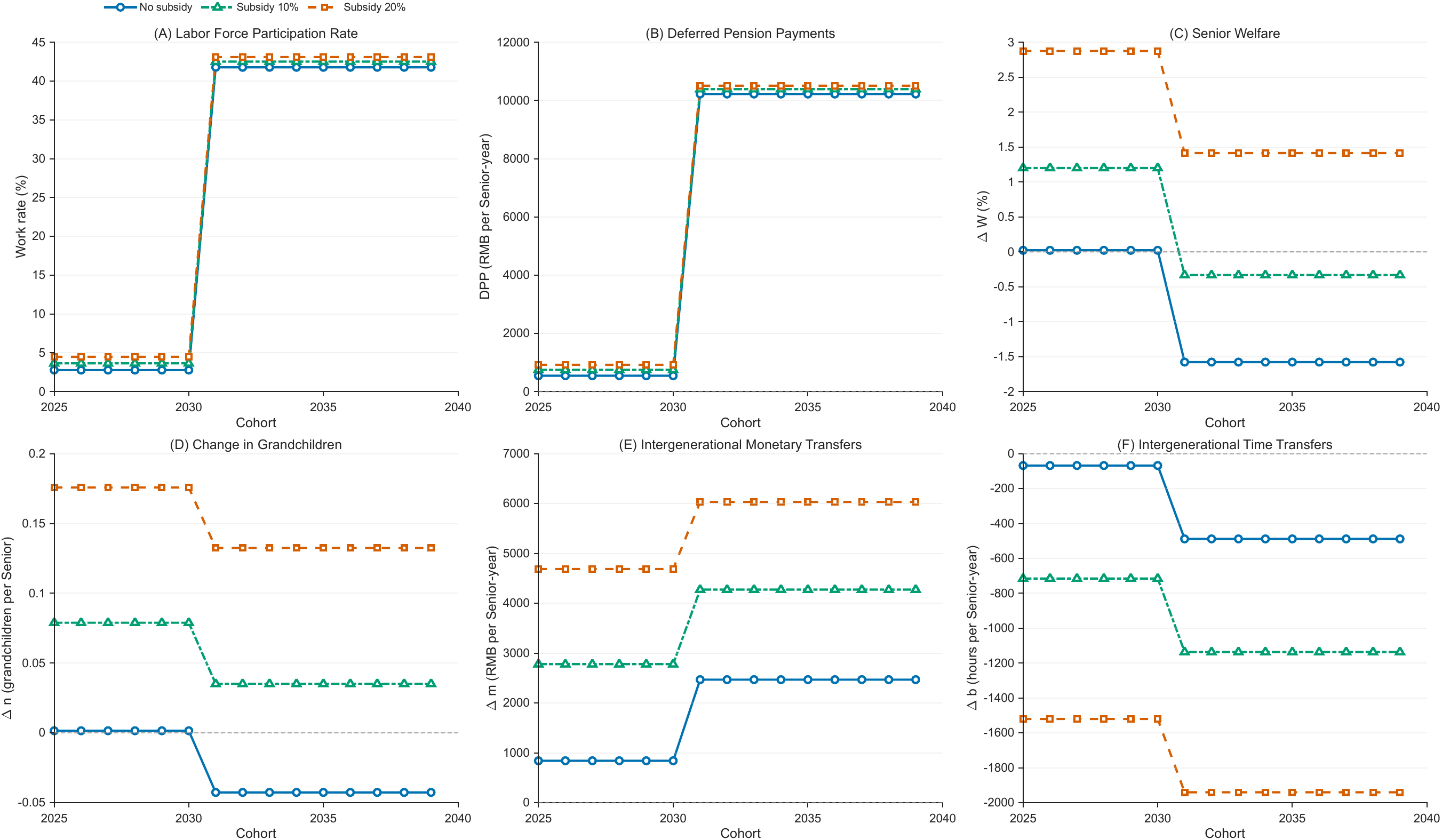}
\caption{Universal Childcare Subsidies under GDR}
\label{fig:cf_subsidy_curves}
\begin{minipage}{0.94\textwidth}
\footnotesize
\textit{Notes:}
This figure reports counterfactual outcomes under GDR combined with universal childcare subsidies of 0, 10\%, and 20\%. The horizontal axis indicates the year in which a Senior reaches the pre-reform statutory retirement age, age 60 for men and age 50 for women. Panel A reports the Senior labor force participation rate, and Panel B reports annual DPP. Panels C--F report changes relative to the benchmark in the mean Senior welfare index, the average number of grandchildren, annual intergenerational monetary transfers, and annual grandparental childcare, respectively. The welfare index is constructed as defined in Appendix~\ref{app:counterfactual}, and its change is expressed in percentage terms. The number of grandchildren is measured as the average number per Senior. DPP and monetary transfers are measured in 2018 RMB per Senior-year, and grandparental childcare is measured in hours per Senior-year. DPP measure annual pension payments deferred at the evaluation age and do not represent a reduction in lifetime pension expenditure. Panels A and B of Table~\ref{tab:cf_childcare_policy_summary} report the corresponding values for the three subsidy rates shown in the figure.
\end{minipage}
\end{figure}

Figure~\ref{fig:cf_subsidy_curves} compares cohort outcomes under no subsidy and universal subsidies of 10\% and 20\%, while Panels A and B of Appendix Table~\ref{tab:cf_childcare_policy_summary} report the corresponding cohort averages. Because the negative effects of GDR on Senior welfare and Junior fertility emerge for the 2031--2039 cohorts, we focus on these cohorts below. Relative to the no-subsidy case, a universal childcare subsidy increases Senior welfare and the number of grandchildren while also slightly increasing Senior labor force participation and DPP. As the subsidy rate rises, Senior welfare and the number of grandchildren continue to increase, while labor force participation and DPP also rise slightly. At the same time, intergenerational monetary transfers increase and grandparental childcare declines further, reflecting substitution away from grandparental childcare as the price of formal childcare falls.

At a childcare subsidy rate of \(s_n^*\approx5.7\%\), the number of grandchildren for the 2031--2039 cohorts equals its pre-reform benchmark, while Senior welfare remains 0.91\% below its pre-reform benchmark. When the subsidy rate rises to \(s_{\mathcal W}^*\approx12.2\%\),  Senior welfare also equals its pre-reform benchmark, while the number of grandchildren is 0.054 above its pre-reform benchmark. Under the baseline assumptions for the fiscal calculation, achieving these two targets requires childcare subsidy expenditures of RMB 836 and RMB 1905 per Senior-year, respectively\footnote{ In the model, total childcare demand is met by the Junior's own childcare, grandparental childcare provided by the Senior, and formal childcare: $ \bar h n_j=b_j+b_s+f_j. $ Because CHARLS does not report the Junior's own childcare time, \(b_j\), we cannot directly observe formal childcare demand. For the fiscal calculation, we therefore assume that formal childcare provides one half of the childcare demand remaining after grandparental childcare and the Junior provides the other half: $ f_j=\lambda(\bar h n_j-b_s), \lambda=0.50. $ Section~\ref{sec:sensitivity} examines how fiscal expenditure changes under alternative values of \(\lambda\).}.

These results show that universal childcare subsidies can offset the negative effects of GDR on average Junior fertility and mean Senior welfare without weakening its employment or DPP gains. Under the baseline assumptions, achieving the respective targets requires subsidy rates of approximately 5.7\% and 12.2\%. The corresponding annual fiscal expenditures of RMB 836 and RMB 1905 per Senior are substantially below the DPP of RMB 10302 and RMB 10406 for the same 2031--2039 cohorts, indicating that the annual subsidy costs are relatively limited compared with annual DPP.

\subsection{Targeting Childcare Support}
\label{sec:cf_targeted}
Section~\ref{sec:cf_subsidies} shows that universal childcare subsidies can offset the adverse effects of GDR on Senior welfare and Junior fertility. We now examine the choice of subsidy coverage. Our analysis of GDR shows that welfare and fertility losses among the 2031--2039 cohorts are concentrated primarily among households with female and lower-income Seniors. Motivated by this heterogeneity, we first restrict eligibility to households with female Seniors and then to those with lower-income female Seniors. We examine whether narrower coverage reduces the fiscal expenditure required to achieve the same mean welfare or fertility target.

\begin{figure}[H]
\centering
\includegraphics[width=\textwidth]{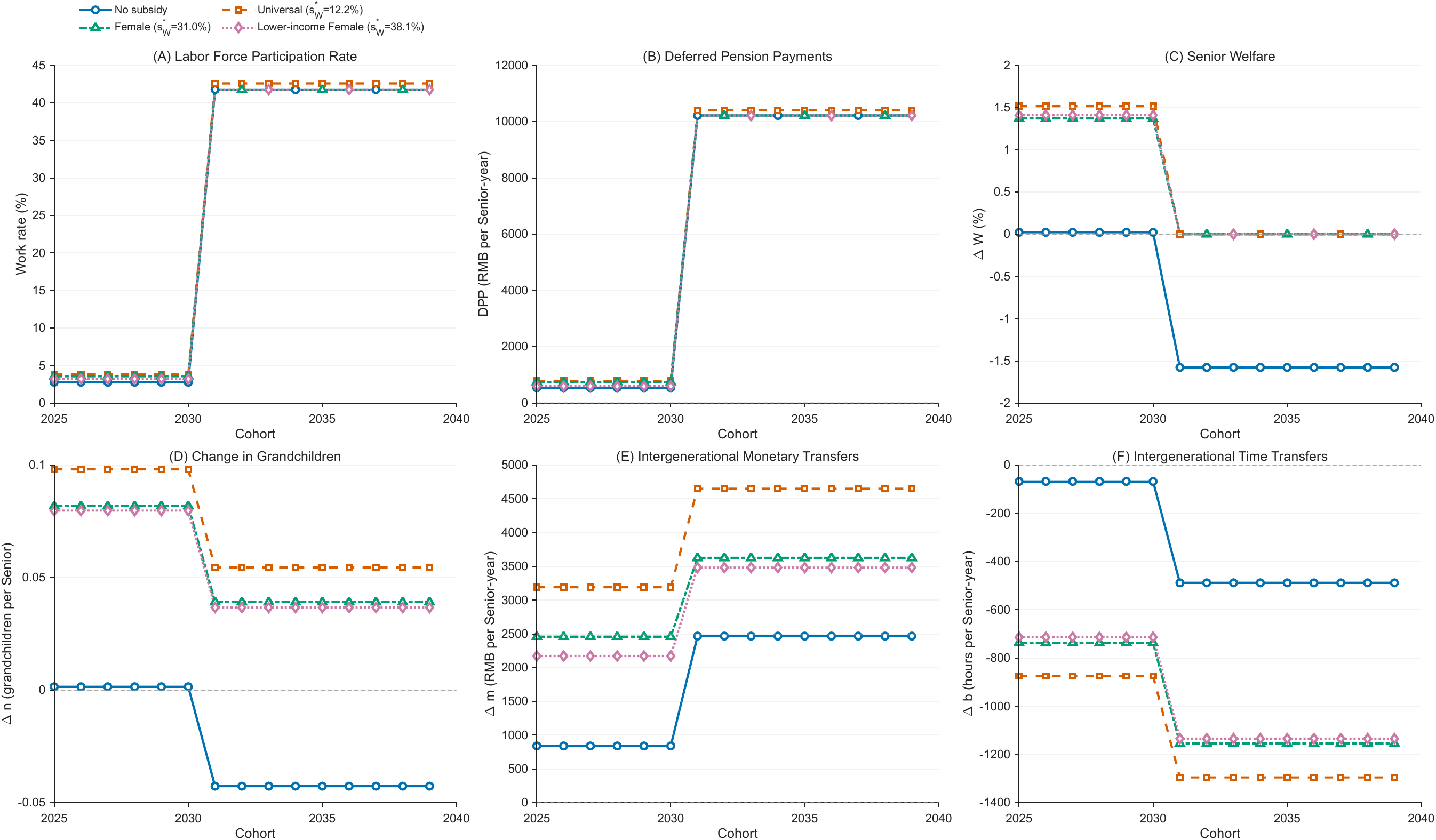}
\caption{Childcare Subsidies by Eligibility Rule: Offsetting the GDR-Induced Loss in Mean Senior Welfare}
\label{fig:cf_targeting_cohorts}
\begin{minipage}{0.94\textwidth}
\footnotesize
\textit{Notes:}
This figure compares counterfactual outcomes under GDR without a childcare subsidy and with subsidies available universally, to households with female Seniors, or to households with lower-income female Seniors. For each scheme, the subsidy rate is set at the level required for full-sample mean Senior welfare among the 2031--2039 cohorts to equal its pre-reform benchmark. The corresponding subsidy rates are 12.2\%, 31.0\%, and 38.1\%, respectively. The horizontal axis indicates the year in which a Senior reaches the pre-reform statutory retirement age. Panels A and B report the Senior labor force participation rate and DPP, respectively. Panels C--F report changes relative to the pre-reform benchmark in Senior welfare, the average number of grandchildren, annual intergenerational monetary transfers, and annual grandparental childcare, respectively. Seniors are classified as lower or higher income relative to the median income within each survey wave $\times$ retirement status cell. DPP and monetary transfers are measured in 2018 RMB per Senior-year, and grandparental childcare is measured in hours per Senior-year. Table~\ref{tab:cf_childcare_policy_summary} reports the corresponding average fiscal expenditures and 2031--2039 cohort averages.
\end{minipage}
\end{figure}

We first examine the subsidy rate and fiscal expenditure required to offset the adverse effect of GDR on mean Senior welfare for the 2031--2039 cohorts. As shown in Figure~\ref{fig:cf_targeting_cohorts} and Table~\ref{tab:cf_childcare_policy_summary}, universal coverage requires a subsidy rate of 12.2\%, with average fiscal expenditure of RMB 1905 per Senior-year. Restricting the subsidy to households with female Seniors raises the required rate to 31.0\% while reducing average fiscal expenditure to RMB 1467 per Senior-year.\footnote{ Fiscal expenditure is averaged over the full sample, with zero expenditure assigned to households ineligible for the subsidy.} Further restricting the subsidy to households with lower-income female Seniors raises the required rate to 38.1\%, while average fiscal expenditure falls further to RMB 1383 per Senior-year. Thus, narrower coverage requires a larger reduction in the childcare price for eligible households to offset the same full-sample mean welfare loss. In our simulations, the reduction in coverage more than offsets the higher subsidy rate, so average fiscal expenditure declines. The subsidy rate alone therefore does not fully capture the fiscal cost of the policy; coverage also matters.

If the objective is to offset the adverse effect of GDR on the average number of grandchildren for the 2031--2039 cohorts, all three schemes require substantially lower subsidy rates than under the welfare target. The required rates are 5.7\%, 19.0\%, and 24.9\% under universal coverage, coverage restricted to households with female Seniors, and coverage restricted to households with lower-income female Seniors, respectively. The corresponding average fiscal expenditures are RMB 836, RMB 800, and RMB 790 per Senior-year. As in the welfare comparison, narrower coverage raises the subsidy rate required for eligible households but reduces average fiscal expenditure across the full sample. However, the reduction in fiscal expenditure is substantially smaller than under the welfare target, indicating that the fiscal savings from targeting depend on the policy objective.

These results reveal a trade-off between subsidy intensity and coverage. Narrowing eligibility requires larger subsidies for eligible households but can reduce the average fiscal expenditure needed to offset the GDR-induced losses in full-sample mean Senior welfare or average Junior fertility. The fiscal advantage of targeting depends on the policy objective.

\section{Sensitivity Analysis}
\label{sec:sensitivity}

The counterfactual analysis shows that the effects of GDR differ substantially across cohorts. For the 2025--2030 cohorts, Seniors can still retire at the pre-reform statutory retirement age, and the reform produces only small increases in Senior employment, mean Senior welfare, and average Junior fertility. For the 2031--2039 cohorts, however, female Seniors whose pre-reform statutory retirement age was 50 face a binding minimum retirement-age constraint. GDR substantially increases Senior employment and DPP while reducing mean Senior welfare and average Junior fertility. The subsequent analysis shows that childcare subsidies can offset these welfare and fertility losses while maintaining the reform's employment gains. We therefore focus the sensitivity analysis on these central counterfactual results for the 2031--2039 cohorts, considering externally calibrated values, assumptions about formal childcare, and alternative income-group classifications.

We first examine whether the main conclusions about the GDR reform depend on the externally calibrated values. In the structural estimation, the price of formal childcare, Senior work time, and childcare requirement per child are calibrated using external information or assumed parameter values. We assess the sensitivity of the counterfactual results to the values assigned to these parameters. The baseline specification sets the formal childcare price at RMB 10.91 per hour and assumes 2000 annual hours for both Senior work time and the childcare requirement per child: $\bar h_s=\bar h=\frac{2000}{5840}\approx0.342.$

Table~\ref{tab:sensitivity_calibration} varies the formal childcare price, Senior work time, and childcare requirement per child separately, re-estimating the structural parameters under each calibration. Although the magnitudes of the policy effects vary, the main conclusions about the GDR reform remain unchanged. Across specifications, the Senior employment rate ranges from 39.5\% to 43.6\%, while DPP range from RMB 9796 to RMB 10636 per Senior-year. The average number of grandchildren remains below its pre-reform benchmark in every specification, with declines ranging from 0.026 to 0.083. Mean Senior welfare also declines in all specifications, by 0.94\%--2.89\%. Thus, the finding that the GDR reform increases Senior employment and DPP while reducing Senior welfare and Junior fertility is not driven by a particular set of externally calibrated values.

\begin{table}[htbp]
\centering
\caption{Sensitivity to Structural Calibration}
\label{tab:sensitivity_calibration}
\begin{threeparttable}
\small
\setlength{\tabcolsep}{4pt}
\begin{tabular*}{\textwidth}{@{\extracolsep{\fill}}lrrrrrrrr@{}}
\toprule
Calibration & $\eta_n$ & $\eta_m$ & $\eta_b$ & $\eta_j$ & Work (\%) & DPP & $\Delta\mathcal W$ (\%) & $\Delta n$ \\
\midrule
Baseline & 0.123 & 0.122 & 0.500 & 0.255 & 41.77 & 10,221 & -1.578 & -0.043 \\
$p=15$ & 0.162 & 0.099 & 0.506 & 0.234 & 40.45 & 9,920 & -2.114 & -0.059 \\
$p=25$ & 0.220 & 0.065 & 0.516 & 0.200 & 39.49 & 9,796 & -2.894 & -0.083 \\
$\bar h_s=0.257$ & 0.233 & 0.170 & 0.343 & 0.254 & 43.56 & 10,636 & -2.165 & -0.056 \\
$\bar h_s=0.428$ & 0.058 & 0.094 & 0.593 & 0.256 & 40.74 & 9,987 & -0.942 & -0.026 \\
$\bar h=0.257$ & 0.147 & 0.116 & 0.482 & 0.255 & 41.77 & 10,221 & -2.508 & -0.068 \\
$\bar h=0.428$ & 0.100 & 0.128 & 0.517 & 0.255 & 41.77 & 10,221 & -1.032 & -0.028 \\
\bottomrule
\end{tabular*}
\begin{tablenotes}[flushleft]
\footnotesize
\item \textit{Notes:} This table reports the sensitivity of the structural estimates and the main counterfactual outcomes for the 2031--2039 cohorts to alternative calibrated values. The baseline sets $p=10.91$ RMB per hour and $\bar h_s=\bar h=2000/5840\approx0.342$. The alternative time inputs are $1500/5840\approx0.257$ and $2500/5840\approx0.428$; $\bar h$ is the childcare requirement per child, as in the theoretical model. Each specification changes one calibrated input at a time and uses the structural parameters estimated under that calibration. Work denotes the Senior employment rate. DPP is measured in 2018 RMB per Senior-year. $\Delta\mathcal W$ is the percentage change in mean Senior welfare relative to the pre-reform retirement benchmark, and $\Delta n$ is the change in the average number of grandchildren relative to the same benchmark.
\end{tablenotes}
\end{threeparttable}
\end{table}
\FloatBarrier

We next examine whether the policy conclusions about childcare subsidies depend on assumptions about formal childcare use and the development of the formal childcare market. The subsidy simulations require assumptions about both the share of childcare provided formally and the extent of search, matching, and information frictions in that market. We assess how these assumptions affect the subsidy rates and fiscal expenditure required to achieve the fertility and welfare targets. The baseline fiscal calculation assumes that formal childcare meets half of the childcare demand remaining after grandparental childcare, so \(\lambda=0.50\). The baseline also sets \(\rho_p=1\), representing a fully developed formal childcare market. Table~\ref{tab:sensitivity_childcare_market} varies these two parameters separately.

Changing the formal childcare share affects fiscal expenditure but not the subsidy rates required to eliminate the negative effects of GDR on average Junior fertility and mean Senior welfare. As \(\lambda\) increases from 0.25 to 1.00, the required rates remain approximately 5.7\% for fertility and 12.2\% for welfare. Annual expenditure rises from RMB 418 to RMB 1672 per Senior for the fertility target, and from RMB 952 to RMB 3810 for the welfare target. When a larger share of the childcare demand remaining after grandparental childcare is met by formal childcare rather than by the Junior, more formal childcare must be subsidized to achieve the same fertility or welfare target. Fiscal expenditure therefore rises even though the required subsidy rate is unchanged.

A less developed formal childcare market increases both the required subsidy rates and fiscal expenditure. When \(\rho_p\) falls from 1 to 0.75 and 0.50, the subsidy rate required to offset the GDR-induced decline in the average number of grandchildren rises to approximately 7.7\% and 11.5\%, respectively. The corresponding rates required to offset the loss in mean Senior welfare rise to approximately 16.2\% and 24.4\%. In a less developed market, search and matching costs and information asymmetry limit the reduction in households' effective childcare costs achieved by a given subsidy. A higher subsidy rate is therefore required to achieve the same fertility or welfare target. These results suggest that improving matching efficiency and reducing information asymmetry could strengthen the effects of childcare subsidies and lower the subsidy rates and fiscal expenditure required to achieve a given policy objective.

\begin{table}[htbp]
\centering
\caption{Sensitivity to Childcare-Market Assumptions}
\label{tab:sensitivity_childcare_market}
\begin{threeparttable}
\small
\setlength{\tabcolsep}{4pt}
\begin{tabular*}{\textwidth}{@{\extracolsep{\fill}}llrrrr@{}}
\toprule
Assumption & Value & \multicolumn{2}{c}{Fertility target} & \multicolumn{2}{c}{Welfare target} \\
\cmidrule(lr){3-4}\cmidrule(lr){5-6}
 & & $s_n^*$ (\%) & Expenditure & $s_{\mathcal W}^*$ (\%) & Expenditure \\
\midrule
Baseline &  & 5.74 & 836 & 12.18 & 1,905 \\
\addlinespace
Formal-care share $\lambda$ & 0.25 & 5.74 & 418 & 12.18 & 952 \\
 & 0.75 & 5.74 & 1,254 & 12.18 & 2,857 \\
 & 1.00 & 5.74 & 1,672 & 12.18 & 3,810 \\
\addlinespace
Market development $\rho_p$ & 0.75 & 7.66 & 1,114 & 16.23 & 2,540 \\
 & 0.50 & 11.49 & 1,672 & 24.35 & 3,810 \\
\bottomrule
\end{tabular*}
\begin{tablenotes}[flushleft]
\footnotesize
\item \textit{Notes:} This table reports the sensitivity of universal childcare subsidies for the 2031--2039 cohorts to alternative assumptions about formal childcare use and formal childcare market development. $s_n^*$ denotes the minimum subsidy rate required for the full-sample average number of grandchildren to equal its pre-reform benchmark, and $s_{\mathcal W}^*$ denotes the corresponding rate for mean Senior welfare. Expenditure is measured in 2018 RMB per Senior-year and averaged over the full sample. The baseline sets $\lambda=0.50$ and assumes a fully developed formal childcare market ($\rho_p=1$). Except for the assumption varied in each row, all other model inputs are held at their baseline values.
\end{tablenotes}
\end{threeparttable}
\end{table}
\FloatBarrier

Finally, we examine whether the income-group classification affects the heterogeneity and targeted-subsidy results. The theoretical model permits a classification based on $p\bar h_s<w_s\bar h_s-P_s.$ This model-based classification, however, places relatively few observations in the higher-income group. In the preceding analysis, Seniors are classified as lower or higher income relative to the median income within each survey wave $\times$ retirement status cell. We refer to this classification as the empirical median split. Here, we repeat the analysis using the model-based split to assess whether the main findings on income heterogeneity and targeted subsidies depend on the classification. Under the model-based split, 9259 of the 9523 observations are classified as lower income, leaving only 264 observations in the higher-income group. Among women, the corresponding sample sizes are 3714 and 37. The empirical median split produces substantially more balanced income groups. Table~\ref{tab:sensitivity_income_definition} compares the effects of the GDR reform and the results for targeted childcare subsidies under the two classifications.

Panel A shows substantial welfare and fertility losses among lower-income female Seniors under both classifications. Under the empirical median split, their mean welfare falls by 5.60\%, and the average number of grandchildren falls by 0.114. Under the model-based split, the corresponding declines are 6.12\% and 0.113. Although group composition differs substantially, the welfare and fertility losses for this group remain similar. The finding of substantial adverse effects among lower-income female Seniors is therefore broadly robust to this alternative income-group classification.

Panel B compares subsidies targeted to households with lower-income female Seniors. Switching from the empirical median split to the model-based split increases the number of eligible observations from 2301 to 3714. The subsidy rate required to offset the GDR-induced decline in the full-sample average number of grandchildren falls from 24.9\% to 19.0\%; the rate required to offset the loss in full-sample mean Senior welfare falls from 38.1\% to 31.0\%. The respective fiscal expenditures change from RMB 790 and RMB 1383 to RMB 799 and RMB 1467 per Senior-year. Thus, the eligibility definition affects the required rates and expenditure, but the comparison with universal subsidies remains unchanged: targeting requires a higher rate for eligible households and lower average fiscal expenditure across the full sample.

The three sets of sensitivity analyses support the main qualitative findings for the 2031--2039 cohorts. Alternative externally calibrated values do not change the finding that GDR raises Senior employment and DPP while reducing mean Senior welfare and average Junior fertility. Assumptions about formal childcare use and formal childcare market development affect the subsidy rates or fiscal expenditure required to offset these losses, but do not alter the conclusion that childcare subsidies can achieve the full-sample mean fertility and welfare targets. The main findings on income heterogeneity and targeted childcare subsidies also remain broadly unchanged under the alternative income-group classification.

\begin{table}[htbp]
\centering
\caption{Sensitivity to the Definition of Lower Income}
\label{tab:sensitivity_income_definition}
\begin{threeparttable}
\footnotesize
\setlength{\tabcolsep}{3pt}
\renewcommand{\arraystretch}{1.08}
\begin{tabular*}{\textwidth}{@{\extracolsep{\fill}}llrrrrr@{}}
\toprule
\multicolumn{7}{l}{\textit{Panel A. GDR Effects, 2031--2039 Cohorts}} \\
\addlinespace
Definition & Group & $N$ & Work (\%) & DPP & $\Delta\mathcal W$ (\%) & $\Delta n$ \\
\midrule
\shortstack[l]{Empirical median\\split} & Lower income & 4,771 & 49.76 & 10,479 & -2.000 & -0.054 \\
 & Higher income & 4,752 & 33.75 & 9,961 & -1.133 & -0.031 \\
 & Lower-income women & 2,301 & 100.00 & 21,339 & -5.596 & -0.114 \\
 & Higher-income women & 1,450 & 100.00 & 30,298 & -7.314 & -0.108 \\
\addlinespace
\shortstack[l]{Model-based\\split} & Lower income & 9,259 & 40.11 & 9,953 & -1.612 & -0.045 \\
 & Higher income & 264 & 100.00 & 19,601 & 3.148 & 0.055 \\
 & Lower-income women & 3,714 & 100.00 & 24,813 & -6.117 & -0.113 \\
 & Higher-income women & 37 & 100.00 & 23,684 & 2.049 & 0.063 \\
\bottomrule
\end{tabular*}
\par\medskip
\begin{tabular*}{\textwidth}{@{\extracolsep{\fill}}llrrrrrr@{}}
\toprule
\multicolumn{8}{l}{\textit{Panel B. Childcare Subsidies for Lower-Income Women, 2031--2039 Cohorts}} \\
\addlinespace
Definition & Subsidy & $N$ & Work (\%) & DPP & $\Delta\mathcal W$ (\%) & $\Delta n$ & Expenditure \\
\midrule
\shortstack[l]{Empirical median\\split} & No subsidy & 2,301 & 41.77 & 10,221 & -1.578 & -0.043 & 0 \\
 & 24.86\% ($s_n^*$) &  & 41.77 & 10,221 & -0.779 & 0.000 & 790 \\
 & 38.11\% ($s_{\mathcal W}^*$) &  & 41.77 & 10,221 & 0.000 & 0.037 & 1,383 \\
\addlinespace
\shortstack[l]{Model-based\\split} & No subsidy & 3,714 & 41.77 & 10,221 & -1.578 & -0.043 & 0 \\
 & 19.01\% ($s_n^*$) &  & 41.77 & 10,221 & -0.801 & 0.000 & 799 \\
 & 31.04\% ($s_{\mathcal W}^*$) &  & 41.77 & 10,221 & 0.000 & 0.039 & 1,467 \\
\bottomrule
\end{tabular*}
\begin{tablenotes}[flushleft]
\footnotesize
\item \textit{Notes:} Panel A compares GDR effects across income groups for the 2031--2039 cohorts. Under the empirical median split, Seniors are classified as lower or higher income relative to the median income within each survey wave $\times$ retirement status cell. The model-based rule defines lower income by $p_0\bar h_s>\widehat w_i  \bar h_s-\widehat P_s$. Panel B compares subsidies targeted to eligible female Seniors under the two definitions. $s_n^*$ and $s_{\mathcal W}^*$ eliminate the full-sample declines in the average number of grandchildren and mean Senior welfare, respectively. In Panel A, $N$ is the number of observations in the group; in Panel B, $N$ is the number eligible for targeting and applies to all three rows of that definition. Panel B outcomes and expenditure are full-sample averages, with zero expenditure assigned to ineligible households. Work denotes Senior employment. $\Delta\mathcal W$ and $\Delta n$ are changes relative to the pre-reform retirement benchmark. DPP and expenditure are in 2018 RMB per Senior-year.
\end{tablenotes}
\end{threeparttable}
\end{table}
\FloatBarrier

\section{Conclusion}
\label{sec:conclusion}

Against the backdrop of population aging and low fertility, extending working lives can increase older individuals’ labor supply and defer current pension payments. However, when grandparents are heavily involved in family care, the effects of retirement policy may extend beyond older individuals themselves and be transmitted to the next generation through the reallocation of resources within the family. In 2025, China began implementing a gradual increase in the statutory retirement age, while introducing flexible early- and delayed-retirement arrangements. This paper examines how retirement and delayed retirement affect the fertility of adult children and the welfare of older individuals through intergenerational monetary and time transfers. To this end, we first develop a theoretical model that incorporates Seniors’ retirement, intergenerational transfers, and the fertility decisions of their adult children, and use the model to characterize the mechanisms through which retirement affects fertility and welfare. We then exploit the institutional variation generated by the pre-reform statutory retirement ages and use a fuzzy regression discontinuity design to identify the causal effects of retirement on the number of grandchildren, intergenerational monetary transfers, and grandparental childcare. Building on these estimates, we estimate the structural parameters of the model and further evaluate the effects of China’s 2025 gradual delayed-retirement reform, alternative retirement arrangements, and childcare subsidies.

We obtain three main findings. First, the RD estimates show that, around the pre-reform statutory retirement-age cutoffs, retirement increases the number of grandchildren by 0.613, reduces annual net monetary transfers to adult children by approximately RMB 4,177, and increases annual grandparental childcare by approximately 786 hours. Given high formal childcare prices, the market replacement value of the additional grandparental childcare exceeds the reduction in monetary transfers, resulting in a positive net effect of retirement on fertility. Subgroup estimates further show that the positive fertility effect of retirement is concentrated primarily among female Seniors and among lower-income Seniors.

Second, the effects of GDR depend on whether the reform imposes binding constraints on Seniors' retirement choices. For cohorts reaching their pre-reform statutory retirement age in 2025--2030, retirement at that age remains feasible, so changes in Senior employment, mean Senior welfare, and average Junior fertility are small. For the 2031--2039 cohorts, the higher minimum retirement age constrains some women whose pre-reform statutory retirement age was 50, requiring some Seniors who would otherwise retire to continue working. Senior labor force participation therefore reaches 41.8\%, and annual deferred pension payments (DPP) reach RMB 10,221 per Senior. Relative to the pre-reform benchmark, however, mean Senior welfare falls by 1.58\% and the average number of grandchildren declines by 0.043. These welfare and fertility losses arise because the additional monetary support generated by continued work is insufficient to cover the market replacement cost of reduced grandparental childcare. Comparisons with alternative retirement regimes further show that, under fully flexible retirement (FFR), mean Senior welfare and average Junior fertility remain close to their pre-reform levels, but employment and annual DPP gains are smaller than under GDR. Mandatory partial retirement (MPR) and mandatory full-time work (MFT) generate larger employment and annual DPP gains than GDR, but also larger welfare and fertility losses.

Third, childcare subsidies reduce the cost of replacing grandparental childcare and can therefore mitigate the fertility and welfare losses caused by GDR while maintaining the gains in Senior employment and annual DPP generated by GDR. In the baseline simulations for the 2031--2039 cohorts, a universal childcare subsidy rate of approximately 5.7\% can offset the negative effect of GDR on full-sample average Junior fertility, while a rate of approximately 12.2\% can offset the negative effect of GDR on mean Senior welfare. The average annual subsidy expenditures required to achieve these two objectives are RMB 836 and RMB 1,905 per Senior, respectively. Restricting subsidies to households with female Seniors or lower-income female Seniors requires higher subsidy rates for eligible households but reduces average fiscal expenditure across the full sample.

These findings have four policy implications. First, evaluations of delayed retirement should consider Senior welfare and Junior fertility alongside current employment and pension payments. Lower fertility may, over the longer run, reduce the future working-age population and pension contribution base. Ignoring the fertility response may therefore lead evaluations to overstate the extent to which delayed retirement can ease the labor supply and pension pressures associated with population aging. Policy evaluations should also examine how benefits and costs are distributed across gender and income groups.

Second, policymakers should retain a degree of retirement choice when raising retirement ages and coordinate retirement ages, work intensity, and pension receipt rules. Our results show that voluntary continued work and continued employment induced by a minimum retirement-age constraint have different implications for welfare and family outcomes. The former increases labor supply primarily by expanding workers' choice sets, whereas the latter may require some workers who would otherwise retire to continue working, generating additional time and welfare costs. Countries seeking to increase older adults' labor supply by extending working lives should therefore avoid relying solely on a uniform increase in the minimum retirement age. Flexible early or delayed retirement, partial retirement, and pension receipt rules linked to working hours can provide the necessary flexibility for workers with different family caregiving responsibilities and income circumstances.

Third, childcare support should be an important complement to delayed retirement, with subsidy eligibility and intensity determined jointly according to explicit compensation objectives. Our findings show that offsetting the negative effect of GDR on average Junior fertility requires a different level of support from offsetting the negative effect of GDR on mean Senior welfare. Achieving a full-sample mean target does not imply that all affected households are fully compensated. Childcare subsidy design should therefore first establish the policy objective and then determine eligibility and subsidy levels according to households' caregiving responsibilities, income, and the extent of the reduction in grandparental childcare. Universal subsidies provide broader coverage, whereas targeted subsidies can reduce average fiscal expenditure but require greater support for eligible households. Countries with limited fiscal resources should explicitly weigh coverage, subsidy intensity, and fiscal costs rather than rely solely on a uniform subsidy rate as the policy instrument.

Fourth, the capacity and accessibility of formal childcare services should be improved alongside childcare subsidies. Subsidies can help households effectively replace the grandparental childcare reduced by delayed retirement only if households can actually access formal childcare services. Our sensitivity analysis shows that a less developed formal childcare market requires greater fiscal support to achieve the same fertility or welfare objective. In countries with limited formal childcare provision and substantial reliance on grandparental childcare, demand-side subsidies should therefore be combined with an expansion of formal childcare services. Increasing service availability and reducing the constraints households face in accessing formal childcare can improve the effectiveness of fiscal subsidies in enabling households to substitute formal childcare for reduced grandparental childcare. Delayed retirement, childcare support for families, and the development of the formal childcare market need to proceed together to increase older adults' labor supply while reducing the adverse consequences of retirement reform for family welfare and fertility decisions.

\clearpage
\appendix

\section{Additional RD Robustness Checks}
\label{app:rd_robustness}

This subsection evaluates both the validity of the RD design and the sensitivity of the baseline estimates to alternative specifications. Overall, the results support the credibility of the identification strategy and the robustness of the main findings, although some demanding specifications naturally reduce precision.

\subsection{Validity of the RD Design}

\begin{figure}[htbp]
\centering
\includegraphics[width=0.72\textwidth]{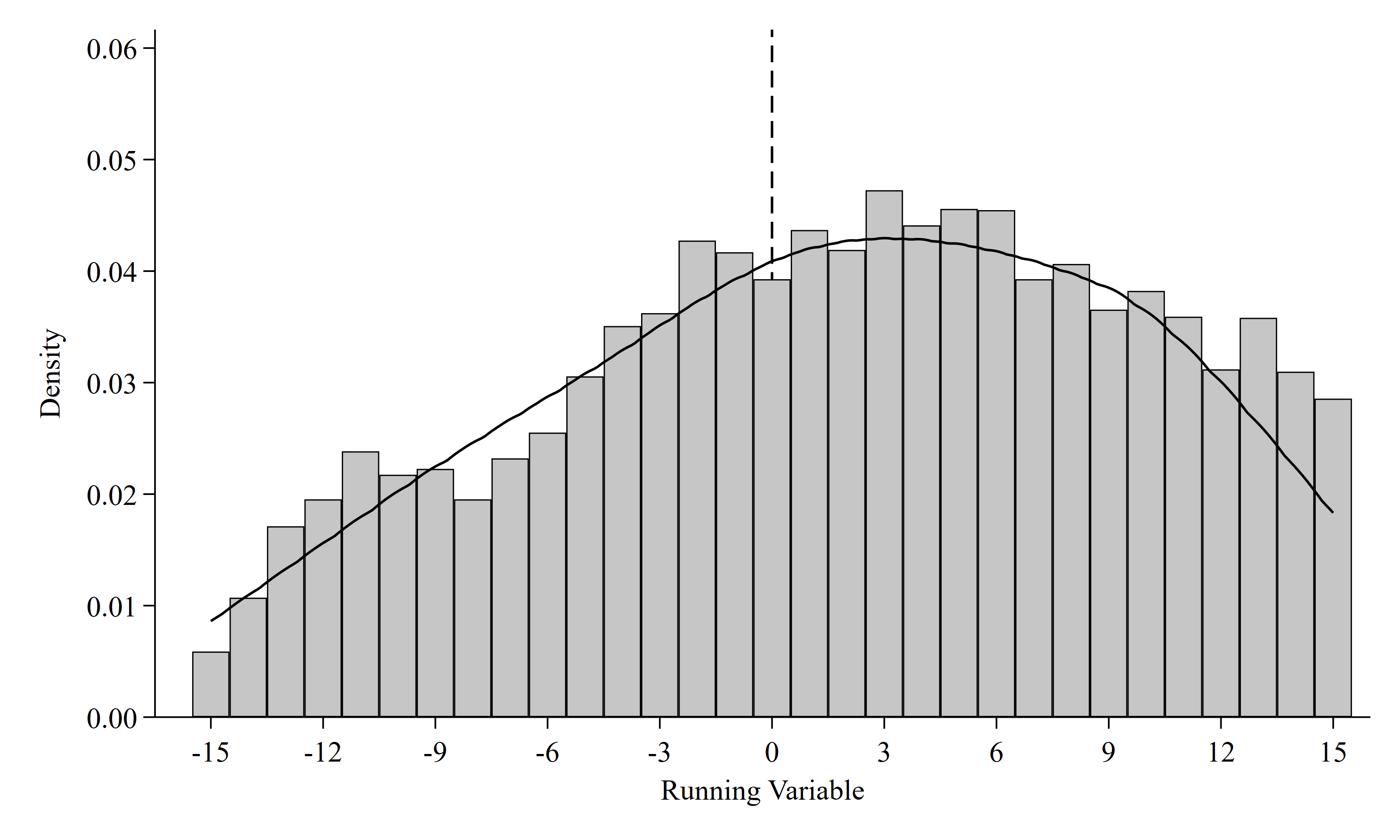}
\caption{Frequency of observations around the statutory retirement age}
\label{fig:running_variable_density}
\begin{minipage}{0.85\textwidth}
\footnotesize
\textit{Notes:} The figure plots the binned empirical density of the running variable, defined as age relative to the relevant statutory retirement cutoff. Bars denote density within one-year bins, and the solid line shows the kernel density estimate. The vertical dashed line marks the cutoff at zero. The absence of visible bunching or missing mass at the threshold is consistent with continuity of the running variable density around the cutoff.
\end{minipage}
\end{figure}

A key identifying assumption in RD designs is that the distribution of the running variable is continuous around the cutoff, so that treatment assignment is locally as good as random in a neighborhood of the threshold. In our setting, since the running variable is age relative to the relevant statutory retirement cutoff, precise manipulation of age is unlikely. A remaining concern is that sample selection or retirement-system coverage may generate discontinuities in the density of the running variable around the cutoff. Figure~\ref{fig:running_variable_density} plots the binned empirical density of the running variable around the cutoff. The figure shows no visible bunching or missing mass at the threshold.

To complement this visual evidence, we implement a density test for the running variable based on the local polynomial density estimation proposed by Cattaneo et al. (2018), using a triangular kernel, a first-order polynomial, and a bandwidth of 5 on both sides of the cutoff. The robust test statistic is \(-0.709\), with a \(p\)-value of 0.478, so the null hypothesis of continuity in the density at the cutoff cannot be rejected.

We next examine whether predetermined covariates are continuous at the statutory retirement cutoff by estimating the baseline RD specification with each covariate as the dependent variable. If the identifying assumptions of the RD design hold, no discontinuity should be observed at the threshold. Table~\ref{tab:covariate_balance} reports the results. Across all specifications, the discontinuities are statistically insignificant: 0.022 for female, 0.002 for rural hukou, 0.008 for urban residence, 0.007 for junior college education or above, and \(-0.015\) for the number of living Juniors. Overall, the absence of significant discontinuities in these predetermined characteristics suggests that observations just below and just above the cutoff are locally comparable, lending support to the validity of the RD design.

\begin{table}[htbp]
\centering
\caption{Continuity of Predetermined Covariates at the Cutoff}
\label{tab:covariate_balance}
\small
\begin{threeparttable}
\setlength{\tabcolsep}{4pt}
\begin{tabular}{lccccc}
\toprule
& \multicolumn{5}{c}{Dependent variables: predetermined covariates} \\
\cmidrule(lr){2-6}
& (1) & (2) & (3) & (4) & (5) \\
& Female & Rural hukou & Urban & \shortstack{Junior college\\or above} & \shortstack{Living\\Juniors} \\
\midrule
Crossing cutoff & 0.022 & 0.002 & 0.008 & 0.007 & -0.015 \\
& (0.022) & (0.312) & (0.011) & (0.024) & (0.016) \\
\addlinespace
Predetermined covariates & No & No & No & No & No \\
Province fixed effects & Yes & Yes & Yes & Yes & Yes \\
Wave fixed effects & Yes & Yes & Yes & Yes & Yes \\
Observations & 9,523 & 9,523 & 9,523 & 9,523 & 9,523 \\
Bandwidth & 5 & 5 & 5 & 5 & 5 \\
Eff. obs. left of cutoff & 1,483 & 1,483 & 1,483 & 1,483 & 1,483 \\
Eff. obs. right of cutoff & 2,059 & 2,059 & 2,059 & 2,059 & 2,059 \\
\bottomrule
\end{tabular}
\begin{tablenotes}
\footnotesize
\item \textit{Notes:} The coefficient row reports the Robust row from the RD covariate-continuity output. Robust standard errors are reported in parentheses. Statistical significance is based on robust confidence intervals centered on bias-corrected estimates, following Calonico et al. (2014, 2019). All specifications use a triangular kernel, a first-order local polynomial, and a symmetric bandwidth of 5 years. Province and wave fixed effects are included as indicated. *** \(p<0.01\), ** \(p<0.05\), * \(p<0.1\).
\end{tablenotes}
\end{threeparttable}
\end{table}

\subsection{Placebo Cutoffs}

A further way to assess the credibility of the RD design is to examine whether similar discontinuities arise at placebo cutoffs away from the true statutory retirement threshold. To this end, we implement a series of placebo tests by shifting the cutoff to alternative values of the running variable and re-estimating the baseline specification.

Table~\ref{tab:placebo_cutoffs} presents the resulting placebo-cutoff estimates for the number of grandchildren. The true-cutoff column reports the baseline main-regression estimate from Table~\ref{tab:main_results}, while the other columns report placebo estimates away from the statutory retirement threshold. The placebo estimates are statistically insignificant. Taken together, these results indicate that the statistically significant discontinuity is concentrated at the statutory retirement cutoff rather than reflecting a smooth age-related trend or other confounding factors.

\begin{table}[htbp]
\centering
\caption{Placebo Cutoff Tests for the Number of Grandchildren}
\label{tab:placebo_cutoffs}
\small
\begin{threeparttable}
\setlength{\tabcolsep}{5pt}
\begin{tabular}{lccccc}
\toprule
& \multicolumn{5}{c}{Dependent variable: Number of grandchildren} \\
\cmidrule(lr){2-6}
& \(c=-2\) & \(c=-1\) & \(c=0\) & \(c=1\) & \(c=2\) \\
\midrule
Retirement & -0.371 & -88.589 & 0.613** & 0.217 & 0.174 \\
& (3.227) & (66.667) & (0.311) & (0.483) & (1.131) \\
\addlinespace
Predetermined covariates & Yes & Yes & Yes & Yes & Yes \\
Province fixed effects & Yes & Yes & Yes & Yes & Yes \\
Wave fixed effects & Yes & Yes & Yes & Yes & Yes \\
Observations & 9,523 & 9,523 & 9,523 & 9,523 & 9,523 \\
Bandwidth & 5 & 5 & 5 & 5 & 5 \\
Eff. obs. left of cutoff & 1,213 & 1,377 & 1,483 & 1,523 & 1,594 \\
Eff. obs. right of cutoff & 1,993 & 2,036 & 2,059 & 2,119 & 2,136 \\
\bottomrule
\end{tabular}
\begin{tablenotes}
\footnotesize
\item \textit{Notes:} The coefficient row reports the Robust row from the placebo-cutoff fuzzy RD output. Robust standard errors are reported in parentheses. Statistical significance is based on robust confidence intervals centered on bias-corrected estimates, following Calonico et al. (2014, 2019). All specifications use a triangular kernel, a first-order local polynomial, and a symmetric bandwidth of 5 years. Predetermined covariates include gender, urban residence, rural hukou, junior college education or above, and the number of living Juniors. Province and wave fixed effects are included as indicated. *** \(p<0.01\), ** \(p<0.05\), * \(p<0.1\).
\end{tablenotes}
\end{threeparttable}
\end{table}

\subsection{Donut RD}

We further conduct a donut RD exercise that excludes observations exactly at the cutoff, \(r_s=0\). Although individuals cannot manipulate their age, observations located exactly at the statutory retirement-age cutoff may be distinctive because retirement procedures, pension claiming, or survey reporting may be concentrated at the statutory age. If these observations are systematically different from nearby observations, they could exert disproportionate influence on the local RD estimates. To examine this concern, we exclude individuals whose running variable is exactly equal to zero and re-estimate both the first-stage and LATE specifications.

The results are reported in Table~\ref{tab:donut_rd}. Excluding observations at the cutoff strengthens the first-stage discontinuity in retirement, from 22.4 percentage points in the baseline specification to 36.5 percentage points. The estimated effect of retirement on the number of grandchildren also remains positive and statistically significant.

\begin{table}[htbp]
\centering
\caption{Donut RD Estimates}
\label{tab:donut_rd}
\small
\begin{threeparttable}
\setlength{\tabcolsep}{7pt}
\begin{tabular}{lcc}
\toprule
& Baseline & Excluding \(r_s=0\) \\
\midrule
\multicolumn{3}{l}{\textit{Dependent variable: Retirement}} \\
Crossing cutoff & 0.224*** & 0.365*** \\
& (0.044) & (0.022) \\
\addlinespace
\multicolumn{3}{l}{\textit{Dependent variable: Number of grandchildren}} \\
Retirement & 0.613** & 1.211** \\
& (0.311) & (0.584) \\
\addlinespace
Predetermined covariates & Yes & Yes \\
Province fixed effects & Yes & Yes \\
Wave fixed effects & Yes & Yes \\
Observations & 9,523 & 9,149 \\
Bandwidth & 5 & 5 \\
Eff. obs. left of cutoff & 1,483 & 1,483 \\
Eff. obs. right of cutoff & 2,059 & 1,685 \\
\bottomrule
\end{tabular}
\begin{tablenotes}
\footnotesize
\item \textit{Notes:} The coefficient rows report the Robust rows from the donut RD outputs that exclude observations exactly at \(r_s=0\). Robust standard errors are reported in parentheses. Statistical significance is based on robust confidence intervals centered on bias-corrected estimates, following Calonico et al. (2014, 2019). All specifications use a triangular kernel, a first-order local polynomial, and a symmetric bandwidth of 5 years. Predetermined covariates include gender, urban residence, rural hukou, junior college education or above, and the number of living Juniors. Province and wave fixed effects are included as indicated. *** \(p<0.01\), ** \(p<0.05\), * \(p<0.1\).
\end{tablenotes}
\end{threeparttable}
\end{table}

\subsection{Bandwidth Sensitivity}

We also vary the bandwidth around the statutory retirement cutoff. Table~\ref{tab:bandwidth_sensitivity} shows that the estimated effect remains positive when we compare the baseline bandwidth of 5 with alternative bandwidths of 4, 6, and 7 years. The estimates remain statistically significant across all reported bandwidths.

\begin{table}[htbp]
\centering
\caption{Bandwidth Sensitivity: Number of Grandchildren}
\label{tab:bandwidth_sensitivity}
\small
\begin{threeparttable}
\setlength{\tabcolsep}{7pt}
\begin{tabular}{lcccc}
\toprule
& \multicolumn{4}{c}{Dependent variable: Number of grandchildren} \\
\cmidrule(lr){2-5}
& Baseline \(h=5\) & \(h=4\) & \(h=6\) & \(h=7\) \\
\midrule
Retirement & 0.613** & 2.659*** & 0.970*** & 0.710** \\
& (0.311) & (0.083) & (0.365) & (0.280) \\
\addlinespace
Predetermined covariates & Yes & Yes & Yes & Yes \\
Province fixed effects & Yes & Yes & Yes & Yes \\
Wave fixed effects & Yes & Yes & Yes & Yes \\
Observations & 9,523 & 9,523 & 9,523 & 9,523 \\
Bandwidth & 5 & 4 & 6 & 7 \\
Eff. obs. left of cutoff & 1,483 & 1,149 & 1,774 & 2,017 \\
Eff. obs. right of cutoff & 2,059 & 1,639 & 2,493 & 2,926 \\
\bottomrule
\end{tabular}
\begin{tablenotes}
\footnotesize
\item \textit{Notes:} The coefficient row reports the Robust row from the bandwidth-sensitivity fuzzy RD output. Robust standard errors are reported in parentheses. Statistical significance is based on robust confidence intervals centered on bias-corrected estimates, following Calonico et al. (2014, 2019). All specifications use a triangular kernel and a first-order local polynomial. Predetermined covariates include gender, urban residence, rural hukou, junior college education or above, and the number of living Juniors. Province and wave fixed effects are included as indicated. *** \(p<0.01\), ** \(p<0.05\), * \(p<0.1\).
\end{tablenotes}
\end{threeparttable}
\end{table}

\subsection{Alternative Specifications}

We assess the robustness of the baseline estimates by considering a range of alternative specifications. Table~\ref{tab:robustness} shows that the estimated effect of retirement on the number of grandchildren remains positive when we use a second-degree polynomial, replace the triangular kernel with a uniform kernel, cluster inference at the wave or household level, or redefine the running variable by assigning women with junior college education or above a retirement threshold of 55 and other women a threshold of 50. All reported estimates are statistically significant, suggesting that the main finding is not driven by a particular modeling choice.

\begin{table}[htbp]
\centering
\caption{Robustness Checks: Alternative Specifications}
\label{tab:robustness}
\scriptsize
\begin{threeparttable}
\setlength{\tabcolsep}{3pt}
\begin{tabular}{lcccccc}
\toprule
& \multicolumn{6}{c}{Dependent variable: Number of grandchildren} \\
\cmidrule(lr){2-7}
& Baseline & \shortstack{Second-degree\\polynomial} & \shortstack{Uniform\\kernel} & \shortstack{Cluster\\by wave} & \shortstack{Cluster by\\household} & \shortstack{Female 50/55\\cutoff} \\
\midrule
Retirement & 0.613** & 1.914*** & 0.692** & 0.746*** & 1.375** & 1.172** \\
& (0.311) & (0.539) & (0.314) & (0.263) & (0.593) & (0.466) \\
\addlinespace
Predetermined covariates & Yes & Yes & Yes & Yes & Yes & Yes \\
Province fixed effects & Yes & Yes & Yes & Yes & Yes & Yes \\
Wave fixed effects & Yes & Yes & Yes & Yes & Yes & Yes \\
Kernel & Triangular & Triangular & Uniform & Triangular & Triangular & Triangular \\
Polynomial order & 1 & 2 & 1 & 1 & 1 & 1 \\
Cluster & Running var. & Running var. & Running var. & Wave & Household & Running var. \\
Observations & 9,523 & 9,523 & 9,523 & 9,523 & 9,523 & 9,523 \\
Bandwidth & 5 & 4 & 4 & 4 & 4 & 5 \\
Eff. obs. left of cutoff & 1,483 & 1,483 & 1,483 & 1,483 & 1,483 & 1,446 \\
Eff. obs. right of cutoff & 2,059 & 2,059 & 2,059 & 2,059 & 2,059 & 2,012 \\
\bottomrule
\end{tabular}
\begin{tablenotes}
\footnotesize
\item \textit{Notes:} The coefficient row reports the Robust row from the alternative-specification fuzzy RD output. Robust standard errors are reported in parentheses. The Female 50/55 cutoff column assigns women with junior college education or above a cutoff of 55 and other women a cutoff of 50, using education as a proxy for cadre status. *** \(p<0.01\), ** \(p<0.05\), * \(p<0.1\).
\end{tablenotes}
\end{threeparttable}
\end{table}

\FloatBarrier
\clearpage

\section{Heterogeneous RD Effects}
\label{app:rd_heterogeneity}

\begin{table}[htbp]
\centering
\caption{Heterogeneous Effects of Retirement on the Number of Grandchildren}
\label{tab:heterogeneity_rd}
\small
\begin{threeparttable}
\setlength{\tabcolsep}{7pt}
\begin{tabular}{lcccc}
\toprule
& \multicolumn{4}{c}{Dependent variable: Number of grandchildren} \\
\cmidrule(lr){2-5}
& (1) & (2) & (3) & (4) \\
& Higher income & Lower income & Female & Male \\
\midrule
Retirement & -0.100 & 1.081*** & 1.194*** & 0.633 \\
& (0.211) & (0.399) & (0.327) & (0.502) \\
\addlinespace
Predetermined covariates & Yes & Yes & Yes & Yes \\
Province fixed effects & Yes & Yes & Yes & Yes \\
Wave fixed effects & Yes & Yes & Yes & Yes \\
Observations & 4,752 & 4,771 & 3,751 & 5,772 \\
Bandwidth & 5 & 5 & 5 & 5 \\
Eff. obs. left of cutoff & 593 & 890 & 707 & 776 \\
Eff. obs. right of cutoff & 952 & 1,107 & 899 & 1,160 \\
\bottomrule
\end{tabular}
\begin{tablenotes}
\footnotesize
\item \textit{Notes:} Robust standard errors are reported in parentheses and clustered by age relative to the statutory retirement cutoff. Statistical significance is based on robust confidence intervals centered on bias-corrected estimates, following Calonico et al. (2014, 2019). All specifications use a triangular kernel and a first-order local polynomial, with symmetric bandwidths as reported. Income groups are defined using the median within each $wave \times retirement$ cell. Predetermined covariates include gender, urban residence, rural hukou, junior college education or above, and the number of living Juniors; gender is constant within the gender subsamples and is omitted from those regressions. Province and wave fixed effects are included in all columns. Observations reports the total number of observations in each subgroup; effective observations are reported separately for each side of the cutoff. *** \(p<0.01\), ** \(p<0.05\), * \(p<0.1\).
\end{tablenotes}
\end{threeparttable}
\end{table}

\FloatBarrier
\clearpage

\section{Additional Counterfactual Results}
\label{app:counterfactual}

This appendix provides implementation details and additional results for the counterfactual exercises in Section~\ref{sec:counterfactual}. We first describe how we construct counterfactual household outcomes and Senior welfare from the estimated model, then report the results underlying the figures and policy comparisons in the main text.

\paragraph{Construction of counterfactual outcomes and welfare.}
For each retirement regime and childcare-subsidy policy, we simulate fertility, intergenerational monetary transfers, and grandparental childcare using equations~\eqref{eq:econ_outcome_n_R}--\eqref{eq:econ_outcome_b_R}. We summarize Senior welfare using a monotone transformation of the Senior's optimized utility, characterized in Section~\ref{subsec:senior_welfare}.

For individual \(i\), let \(r\) denote the retirement regime and \(p_i\) the childcare price paid by the household under the policy. We define the welfare index as
\begin{equation}
\mathcal W_i^r(p_i)
=
S_i^r(p_i)\,
p_i^{-(\widehat\eta_n+\widehat\eta_b)},
\label{eq:welfare_index_cf}
\end{equation}
where
\begin{equation}
S_i^r(p_i)
=
Y_i+I_i^r+p_i\Omega_i^r
\label{eq:effective_resource_cf}
\end{equation}
is the effective resource index. Here, \(Y_i\) denotes the component of the Junior household's resources not provided by the Senior, \(I_i^r\) denotes the Senior's income under regime \(r\), and \(\Omega_i^r\) denotes the Senior's non-work time available for leisure and grandparental childcare. We hold \(Y_i\) fixed at its recovered baseline value across counterfactual retirement regimes and childcare prices. Retirement policies therefore affect effective resources through the Senior's income and available time, while childcare subsidies also operate through the childcare price paid by the household.

\paragraph{Recovery of baseline resources.}
The counterfactual exercises are anchored to each household's observed baseline allocation. Let \(p_0\) denote the baseline childcare price, and let \(n_i^{obs}\), \(m_i^{obs}\), \(b_i^{obs}\), and \(R_i^{obs}\) denote the observed number of grandchildren, monetary transfers, grandparental childcare, and retirement status, respectively. We recover the baseline effective resource index from the model's fertility equation as
\begin{equation}
S_i^{obs}
=
\frac{p_0\bar h}{\widehat\eta_n}n_i^{obs}.
\label{eq:Sobs_recovery}
\end{equation}
We then recover the Senior's income and available time from the monetary- and time-transfer equations:
\begin{align}
I_i^{obs}
&=
m_i^{obs}+\widehat\eta_m S_i^{obs},
\label{eq:Iobs_recovery}
\\
\Omega_i^{obs}
&=
b_i^{obs}
+
\frac{\widehat\eta_b}{p_0}S_i^{obs}.
\label{eq:Omegaobs_recovery}
\end{align}

Let
\[
\Delta I_i=\widehat P_i-\widehat W_i
\]
denote the change in the Senior's income associated with retirement. We recover the corresponding primitives in the work state as
\begin{align}
I_i^0
&=
I_i^{obs}-R_i^{obs}\Delta I_i,
\label{eq:I0_recovery}
\\
\Omega_i^0
&=
\Omega_i^{obs}-R_i^{obs}\bar h_s,
\label{eq:Omega0_recovery}
\end{align}
and recover the Junior household's residual resource component as
\begin{equation}
Y_i
=
S_i^{obs}-I_i^{obs}-p_0\Omega_i^{obs}.
\label{eq:Y_recovery}
\end{equation}
This procedure ensures that the recovered primitives reproduce each household's observed allocation at the baseline childcare price.

For any counterfactual regime \(r\), we construct the Senior's income and available time as
\begin{align}
I_i^r
&=
I_i^0+R_i^r\Delta I_i,
\label{eq:cf_income_state}
\\
\Omega_i^r
&=
\Omega_i^0+R_i^r\bar h_s.
\label{eq:cf_time_state}
\end{align}
Substituting equations~\eqref{eq:cf_income_state} and~\eqref{eq:cf_time_state} into equation~\eqref{eq:effective_resource_cf} yields the effective resources available to the household under each counterfactual retirement and childcare policy.

\paragraph{Retirement choice.}
Whenever retirement is feasible, the Senior chooses between full retirement and continued employment by comparing welfare in the two states. Because the welfare index in equation~\eqref{eq:welfare_index_cf} is monotone in effective resources at a given childcare price, the optimal retirement decision can be written as
\begin{equation}
R_i^\ast(p_i)
=
\mathbf 1
\left\{
\widehat P_i-\widehat W_i+p_i\bar h_s\geq0
\right\}.
\label{eq:retirement_choice_rule}
\end{equation}
Thus, the Senior retires when the value of the additional non-work time released by retirement is at least as large as the associated income loss. When the minimum retirement age imposed by policy binds, however, this choice is unavailable, and the Senior must remain employed until becoming eligible to retire.

\paragraph{Welfare measurement.}
For each counterfactual exercise, we evaluate welfare on a fixed common support across the policies compared. Let \(\mathcal C\) denote the relevant common support and \(\omega_i\) the sampling weight. We calculate the weighted mean welfare index as
\begin{equation}
\overline{\mathcal W}^{\,r}(p)
=
\frac{
\sum_{i\in\mathcal C}
\omega_i\mathcal W_i^r(p_i)
}{
\sum_{i\in\mathcal C}\omega_i
}.
\label{eq:mean_welfare_cf}
\end{equation}
The reported welfare effect is the percentage change in this mean index relative to the benchmark of mandatory retirement at the pre-reform statutory retirement age and the baseline childcare price:
\begin{equation}
\Delta \mathcal W^r(p)
=
100\times
\left[
\frac{
\overline{\mathcal W}^{\,r}(p)
}{
\overline{\mathcal W}^{\,old}(p_0)
}
-1
\right].
\label{eq:welfare_pct_cf}
\end{equation}

\FloatBarrier
\clearpage
\subsection{Gradual and Alternative Retirement Regimes}
\label{app:retirement_regimes}

\begin{figure}[H]
    \centering
    \includegraphics[width=\textwidth]{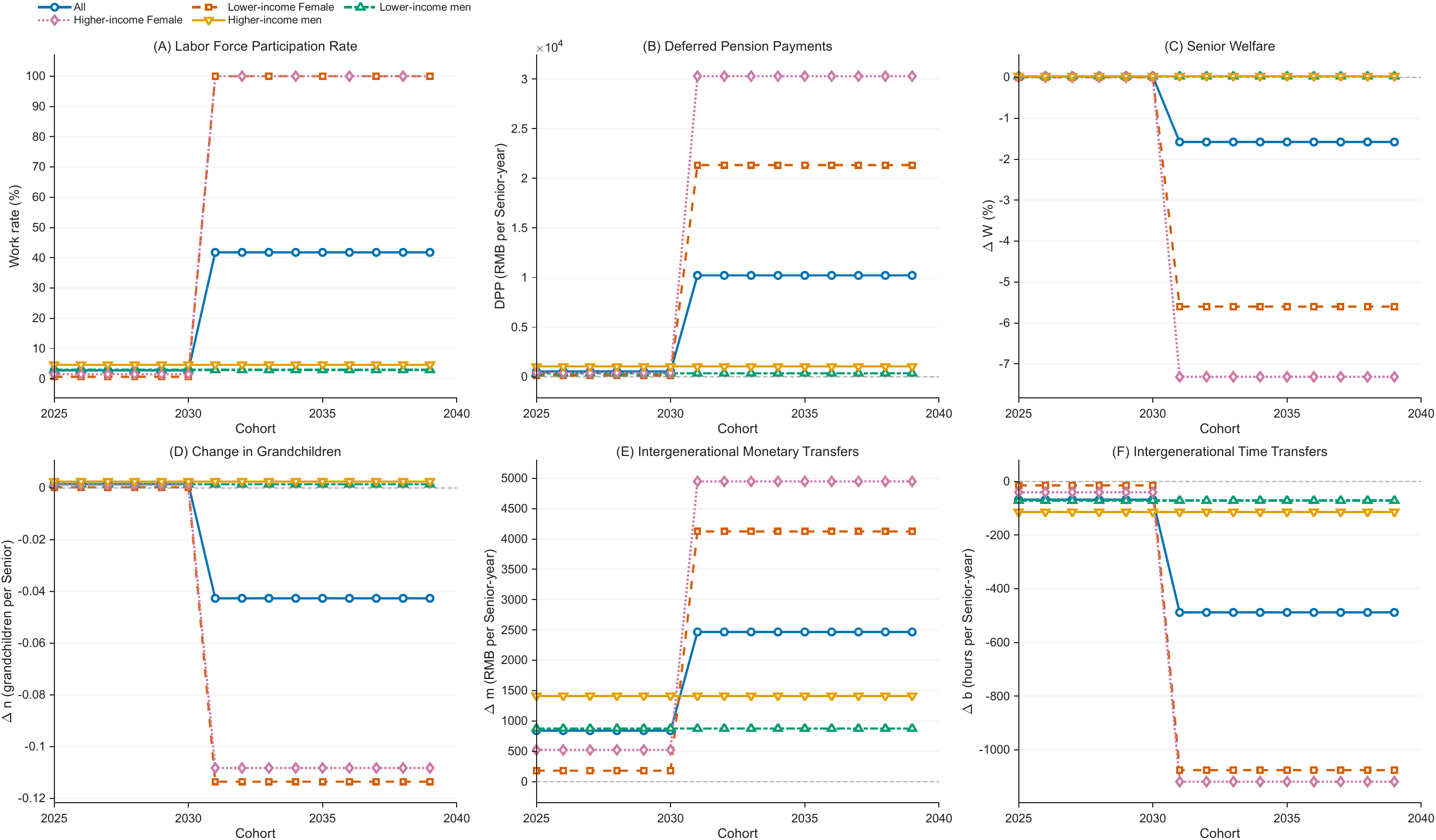}
    \caption{GDR: Cohort Effects by Gender and Income}
    \label{fig:cf_gradual_broad}
    \begin{minipage}{0.94\textwidth}
        \footnotesize
        \textit{Notes:}
        The figure reports counterfactual outcomes under the GDR reform for cohorts reaching the pre-reform statutory retirement age between 2025 and 2039. The five series correspond to the full sample, lower-income women, lower-income men, higher-income women, and higher-income men. Panel A reports the labor force participation rate; Panel B reports DPP; Panels C--F report changes in Senior welfare, the number of grandchildren, annual intergenerational monetary transfers, and annual grandparental childcare, respectively, relative to the pre-reform benchmark. This benchmark assumes that all Seniors retire at the statutory retirement ages applicable to our sample before the reform: 60 for men and 50 for women. Senior welfare is expressed as a percentage change, monetary transfers and DPP are measured in RMB per Senior-year, and grandparental childcare is measured in hours per Senior-year. Seniors are classified as lower or higher income relative to the median income within each survey wave $\times$ retirement status cell. The appendix provides details of the simulation, additional discussion, and results.
    \end{minipage}
\end{figure}

\clearpage

\begingroup

\setstretch{1.0}

\fontsize{9}{11}\selectfont

\setlength{\tabcolsep}{3pt}

\renewcommand{\arraystretch}{1.13}

\setlength{\LTleft}{\fill}

\setlength{\LTright}{\fill}

\setlength{\LTcapwidth}{\textwidth}

\setlength{\LTpost}{5pt}

\begin{xltabular}{\textwidth}{@{}>{\raggedright\arraybackslash}Xrrrrrrr@{}}

\caption{Alternative Retirement Regimes: 2025--2039 Cohorts}
\label{tab:cf_retirement_regimes}\\

\toprule

Group & Work (\%) & DPP & \(\Delta\mathcal W\) (\%) & \(n^{old}(R=1)\) & \(\Delta n\) & \(\Delta m\) & \(\Delta b\) \\

\midrule

\endfirsthead

\multicolumn{8}{c}{\tablename\ \thetable\ -- continued}\\

\toprule

Group & Work (\%) & DPP & \(\Delta\mathcal W\) (\%) & \(n^{old}(R=1)\) & \(\Delta n\) & \(\Delta m\) & \(\Delta b\) \\

\midrule

\endhead

\midrule

\multicolumn{8}{r}{Continued on next page}\\

\endfoot

\bottomrule

\endlastfoot

\multicolumn{8}{l}{\textit{Panel A. FFR}}\\[3pt]

All & 2.77 & 543 & 0.022 & 1.505 & 0.002 & 840 & \(-68\) \\
Men & 3.93 & 745 & 0.029 & 1.826 & 0.002 & 1,184 & \(-96\) \\
Women & 0.99 & 234 & 0.004 & 1.011 & 0.001 & 313 & \(-25\)\\
Lower income & 1.84 & 248 & 0.020 & 1.538 & 0.001 & 540 & \(-44\) \\
Higher income & 3.70 & 840 & 0.024 & 1.472 & 0.002 & 1,142 & \(-91\) \\
Lower-income men & 2.96 & 362 & 0.030 & 1.893 & 0.001 & 876 & \(-71\) \\
Higher-income men & 4.66 & 1,031 & 0.028 & 1.776 & 0.003 & 1,414 & \(-114\) \\
Lower-income women & 0.65 & 126 & 0.004 & 1.156 & 0.000 & 180 & \(-15\) \\
Higher-income women & 1.52 & 405 & 0.004 & 0.782 & 0.001 & 523 & \(-40\) \\

\addlinespace[6pt]

\multicolumn{8}{l}{\textit{Panel B. MPR}}\\[3pt]

All & 100.00 & 27,450 & \(-5.484\) & 1.505 & \(-0.130\) & \(-9,322\) & 57 \\
Men & 100.00 & 29,171 & \(-5.100\) & 1.826 & \(-0.133\) & \(-9,750\) & 79 \\
Women & 100.00 & 24,802 & \(-6.564\) & 1.011 & \(-0.126\) & \(-8,664\) & 22\\
Lower income & 100.00 & 22,514 & \(-4.783\) & 1.538 & \(-0.118\) & \(-7,419\) & \(-43\) \\
Higher income & 100.00 & 32,406 & \(-6.222\) & 1.472 & \(-0.142\) & \(-11,233\) & 156 \\
Lower-income men & 100.00 & 23,608 & \(-4.310\) & 1.893 & \(-0.118\) & \(-7,526\) & \(-37\) \\
Higher-income men & 100.00 & 33,332 & \(-5.732\) & 1.776 & \(-0.143\) & \(-11,413\) & 166 \\
Lower-income women & 100.00 & 21,339 & \(-5.621\) & 1.156 & \(-0.117\) & \(-7,304\) & \(-49\) \\
Higher-income women & 100.00 & 30,298 & \(-8.828\) & 0.782 & \(-0.139\) & \(-10,823\) & 135 \\

\addlinespace[6pt]

\multicolumn{8}{l}{\textit{Panel C. MFT}}\\[3pt]

All & 100.00 & 27,450 & \(-4.629\) & 1.505 & \(-0.105\) & 5,452 & \(-1,145\) \\
Men & 100.00 & 29,171 & \(-4.106\) & 1.826 & \(-0.101\) & 6,108 & \(-1,179\) \\
Women & 100.00 & 24,802 & \(-6.101\) & 1.011 & \(-0.112\) & 4,443 & \(-1,092\)\\
Lower income & 100.00 & 22,514 & \(-4.506\) & 1.538 & \(-0.108\) & 4,925 & \(-1,118\) \\
Higher income & 100.00 & 32,406 & \(-4.757\) & 1.472 & \(-0.102\) & 5,981 & \(-1,173\) \\
Lower-income men & 100.00 & 23,608 & \(-3.891\) & 1.893 & \(-0.104\) & 5,672 & \(-1,157\) \\
Higher-income men & 100.00 & 33,332 & \(-4.277\) & 1.776 & \(-0.099\) & 6,434 & \(-1,196\) \\
Lower-income women & 100.00 & 21,339 & \(-5.596\) & 1.156 & \(-0.114\) & 4,124 & \(-1,076\) \\
Higher-income women & 100.00 & 30,298 & \(-7.314\) & 0.782 & \(-0.108\) & 4,949 & \(-1,119\) \\

\end{xltabular}

\noindent\begin{minipage}{\textwidth}

\fontsize{9}{11}\selectfont

\textit{Notes:} Entries are equally weighted means across cohorts reaching the pre-reform statutory retirement age between 2025 and 2039. Work denotes the share of Seniors who remain employed. Under FFR and MFT, employed Seniors work eight hours per day; under MPR, all Seniors work four hours per day, corresponding to a 50\% full-time-equivalent employment rate. Pension benefits begin upon full retirement in all three regimes. \(n^{old}(R=1)\) denotes the simulated number of grandchildren in the pre-reform benchmark, which assumes that all Seniors retire at the statutory retirement ages applicable to our sample before the reform: 60 for men and 50 for women. \(\Delta n\), \(\Delta m\), and \(\Delta b\) denote changes in the number of grandchildren, intergenerational monetary transfers, and grandparental childcare relative to this benchmark. \(\Delta\mathcal W\) denotes the percentage change in mean Senior welfare. DPP and \(\Delta m\) are measured in RMB per Senior-year, and \(\Delta b\) in hours per Senior-year. Welfare comparisons use a fixed common support across retirement regimes. Seniors are classified as lower or higher income relative to the median income within each survey wave $\times$ retirement status cell.

\end{minipage}

\endgroup

\FloatBarrier

\clearpage
\subsection{Childcare Subsidies and Policy Targeting}
\label{app:childcare_policy}

\begingroup
\setstretch{1.05}
\scriptsize
\setlength{\tabcolsep}{2.0pt}
\renewcommand{\arraystretch}{1.00}
\setlength{\LTleft}{\fill}
\setlength{\LTright}{\fill}
\setlength{\LTcapwidth}{\textwidth}
\setlength{\LTpost}{4pt}
\begin{xltabular}{\textwidth}{@{}>{\raggedright\arraybackslash}Xrrrrrrrr@{}}
\caption{Childcare Subsidies and Policy Targeting}\label{tab:cf_childcare_policy_summary}\\
\toprule
Policy / group & Work (\%) & DPP & \(\Delta\mathcal W\) (\%) & \(n^{old}(R=1)\) & \(\Delta n\) & \(\Delta m\) & \(\Delta b\) & Expenditure \\
\midrule
\endfirsthead
\multicolumn{9}{c}{\tablename\ \thetable\ -- continued}\\
\toprule
Policy / group & Work (\%) & DPP & \(\Delta\mathcal W\) (\%) & \(n^{old}(R=1)\) & \(\Delta n\) & \(\Delta m\) & \(\Delta b\) & Expenditure \\
\midrule
\endhead
\midrule
\multicolumn{9}{r}{\footnotesize Continued on next page}\\
\endfoot
\bottomrule
\endlastfoot

\multicolumn{9}{l}{\textit{Panel A. Universal childcare subsidies: 2025--2030 cohorts (period average)}}\\*
No subsidy & 2.80 & 543 & 0.020 & 1.505 & 0.002 & 840 & \(-68\) & 0 \\
5.74\% (\(s_n^*\)) & 3.30 & 646 & 0.650 & 1.505 & 0.044 & 1,954 & \(-423\) & 828 \\
10.00\% subsidy & 3.60 & 747 & 1.200 & 1.505 & 0.079 & 2,778 & \(-715\) & 1,521 \\
12.18\% (\(s_{\mathcal W}^*\)) & 3.80 & 788 & 1.520 & 1.505 & 0.098 & 3,191 & \(-875\) & 1,900 \\
20.00\% subsidy & 4.50 & 921 & 2.870 & 1.505 & 0.176 & 4,688 & \(-1,521\) & 3,397 \\
\addlinespace

\multicolumn{9}{l}{\textit{Panel B. Universal childcare subsidies: 2031--2039 cohorts (period average)}}\\*
No subsidy & 41.80 & 10,221 & \(-1.580\) & 1.505 & \(-0.043\) & 2,467 & \(-488\) & 0 \\
5.74\% (\(s_n^*\)) & 42.20 & 10,302 & \(-0.910\) & 1.505 & 0.000 & 3,506 & \(-844\) & 836 \\
10.00\% subsidy & 42.50 & 10,387 & \(-0.330\) & 1.505 & 0.035 & 4,273 & \(-1,136\) & 1,528 \\
12.18\% (\(s_{\mathcal W}^*\)) & 42.60 & 10,406 & \(0.000\) & 1.505 & 0.054 & 4,649 & \(-1,295\) & 1,905 \\
20.00\% subsidy & 43.10 & 10,505 & 1.410 & 1.505 & 0.133 & 6,033 & \(-1,941\) & 3,387 \\
\addlinespace

\multicolumn{9}{l}{\textit{Panel C. Women-only childcare subsidies: 2025--2030 cohorts (period average)}}\\*
No subsidy & 2.80 & 543 & 0.020 & 1.505 & 0.002 & 840 & \(-68\) & 0 \\
10.00\% subsidy & 2.90 & 581 & 0.310 & 1.505 & 0.021 & 1,346 & \(-232\) & 378 \\
18.97\% (\(s_n^*\)) & 3.10 & 630 & 0.670 & 1.505 & 0.043 & 1,812 & \(-415\) & 806 \\
20.00\% subsidy & 3.20 & 636 & 0.710 & 1.505 & 0.046 & 1,870 & \(-439\) & 861 \\
31.00\% (\(s_{\mathcal W}^*\)) & 3.60 & 746 & 1.370 & 1.505 & 0.082 & 2,459 & \(-737\) & 1,509 \\
\addlinespace

\multicolumn{9}{l}{\textit{Panel D. Women-only childcare subsidies: 2031--2039 cohorts (period average)}}\\*
No subsidy & 41.80 & 10,221 & \(-1.580\) & 1.505 & \(-0.043\) & 2,467 & \(-488\) & 0 \\
10.00\% subsidy & 41.80 & 10,221 & \(-1.230\) & 1.505 & \(-0.022\) & 2,841 & \(-653\) & 386 \\
18.97\% (\(s_n^*\)) & 41.80 & 10,221 & \(-0.800\) & 1.505 & 0.000 & 3,177 & \(-835\) & 800 \\
20.00\% subsidy & 41.80 & 10,221 & \(-0.740\) & 1.505 & 0.003 & 3,215 & \(-859\) & 851 \\
31.00\% (\(s_{\mathcal W}^*\)) & 41.80 & 10,221 & 0.000 & 1.505 & 0.039 & 3,626 & \(-1,154\) & 1,467 \\
\addlinespace

\multicolumn{9}{l}{\textit{Panel E. Lower-income-women childcare subsidies: 2025--2030 cohorts (period average)}}\\*
No subsidy & 2.80 & 543 & 0.020 & 1.505 & 0.002 & 840 & \(-68\) & 0 \\
10.00\% subsidy & 2.80 & 553 & 0.230 & 1.505 & 0.016 & 1,184 & \(-184\) & 272 \\
20.00\% subsidy & 2.90 & 563 & 0.520 & 1.505 & 0.033 & 1,528 & \(-329\) & 615 \\
24.86\% (\(s_n^*\)) & 3.00 & 574 & 0.700 & 1.505 & 0.044 & 1,703 & \(-414\) & 807 \\
38.11\% (\(s_{\mathcal W}^*\)) & 3.20 & 599 & 1.410 & 1.505 & 0.080 & 2,172 & \(-714\) & 1,424 \\
\addlinespace

\multicolumn{9}{l}{\textit{Panel F. Lower-income-women childcare subsidies: 2031--2039 cohorts (period average)}}\\*
No subsidy & 41.80 & 10,221 & \(-1.580\) & 1.505 & \(-0.043\) & 2,467 & \(-488\) & 0 \\
10.00\% subsidy & 41.80 & 10,221 & \(-1.330\) & 1.505 & \(-0.028\) & 2,734 & \(-605\) & 276 \\
20.00\% subsidy & 41.80 & 10,221 & \(-0.990\) & 1.505 & \(-0.010\) & 3,001 & \(-751\) & 607 \\
24.86\% (\(s_n^*\)) & 41.80 & 10,221 & \(-0.780\) & 1.505 & 0.000 & 3,130 & \(-835\) & 790 \\
38.11\% (\(s_{\mathcal W}^*\)) & 41.80 & 10,221 & 0.000 & 1.505 & 0.037 & 3,483 & \(-1,134\) & 1,383 \\
\addlinespace

\end{xltabular}
\noindent\begin{minipage}{\textwidth}
\scriptsize
\textit{Notes:} Panels A and B report universal childcare subsidies; Panels C and D report subsidies restricted to households with female Seniors; and Panels E and F report subsidies restricted to households with lower-income female Seniors. Each pair of panels reports results for the 2025--2030 and 2031--2039 cohorts, respectively. Seniors are classified as lower or higher income relative to the median income within each survey wave $\times$ retirement status cell. Work denotes the Senior labor force participation rate. \(n^{old}(R=1)\) denotes the average number of grandchildren under the benchmark. \(\Delta n\), \(\Delta m\), and \(\Delta b\) denote changes relative to the benchmark in the average number of grandchildren, annual intergenerational monetary transfers, and annual grandparental childcare, respectively. \(\Delta\mathcal W\) denotes the percentage change in the mean Senior welfare index defined in Appendix~\ref{app:counterfactual}. \(s_n^*\) and \(s_{\mathcal W}^*\) denote the subsidy rates required to offset the negative effects of GDR on the number of grandchildren and Senior welfare, respectively, for the 2031--2039 cohorts. Childcare subsidy expenditure is averaged over the full sample, with zero expenditure assigned to ineligible households. Subsidy expenditure, DPP, and \(\Delta m\) are measured in 2018 RMB per Senior-year, and \(\Delta b\) in hours per Senior-year. DPP measure annual pension payments deferred at the evaluation age and do not represent a reduction in lifetime pension expenditure.
\end{minipage}
\endgroup

\FloatBarrier
\clearpage
\section{Official Schedules for the Gradual Increase in the Statutory Retirement Age}
\label{app:official_schedules}

\begingroup
\scriptsize
\begin{singlespace}
\setlength{\LTleft}{\fill}
\setlength{\LTright}{\fill}
\setlength{\tabcolsep}{5pt}
\renewcommand{\arraystretch}{0.94}

\begin{longtable}{@{}rlll@{}}
\caption{Official Statutory Retirement-Age Schedule for Male Employees: One-Month Increase per Four Birth Months}
\label{tab:official_schedule_men}\\
\toprule
Delay (months) & Birth month(s) & Reformed statutory retirement age & Statutory retirement month(s) \\
\midrule
\endfirsthead
\multicolumn{4}{@{}l}{\textit{Table \thetable\ continued}}\\
\toprule
Delay (months) & Birth month(s) & Reformed statutory retirement age & Statutory retirement month(s) \\
\midrule
\endhead
\midrule
\multicolumn{4}{r@{}}{\textit{Continued on next page}}\\
\endfoot
\bottomrule
\endlastfoot
1 & January--April 1965 & 60 years, 1 month & February--May 2025 \\
2 & May--August 1965 & 60 years, 2 months & July--October 2025 \\
3 & September--December 1965 & 60 years, 3 months & December 2025--March 2026 \\
4 & January--April 1966 & 60 years, 4 months & May--August 2026 \\
5 & May--August 1966 & 60 years, 5 months & October 2026--January 2027 \\
6 & September--December 1966 & 60 years, 6 months & March--June 2027 \\
7 & January--April 1967 & 60 years, 7 months & August--November 2027 \\
8 & May--August 1967 & 60 years, 8 months & January--April 2028 \\
9 & September--December 1967 & 60 years, 9 months & June--September 2028 \\
10 & January--April 1968 & 60 years, 10 months & November 2028--February 2029 \\
11 & May--August 1968 & 60 years, 11 months & April--July 2029 \\
12 & September--December 1968 & 61 years & September--December 2029 \\
13 & January--April 1969 & 61 years, 1 month & February--May 2030 \\
14 & May--August 1969 & 61 years, 2 months & July--October 2030 \\
15 & September--December 1969 & 61 years, 3 months & December 2030--March 2031 \\
16 & January--April 1970 & 61 years, 4 months & May--August 2031 \\
17 & May--August 1970 & 61 years, 5 months & October 2031--January 2032 \\
18 & September--December 1970 & 61 years, 6 months & March--June 2032 \\
19 & January--April 1971 & 61 years, 7 months & August--November 2032 \\
20 & May--August 1971 & 61 years, 8 months & January--April 2033 \\
21 & September--December 1971 & 61 years, 9 months & June--September 2033 \\
22 & January--April 1972 & 61 years, 10 months & November 2033--February 2034 \\
23 & May--August 1972 & 61 years, 11 months & April--July 2034 \\
24 & September--December 1972 & 62 years & September--December 2034 \\
25 & January--April 1973 & 62 years, 1 month & February--May 2035 \\
26 & May--August 1973 & 62 years, 2 months & July--October 2035 \\
27 & September--December 1973 & 62 years, 3 months & December 2035--March 2036 \\
28 & January--April 1974 & 62 years, 4 months & May--August 2036 \\
29 & May--August 1974 & 62 years, 5 months & October 2036--January 2037 \\
30 & September--December 1974 & 62 years, 6 months & March--June 2037 \\
31 & January--April 1975 & 62 years, 7 months & August--November 2037 \\
32 & May--August 1975 & 62 years, 8 months & January--April 2038 \\
33 & September--December 1975 & 62 years, 9 months & June--September 2038 \\
34 & January--April 1976 & 62 years, 10 months & November 2038--February 2039 \\
35 & May--August 1976 & 62 years, 11 months & April--July 2039 \\
36 & September--December 1976 & 63 years & September--December 2039 \\
\end{longtable}

\begin{longtable}{@{}rlll@{}}
\caption{Official Statutory Retirement-Age Schedule for Female Employees Previously Subject to Age 55: One-Month Increase per Four Birth Months}
\label{tab:official_schedule_women55}\\
\toprule
Delay (months) & Birth month(s) & Reformed statutory retirement age & Statutory retirement month(s) \\
\midrule
\endfirsthead
\multicolumn{4}{@{}l}{\textit{Table \thetable\ continued}}\\
\toprule
Delay (months) & Birth month(s) & Reformed statutory retirement age & Statutory retirement month(s) \\
\midrule
\endhead
\midrule
\multicolumn{4}{r@{}}{\textit{Continued on next page}}\\
\endfoot
\bottomrule
\endlastfoot
1 & January--April 1970 & 55 years, 1 month & February--May 2025 \\
2 & May--August 1970 & 55 years, 2 months & July--October 2025 \\
3 & September--December 1970 & 55 years, 3 months & December 2025--March 2026 \\
4 & January--April 1971 & 55 years, 4 months & May--August 2026 \\
5 & May--August 1971 & 55 years, 5 months & October 2026--January 2027 \\
6 & September--December 1971 & 55 years, 6 months & March--June 2027 \\
7 & January--April 1972 & 55 years, 7 months & August--November 2027 \\
8 & May--August 1972 & 55 years, 8 months & January--April 2028 \\
9 & September--December 1972 & 55 years, 9 months & June--September 2028 \\
10 & January--April 1973 & 55 years, 10 months & November 2028--February 2029 \\
11 & May--August 1973 & 55 years, 11 months & April--July 2029 \\
12 & September--December 1973 & 56 years & September--December 2029 \\
13 & January--April 1974 & 56 years, 1 month & February--May 2030 \\
14 & May--August 1974 & 56 years, 2 months & July--October 2030 \\
15 & September--December 1974 & 56 years, 3 months & December 2030--March 2031 \\
16 & January--April 1975 & 56 years, 4 months & May--August 2031 \\
17 & May--August 1975 & 56 years, 5 months & October 2031--January 2032 \\
18 & September--December 1975 & 56 years, 6 months & March--June 2032 \\
19 & January--April 1976 & 56 years, 7 months & August--November 2032 \\
20 & May--August 1976 & 56 years, 8 months & January--April 2033 \\
21 & September--December 1976 & 56 years, 9 months & June--September 2033 \\
22 & January--April 1977 & 56 years, 10 months & November 2033--February 2034 \\
23 & May--August 1977 & 56 years, 11 months & April--July 2034 \\
24 & September--December 1977 & 57 years & September--December 2034 \\
25 & January--April 1978 & 57 years, 1 month & February--May 2035 \\
26 & May--August 1978 & 57 years, 2 months & July--October 2035 \\
27 & September--December 1978 & 57 years, 3 months & December 2035--March 2036 \\
28 & January--April 1979 & 57 years, 4 months & May--August 2036 \\
29 & May--August 1979 & 57 years, 5 months & October 2036--January 2037 \\
30 & September--December 1979 & 57 years, 6 months & March--June 2037 \\
31 & January--April 1980 & 57 years, 7 months & August--November 2037 \\
32 & May--August 1980 & 57 years, 8 months & January--April 2038 \\
33 & September--December 1980 & 57 years, 9 months & June--September 2038 \\
34 & January--April 1981 & 57 years, 10 months & November 2038--February 2039 \\
35 & May--August 1981 & 57 years, 11 months & April--July 2039 \\
36 & September--December 1981 & 58 years & September--December 2039 \\
\end{longtable}

\begin{longtable}{@{}rlll@{}}
\caption{Official Statutory Retirement-Age Schedule for Female Employees Previously Subject to Age 50: One-Month Increase per Two Birth Months}
\label{tab:official_schedule_women50}\\
\toprule
Delay (months) & Birth month(s) & Reformed statutory retirement age & Statutory retirement month(s) \\
\midrule
\endfirsthead
\multicolumn{4}{@{}l}{\textit{Table \thetable\ continued}}\\
\toprule
Delay (months) & Birth month(s) & Reformed statutory retirement age & Statutory retirement month(s) \\
\midrule
\endhead
\midrule
\multicolumn{4}{r@{}}{\textit{Continued on next page}}\\
\endfoot
\bottomrule
\endlastfoot
1 & January--February 1975 & 50 years, 1 month & February--March 2025 \\
2 & March--April 1975 & 50 years, 2 months & May--June 2025 \\
3 & May--June 1975 & 50 years, 3 months & August--September 2025 \\
4 & July--August 1975 & 50 years, 4 months & November--December 2025 \\
5 & September--October 1975 & 50 years, 5 months & February--March 2026 \\
6 & November--December 1975 & 50 years, 6 months & May--June 2026 \\
7 & January--February 1976 & 50 years, 7 months & August--September 2026 \\
8 & March--April 1976 & 50 years, 8 months & November--December 2026 \\
9 & May--June 1976 & 50 years, 9 months & February--March 2027 \\
10 & July--August 1976 & 50 years, 10 months & May--June 2027 \\
11 & September--October 1976 & 50 years, 11 months & August--September 2027 \\
12 & November--December 1976 & 51 years & November--December 2027 \\
13 & January--February 1977 & 51 years, 1 month & February--March 2028 \\
14 & March--April 1977 & 51 years, 2 months & May--June 2028 \\
15 & May--June 1977 & 51 years, 3 months & August--September 2028 \\
16 & July--August 1977 & 51 years, 4 months & November--December 2028 \\
17 & September--October 1977 & 51 years, 5 months & February--March 2029 \\
18 & November--December 1977 & 51 years, 6 months & May--June 2029 \\
19 & January--February 1978 & 51 years, 7 months & August--September 2029 \\
20 & March--April 1978 & 51 years, 8 months & November--December 2029 \\
21 & May--June 1978 & 51 years, 9 months & February--March 2030 \\
22 & July--August 1978 & 51 years, 10 months & May--June 2030 \\
23 & September--October 1978 & 51 years, 11 months & August--September 2030 \\
24 & November--December 1978 & 52 years & November--December 2030 \\
25 & January--February 1979 & 52 years, 1 month & February--March 2031 \\
26 & March--April 1979 & 52 years, 2 months & May--June 2031 \\
27 & May--June 1979 & 52 years, 3 months & August--September 2031 \\
28 & July--August 1979 & 52 years, 4 months & November--December 2031 \\
29 & September--October 1979 & 52 years, 5 months & February--March 2032 \\
30 & November--December 1979 & 52 years, 6 months & May--June 2032 \\
31 & January--February 1980 & 52 years, 7 months & August--September 2032 \\
32 & March--April 1980 & 52 years, 8 months & November--December 2032 \\
33 & May--June 1980 & 52 years, 9 months & February--March 2033 \\
34 & July--August 1980 & 52 years, 10 months & May--June 2033 \\
35 & September--October 1980 & 52 years, 11 months & August--September 2033 \\
36 & November--December 1980 & 53 years & November--December 2033 \\
37 & January--February 1981 & 53 years, 1 month & February--March 2034 \\
38 & March--April 1981 & 53 years, 2 months & May--June 2034 \\
39 & May--June 1981 & 53 years, 3 months & August--September 2034 \\
40 & July--August 1981 & 53 years, 4 months & November--December 2034 \\
41 & September--October 1981 & 53 years, 5 months & February--March 2035 \\
42 & November--December 1981 & 53 years, 6 months & May--June 2035 \\
43 & January--February 1982 & 53 years, 7 months & August--September 2035 \\
44 & March--April 1982 & 53 years, 8 months & November--December 2035 \\
45 & May--June 1982 & 53 years, 9 months & February--March 2036 \\
46 & July--August 1982 & 53 years, 10 months & May--June 2036 \\
47 & September--October 1982 & 53 years, 11 months & August--September 2036 \\
48 & November--December 1982 & 54 years & November--December 2036 \\
49 & January--February 1983 & 54 years, 1 month & February--March 2037 \\
50 & March--April 1983 & 54 years, 2 months & May--June 2037 \\
51 & May--June 1983 & 54 years, 3 months & August--September 2037 \\
52 & July--August 1983 & 54 years, 4 months & November--December 2037 \\
53 & September--October 1983 & 54 years, 5 months & February--March 2038 \\
54 & November--December 1983 & 54 years, 6 months & May--June 2038 \\
55 & January--February 1984 & 54 years, 7 months & August--September 2038 \\
56 & March--April 1984 & 54 years, 8 months & November--December 2038 \\
57 & May--June 1984 & 54 years, 9 months & February--March 2039 \\
58 & July--August 1984 & 54 years, 10 months & May--June 2039 \\
59 & September--October 1984 & 54 years, 11 months & August--September 2039 \\
60 & November--December 1984 & 55 years & November--December 2039 \\
\end{longtable}

\end{singlespace}
\endgroup

\clearpage
\bibliographystyle{agsm}
\bibliography{references}

\end{document}